%% file: article.tex
\documentclass{article}

\usepackage{epubtk}
\usepackage{graphicx}
\usepackage{amsfonts}
\usepackage{amsmath}
\usepackage{amssymb}
\usepackage{bm}
\usepackage{dcolumn}
\usepackage{latexsym}
\usepackage{rotating}
\usepackage{eucal}
\usepackage{booktabs}
\usepackage{xspace} 

\newcommand\be{\begin{equation}}
\newcommand\ba{\begin{eqnarray}}
\newcommand\bal{\begin{align}}
\newcommand\eal{\end{align}}
\newcommand\ee{\end{equation}}
\newcommand\ea{\end{eqnarray}}

\begin{document}

\title{Gravitational-Wave Tests of General Relativity with
  Ground-Based Detectors and Pulsar-Timing Arrays}

\def\headertitle{Gravitational-Wave Tests of GR with
  Ground-Based Detectors and Pulsar-Timing Arrays}

\author{%
\epubtkAuthorData{Nicol\'as Yunes}{%
Department of Physics, \\
Montana State University, \\
Bozeman, MO 59717, USA}
{nyunes@physics.montana.edu}
{http://www.physics.montana.edu/faculty/yunes/public_html/home.html}
\and
\epubtkAuthorData{Xavier Siemens}{%
Center for Gravitation, Cosmology, and Astrophysics\\
Department of Physics,  \\
University of Wisconsin-Milwaukee, P.\ O.\ Box 413,\\
Milwaukee, WI 53201, USA}
{siemens@gravity.phys.uwm.edu}
{http://www.lsc-group.phys.uwm.edu/~siemens/}
}

\date{}
\maketitle

%--------------------------------------------------------------------------------
\begin{abstract}
This review is focused on tests of Einstein's theory of General
Relativity with gravitational waves that are detectable by
ground-based interferometers and pulsar-timing experiments. Einstein's
theory has been greatly constrained in the quasi-linear,
quasi-stationary regime, where gravity is weak and velocities are
small. Gravitational waves will allow us to probe a complimentary, yet
previously unexplored regime: the non-linear and dynamical
\emph{strong-field regime}. Such a regime is, for example, applicable
to compact binaries coalescing, where characteristic velocities can
reach fifty percent the speed of light and gravitational fields are large and dynamical. 
This review begins with the theoretical basis and the predicted
gravitational-wave observables of modified gravity theories. The
review continues with a brief description of the detectors, including
both gravitational-wave interferometers and pulsar-timing arrays,
leading to a discussion of the data analysis formalism that is
applicable for such tests. The review ends with a discussion of
gravitational-wave tests for compact binary systems.
\end{abstract}
%--------------------------------------------------------------------------------

\epubtkKeywords{General relativity, Gravitational waves, Pulsar
  timing, Experimental tests, Observational test, Scalar-tensor
  gravity, Brans--Dicke theory, String theory, Quantum gravity,
  Quadratic gravity, Massive gravity, Alternative theories, Modified
  gravity, Chern--Simons theory, Parity violation, Compact binaries,
  Non-commutative geometry, Extra dimensions, post-Einsteinian
  framework}
%% \epubtkKeywords{General Relativity, Gravitational Waves, Pulsar Timing, Experimental Tests, 
%% Observational Test, Scalar Tensor, Brans-Dicke, String Theory, Quantum Gravity, 
%% Quadratic Gravity, Massive Gravity, Alternative Theories, Modified Gravity, Chern-Simons, 
%% Parity Violation, Compact Binaries, Non-Commutative Geometry, Extra Dimensions, Bayesian, Fisher,
%% Coalescence, Burst, Stochastic, post-Einsteinian, Ringdown}

%--------------------------------------------------------------------------------
%TOC
\newpage
\tableofcontents

\newpage
\section{Introduction}
% Xavi + Nico -> DONE
\label{section:introduction}
\input{introduction}

%--------------------------------------------------------------------------------
\newpage
\section{Alternative Theories of Gravity}
% Nico -> DONE
\label{section:alt-theories}
\input{alt-theories}

%--------------------------------------------------------------------------------
\newpage
% Xavi
\section{Detectors}
\label{section:detectors}
\input{detectors}

%--------------------------------------------------------------------------------
\newpage
\section{Testing Techniques}
% Xavi
\label{section:techniques}
\input{techniques}

%--------------------------------------------------------------------------------
\newpage
\section{Compact Binary Tests}
% Nico
\label{section:binary-sys-tests}
\input{binary-sys-tests}

%--------------------------------------------------------------------------------
\newpage
\section{Musings About the Future}
% Xavi + Nico
\label{section:musings}
\input{musings}

%==============================================================================
\newpage
\section{Acknowledgements}
\label{section:acknowledgements}

We would like to thank Emanuele Berti, Vitor Cardoso, William Nelson,
Bangalore Sathyaprakash, and Leo Stein for many discussions. We would
also like to thank Laura Sampson and Tyson Littenberg for helping us
write parts of the data analysis sections. Finally, we would like to
thank Matt Adams, Katerina Chatziioannou, Tyson Littenberg, Laura
Sampson, and Kent Yagi for proof-reading earlier versions of this
manuscript.  Nicolas Yunes would like to acknowledge support from NSF
grant PHY-1114374 and NASA grant NNX11AI49G, under sub-award
00001944. 
Xavier Siemens would like to acknowledge support from the NSF CAREER
award number 0955929, the 
PIRE award number 0968126, and award number 0970074. 

%==============================================================================
% Bibliography
%==============================================================================
\newpage
\bibliography{refs}

\end{document}

%% file: introduction.tex
% Nico and Xavi write this
%
%   1.1 The Importance of Testing
%   1.2 Testing General Relativity versus Testing Alternative Theories
%   1.3 Gravitational Wave Tests versus Other Tests of General Relativity	
%------------------------------------------------------------------------------------------------------------------------------
\subsection{The importance of testing}

The era of precision gravitational wave astrophysics is at our doorstep. With it, a plethora of previously unavailable information will flood in, allowing for unprecedented astrophysical measurements and tests of fundamental theories. Nobody would question the importance of more precise astrophysical measurements, but one may wonder whether fundamental tests are truly necessary, considering the many successes of Einstein's theory of General Relativity. Indeed, General Relativity has passed many tests, including Solar System ones, binary pulsar ones and cosmological ones (for a recent review, see~\cite{lrr-2006-3,lrr-2008-9}).  

What all of these tests have in common is that they sample the quasi-stationary, quasi-linear {\emph{weak field}} regime of General Relativity. That is, they sample the regime of spacetime where the gravitational field is weak relative to the mass-energy of the system, the characteristic velocities of gravitating bodies are small relative to the speed of light, and the gravitational field is stationary or quasi-stationary relative to the characteristic size of the system. A direct consequence of this is that gravitational waves emitted by weakly-gravitating, quasi-stationary sources are necessarily extremely weak. To make this more concrete, let us define the gravitational compactness as a measure of the strength of the gravitational field:
\begin{equation}
{\cal{C}} = {\frac{{\cal{M}}}{\cal{R}}}\,,
\end{equation}
where ${\cal{M}}$ is the characteristic mass of the system, ${\cal{R}}$ is the characteristic length scale {\emph{associated with gravitational radiation}}, and henceforth we set $G = c = 1$. For binary systems, the orbital separation serves as this characteristic length scale. The strength of gravitational waves and the mutual gravitational interaction between bodies scale linearly with this compactness measure. Let us also define the characteristic velocities of such a system ${\cal{V}}$ as a quantity related to the rate of change of the gravitational field in the center of mass frame. We can then more formally define the weak field as the region of spacetime where the following two conditions are simultaneously satisfied:
\begin{equation}
{\bf{Weak}}\,{\bf{Field}}: \quad {\cal{C}} \ll 1\,, \quad {\cal{V}} \ll 1\,.
\label{eq:weak-field-def}
\end{equation}
By similarity, the {\emph{strong field}} is defined as the region of spacetime where both conditions in Eq.~\eqref{eq:weak-field-def} are not valid simultaneously\epubtkFootnote{Notice that ``strong-field'' is not synonymous with Planck scale physics in this context. In fact, a stationary black hole would not serve as a probe of the strong-field, even if one were to somehow acquire information about the gravitational potential close to the singularity. This is because any such observation would necessarily be lacking information about the dynamical sector of the gravitational interaction. Planck scale physics is perhaps more closely related to strong-curvature physics.}.

Let us provide some examples. For the Earth-Sun system, ${\cal{M}}$ is essentially the mass of the sun, while ${\cal{R}}$ is the orbital separation, which leads to ${\cal{C}} \approx 9.8 \times 10^{-9}$ and ${\cal{V}} \approx 9.9 \times 10^{-5}$. Even if an object were in a circular orbit at the surface of the Sun, its gravitational compactness would be ${\cal{O}}(10^{-6})$ and its characteristic velocity ${\cal{O}}(10^{-3})$. We thus conclude that all Solar System experiments are necessarily sampling the weak field regime of gravity. Similarly, for the double binary pulsar J0737-3039~\cite{Lyne:2004cj,Kramer:2006nb}, ${\cal{C}} \approx 6 \times 10^{-6}$ and ${\cal{V}} \approx 2 \times 10^{-3}$, where we have set the characteristic length ${\cal{R}}$ to the orbital separation via ${\cal{R}} \approx [M P^{2}/(4 \pi^{2})]^{1/3} \approx 10^{6} \, {\rm{km}}$, where $P = 0.1 \; {\rm{days}}$ is the orbital period and $M \approx 3 M_{\odot}$ is the total mass. Although neutron stars are sources of strong gravity (the ratio of their mass to their radius is of order one tenth), binary pulsars are most sensitive to the quasi-static part of the post-Newtonian effective potential or to the leading-order (so-called Newtonian piece) of the radiation-reaction force. On the other hand, in compact binary coalescence the gravitational compactness and the characteristic speed can reach values much closer to unity. Therefore, although in much of the pulsar-timing literature binary pulsar timing is said to allow for strong-field tests of gravity, gravitational information during compact binary coalescence would be a much stronger-field test.  

Even though current data does not give us access to the full non-linear and dynamical regime of General Relativity, Solar System tests and binary pulsar observations have served (and will continue to serve) an invaluable role in testing Einstein's theory. Solar system tests effectively cured an outbreak of modified gravity theories in the 1970s and 1980s, as summarized for example in~\cite{lrr-2006-3}. Binary pulsars were crucial as the first indirect detectors of gravitational waves, and later to kill certain theories, like Rosen's bimetric gravity~\cite{Rosen:1974ua}, and heavily constrain others that predict dipolar energy loss, as we will see in Sections~\ref{section:alt-theories} and~\ref{section:binary-sys-tests}. Binary pulsars are probes of General Relativity in a certain sector of the strong-field: in the dynamical but quasi-linear sector, verifying that compact objects move as described by a perturbative, post-Newtonian analysis to leading-order. Binary pulsars can be used to test General Relativity in the ``strong-field'' in the only sense that they probe non-linear stellar-structure effects, but they can say very little to nothing about non-linear radiative effects. Similarly, future electromagnetic observations of black hole accretion disks may probe General Relativity in another strong-field sector: the non-linear but fully stationary regime, verifying that black holes are described by the Kerr metric. As of the writing of this review article, only gravitational waves will allow for tests of General Relativity in the full strong-field regime, where gravity is both heavily non-linear and inherently dynamical. 

No experiments exist to date that validate Einstein's theory of General Relativity in the highly-dynamical, strong field region. Due to previous successes of General Relativity, one might consider such validation unnecessary. However, as most scientists would agree, the role of science is to {\emph{predict and verify}} and not to assume without proof. Moreover, the incompleteness of General Relativity in the quantum regime, together with the somewhat unsatisfactory requirement of the dark sector of cosmology (including dark energy and dark matter), have prompted more than one physicist to consider deviations from General Relativity more seriously. Gravitational waves will soon allow us to verify Einstein's theory in a regime previously inaccessible to us, and as such, these tests are invaluable.   

In many areas of physics, however, General Relativity is so ingrained that questioning its validity (even in a regime where Einstein's theory has not yet been validated) is synonymous with heresy. Dimensional arguments are usually employed to argue that any quantum gravitational correction will necessarily and unavoidably be unobservable with gravitational waves, as the former are expected at a (Planck) scale that is inaccessible to gravitational wave detectors. This rationalization is dangerous, as it introduces a theoretical bias in the analysis of new observations of the Universe, thus quenching the potential for new discoveries. For example, if astrophysicists had followed such a rationalization when studying supernova data, they would not have discovered that the Universe is expanding. Dimensional arguments suggest that the cosmological constant is over 100 orders of magnitude larger than the value required to agree with observations. When observing the Universe for the first time in a completely new way, it seems more conservative to remain {\emph{agnostic}} about what is expected and what is not, thus allowing the data itself to guide our efforts toward theoretically understanding the gravitational interaction. 

%------------------------------------------------------------------------------------------------------------------------------
\subsection{Testing General Relativity versus testing alternative theories}

When {\emph{testing General Relativity}}, one considers Einstein's theory as a {\emph{null hypothesis}} and searches for generic deviations. On the other hand, when {\emph{testing alternative theories}} one starts from a particular modified gravity model, develops its equations and solutions and then predicts certain observables that then might or might not agree with experiment. Similarly, one may define a {\emph{bottom-up}} approach versus a {\emph{top-down}} approach. In the former, one starts from some observables in an attempt to discover fundamental symmetries that may lead to a more complete theory, as was done when constructing the standard model of elementary particles. On the other hand, a top-down approach starts from some fundamental theory and then derives its consequence.

Both approaches possess strengths and weaknesses. In the top-down approach one has complete control over the theory under study, being able to write down the full equations of motion, answer questions about well-posedness and stability of solutions, and predict observables. But as we shall see in Section~\ref{section:alt-theories}, carrying out such an approach can be quixotic within any one model. What is worse, the lack of a complete and compelling alternative to General Relativity makes choosing a particular modified theory difficult.

Given this, one might wish to attempt a bottom-up approach, where one considers a set of principles one wishes to test without explicit mention to any particular theory. One usually starts by assuming General Relativity as a null-hypothesis and then considers deformations away from General Relativity. The hope is that experiments will be sensitive to such deformations, thus either constraining the size of the deformations or pointing toward a possible inconsistency. But if experiments do confirm a General Relativity deviation, a bottom-up approach fails at providing a given particular action from which to derive such a deformation. In fact, there can be several actions that lead to similar deformations, all of which can be consistent with the data within its experimental uncertainties. 

Nonetheless, both approaches are complementary. The bottom-up approach draws inspiration from particular examples carried out in the top-down approach. Given a plausible measured deviation from General Relativity within a bottom-up approach, one will still need to understand what plausible top-down theories can lead to such deviations. From this standpoint, then, both approaches are intrinsically intertwined and worth pursuing. 

%------------------------------------------------------------------------------------------------------------------------------
\subsection{Gravitational-wave tests versus other tests of General Relativity}

Gravitational wave tests differ from other tests of General Relativity in many ways. Perhaps one of the most important differences is the spacetime regime gravitational waves sample. Indeed, as already mentioned, gravitational waves have access to the most extreme gravitational environments in Nature. Moreover, gravitational waves travel essentially unimpeded from their source to Earth, and thus, they do not suffer from issues associated with obscuration. Gravitational waves also exist in the absence of luminous matter, thus allowing us to observe electromagnetically dark objects, such as black hole inspirals.

This last point is particularly important as gravitational waves from inspiral black hole binaries are one of the cleanest astrophysical systems in Nature. In the last stages of inspiral, when such gravitational waves would be detectable by ground-based interferometers, the evolution of binary black holes is essentially unaffected by any other matter or electromagnetic fields present in the system. As such, one does not need to deal with uncertainties associated with astrophysical matter. Unlike other tests of General Relativity, such as those attempted with accretion disk observations, binary black hole gravitational wave tests may well be the cleanest probes of Einstein's theory.  

Of course, what is an advantage here, can be also a huge disadvantage in another context. Gravitational waves from compact binaries are intrinsically transient (they turn on for a certain amount of time and then shut off). This is unlike binary pulsar systems, for which astrophysicists have already collected tens of years of data. Moreover, gravitational wave tests rely on specific detections that cannot be anticipated before hand. This is in contrast to Earth-based laboratory experiments, where one has complete control over the experimental setup. Finally, the intrinsic feebleness of gravitational waves makes detection a very difficult task that requires complex data analysis algorithms to extract signals from the noise. As such, gravitational wave tests are limited by the signal-to-noise ratio and affected by systematics associated with the modeling of the waves, issues that are not as important in other loud astrophysical systems.  

%------------------------------------------------------------------------------------------------------------------------------
\subsection{Ground-based vs space-based and interferometers vs pulsar timing}

This review article focuses only on ground-based detectors, by which we mean both gravitational wave interferometers, such as the Laser Interferometer Gravitational Observatory (LIGO)~\cite{Abramovici:1992ah,Abbott:2007kv,Harry:2010zz}, Virgo~\cite{Acernese:2005yh,Acernese:2007zze} and the Einstein Telescope (ET)~\cite{Punturo:2010zz,Sathyaprakash:2012jk}, as well as pulsar timing arrays (for a recent review of gravitational wave tests of General Relativity with space-based detectors, see~\cite{Gair:2012nm,Yagi:2013du}). Ground-based detectors have the limitation of being contaminated by man-made and Nature-made noise, such as ground and air traffic, logging, earthquakes, ocean tides and waves, which are clearly absent in space-based detectors. Ground-based detectors, however, have the clear benefit that they can be continuously upgraded and repaired in case of malfunction, which is obviously not possible with space-based detectors. 

As far as tests of General Relativity are concerned, there is a drastic difference in space-based and ground-based detectors: the gravitational wave frequencies these detectors are sensitive to. For various reasons that we will not go into, space-based interferometers are likely to have million kilometer long arms, and thus, be sensitive in the milli-Hz band. On the other hand, ground-based interferometers are bounded to the surface and curvature of the Earth, and thus, they have kilometer-long arms and are sensitive in the deca- and hecta-Hz band. Different types of interferometers are then sensitive to different types of gravitational wave sources. For example, when considering binary coalescences, ground-based interferometers are sensitive to late inspirals and mergers of neutron stars and stellar-mass black holes, while space-based detectors will be sensitive to supermassive black hole binaries with masses around $10^{5} M_{\odot}$.

The impact of a different population of sources in tests of General Relativity depends on the particular modified gravity theory considered. When studying quadratic gravity theories, as we will see in Section~\ref{section:alt-theories}, the Einstein-Hilbert action is modified by introducing higher order curvature operators, which are naturally suppressed by powers of the inverse of the radius of curvature. Thus, space-based detectors will not be ideal at constraining these theories, as the radius of curvature of supermassive black holes is much larger than that of stellar-mass black holes at merger. Moreover, space-based detectors will not be sensitive to neutron star binary coalescences; they are sensitive to supermassive black hole/neutron star coalescences, where the radius of curvature of the system is controlled by the supermassive black hole. 

On the other hand, space-based detectors are unique in their potential to probe the spacetime geometry of supermassive black holes through gravitational waves emitted during extreme mass-ratio inspirals. These inspirals consist of a stellar-mass compact object in a generic decaying orbit around a supermassive black hole. Such inspirals produce millions of cycles of gravitational waves in the sensitivity band of space-based detectors (in fact, they can easily out-live the observation time!). Therefore, even small changes to the radiation-reaction force, or to the background geometry, can lead to noticeable effects in the waveform observable and thus strong tests of General Relativity, albeit constrained to the radius of curvature of the supermassive black hole. For recent work on such systems and tests, see~\cite{AmaroSeoane:2007aw,Ryan:1995wh,Ryan:1997hg,Kesden:2004qx,Glampedakis:2005cf,Barack:2006pq,Li:2007qu,Gair:2007kr,Sopuerta:2009iy,Yunes:2009ry,Apostolatos:2009vu,LukesGerakopoulos:2010rc,Gair:2011ym,Contopoulos:2011dz,Canizares:2012ji,Gair:2012nm}.

Space-based detectors also have the advantage of range, which is particularly important when considering theories where gravitons do not travel at light speed~\cite{Mirshekari:2011yq}. Space-based detectors have a horizon distance much larger than ground-based detectors; the former can see black hole mergers to redshifts of order 10 if there are any at such early times in the Universe, while the latter are confined to events within redshift 1. Gravitational waves emitted from distant regions in spacetime need a longer time to propagate from the source to the detectors. Thus, theories that modify the propagation of gravitational waves will be best constrained by space-based type systems. Of course, such theories are also likely to modify the generation of gravitational waves, which ground-based detectors should also be sensitive to.  

Another important difference between detectors is in their response to an impinging gravitational wave. Ground-based detectors, as we will see in Section~\ref{section:detectors}, cannot separate between the two possible scalar modes (the longitudinal and the breathing modes) of metric theories of gravity, due to an intrinsic degeneracy in the response functions. Space-based detectors in principle also possess this degeneracy, but they may be able to break it through Doppler modulation if the interferometer orbits the Sun. Pulsar timing arrays, on the other hand, lack this degeneracy altogether, and thus, they can in principle constrain the existence of both modes independently. 

Pulsar timing arrays differ from interferometers in their potential to test General Relativity mostly by the frequency space they are most sensitive to. The latter can observe the late inspiral and merger of compact binaries, while the former is restricted to the very early inspiral. This is why pulsar timing arrays do not need very accurate waveform templates that account for the highly-dynamical and non-linear nature of gravity to detect gravitational waves; leading-order quadrupole waveforms are sufficient~\cite{Corbin:2010kt}. In turn, this implies that pulsar timing arrays cannot constrain theories that only deviate significantly from General Relativity in the late inspiral, while they are exceptionally well-suited for constraining low-frequency deviations.  

We therefore see a complementarity emerging: different detectors can test General Relativity in different complementary regimes:
\begin{itemize}
\item Ground-based detectors are best at constraining higher-curvature type modified theories that deviate from General Relativity the most in the late inspiral and merger phase. 
\item Space-based detectors are best at constraining modified graviton dispersion relations and the geometry of supermassive compact objects.
\item Pulsar-timing arrays are best at independently constraining the existence of both scalar modes and any deviation from General Relativity that dominates at low orbital frequencies. 
\end{itemize}
Through the simultaneous implementation of all these tests, General Relativity can be put on a much firmer footing in all phases of the strong-field regime.   

%------------------------------------------------------------------------------------------------------------------------------
\subsection{Notation and conventions}

We follow mainly the notation of~\cite{Misner:1973cw}, where Greek indices stand for spacetime coordinates and spatial indices in the middle of the alphabet $(i,j,k,\ldots)$ for spatial indices. Parenthesis and square brackets in index lists stand for symmetrization and anti-symmetrization respectively, e.g.~$A_{(\mu \nu)} = (A_{\mu \nu} + A_{\nu \mu})/2$ and  $A_{[\mu \nu]} = (A_{\mu \nu} - A_{\nu \mu})/2$. Partial derivatives with respect to spacetime and spatial coordinates are denoted $\partial_{\mu} A = A_{,\mu}$ and $\partial_{i} A = A_{,i}$ respectively. Covariant differentiation is denoted $\nabla_{\mu} A = A_{;\mu}$, multiple covariant derivatives $\nabla^{\mu \nu \ldots} = \nabla^{\mu} \nabla^{\nu} \ldots$, and the curved spacetime D'Alembertian $\square A = \nabla_{\mu} \nabla^{\mu} A$. The determinant of the metric $g_{\mu \nu}$ is $g$, $R_{\mu \nu \delta \sigma}$ is the Riemann tensor, $R_{\mu \nu}$ is the Ricci tensor, $R$ is the Ricci scalar and $G_{\mu \nu}$ is the Einstein tensor. The Levi-Civita tensor and symbol are $\epsilon^{\mu \nu \delta \sigma}$ and $\bar{\epsilon}^{\mu \nu \delta \sigma}$ respectively, with $\bar{\epsilon}^{0123} = +1$ in an orthonormal, positively oriented frame. We use geometric units ($G = c = 1$) and the Einstein summation convention is implied. 

We will be mostly concerned with {\emph{metric theories}}, where gravitational radiation is only defined much farther than a gravitational wave wavelength from the source. In this far- or radiation-zone, the metric tensor can be decomposed as
\begin{equation}
g_{\mu \nu} = \eta_{\mu \nu} + h_{\mu \nu}\,,
\label{eq:hab-def}
\end{equation}
with $\eta_{\mu \nu}$ the Minkowski metric and $h_{\mu \nu}$ the metric perturbation. If the theory considered has additional fields $\phi$, these can also be decomposed in the far-zone as
\begin{equation}
\phi = \phi_{0} + \psi\,,
\label{eq:psi-def}
\end{equation}
with $\phi_{0}$ the background value of the field and $\psi$ a perturbation. With such a decomposition, the field equations for the metric will usually be wave equations for the metric perturbation and for the field perturbation, in a suitable gauge.

%% file: alt-theories.tex
%2. Alternative Theories of Gravity
%   2.1 Desirable Theoretical Properties
%   2.2 Explored Theories
%       2.2.1 Scalar Tensor Theories
%       2.2.2 Massive Graviton Theories and Lorentz Violation
%       2.2.3 Modified Quadratic Gravity
%       2.2.4 Variable G Theories and Extra Dimensions
%       2.2.5 Non-Commutative Geometry
%       2.2.6 Parity Violation
%   2.3 Currently Unexplored Theories [F(R) and its variations, Einstein-Aether, TeVeS, LQG, ST]
%
%------------------------------------------------------------------------------------------------------------------------------

In this section, we discuss the many possible alternative theories that have been studied so far in the context of gravitational wave tests. We begin with a description of the theoretically desirable properties that such theories must have. We then proceed with a review of the theories so far explored as far as gravitational waves are concerned. We will leave out the description of many theories in this chapter, especially those which currently lack a gravitational wave analysis. We will conclude with a brief description of unexplored theories as possible avenues for future research.

%------------------------------------------------------------------------------------------------------------------------------
\subsection{Desirable theoretical properties}
\label{sec:properties}

The space of possible theories is infinite, and thus, one is tempted to reduce it by considering a subspace that satisfies a certain number of properties. Although the number and details of such properties depend on the theorist's taste, there is at least one {\emph{fundamental property}} that all scientists would agree on:
\begin{enumerate}
\item {\bf{Precision Tests}}. The theory must produce predictions that pass all Solar System, binary pulsar, cosmological and experimental tests that have been carried out so far. 
\end{enumerate}
This requirement can be further divided into the following:
\begin{enumerate}
\item[]
\begin{enumerate}
\item[1.a] {\bf{General Relativity Limit}}. There must exist some limit, continuous or discontinuous, such as the weak-field one, in which the predictions of the theory are consistent with those of General Relativity within experimental precision. 
\item[1.b] {\bf{Existence of Known Solutions}}~\cite{walds-presentation}. The theory must admit solutions that correspond to observed phenomena, including but not limited to (nearly) flat spacetime, (nearly) Newtonian stars, and cosmological solutions. 
\item[1.c] {\bf{Stability of Solutions}}~\cite{walds-presentation}. The special solutions described in property (1.b) must be stable to small perturbations on timescales smaller than the age of the Universe. For example, perturbations to (nearly) Newtonian stars, such as impact by asteroids, should not render such solutions unstable.
\end{enumerate}
\end{enumerate}
Of course, these properties  are not all necessarily independent, as the existence of a weak-field limit usually also implies the existence of known solutions. On the other hand, the mere existence of solutions does not necessarily imply that these are stable. 

In addition to these fundamental requirements, one might also wish to require that any new modified gravity theory possesses certain {\emph{theoretical properties}}. These properties will vary depending on the theorist, but the two most common ones are listed below:
\begin{enumerate}
\item[2.] {\bf{Well-motivated from Fundamental Physics}}. There must be some fundamental theory or principle from which the modified theory (effective or not) derives. This fundamental theory would solve some fundamental problem in physics, such as late time acceleration or the incompatibility between quantum mechanics and General Relativity. 
\item[3.] {\bf{Well-posed Initial Value Formulation}}~\cite{walds-presentation}. A wide class of freely specifiable initial data must exist, such that there is a uniquely determined solution to the modified field equations that depends continuously on this data.
\end{enumerate} 
The second property goes without saying at some level, as one expects modified gravity theory constructions to be motivated from some (perhaps yet incomplete) quantum gravitational description of nature. As for the third property, the continuity requirement is necessary because otherwise the theory would lose predictive power, given that initial conditions can only be measured to a finite accuracy. Moreover, small changes in the initial data should not lead to solutions outside the causal future of the data; that is, causality must be preserved. Section~\ref{well-posed} expands on this well-posedness property further. 

One might be concerned that property (2) automatically implies that any predicted deviation to astrophysical observables will be too small to be detectable. This argument usually goes as follows. Any quantum gravitational correction to the action will ``naturally'' introduce at least one new scale, and this, by dimensional analysis, must be the Planck scale. Since this scale is usually assumed to be larger than 1 TeV in natural units (or $10^{-35}$ meters in geometric units), gravitational wave observations will never be able to observe quantum gravitational modifications (see e.g.~\cite{Dubovsky:2007zi} for a similar argument). Although this might be true, in our view such arguments can be extremely dangerous, since they induce a certain theoretical bias in the search for new phenomena. For example, let us consider the supernova observations of the late time expansion of the universe that led to the discovery of the cosmological constant. The above argument certainly fails for the cosmological constant, which on dimensional arguments is over 100 orders of magnitude too small. If the supernova teams had respected this argument, they would not have searched for a cosmological constant in their data. Today, we try to explain our way out of the failure of such dimensional arguments by claiming that there must be some exquisite cancelation that renders the cosmological constant small; but this, of course, came only {\emph{after}} the constant had been measured. One is not trying to argue here that cancelations of this type are common and that quantum gravitational modifications are necessarily expected in gravitational wave observations. Rather, we are arguing that one should remain {\emph{agnostic}} about what is expected and what is not, and allow oneself to be surprised without quenching the potential for new discoveries that will accompany the new era of gravitational wave astrophysics. 

One last property that we will wish to consider for the purposes of this review is the following:
\begin{enumerate}
\item[4.] {\bf{Strong Field Inconsistency}}. The theory must lead to observable deviations from General Relativity in the strong-field regime.
\end{enumerate}
Many modified gravity models have been proposed that pose {\emph{infrared}} or cosmological modifications to General Relativity, aimed at explaining certain astrophysical or cosmological observables, like the late expansion of the Universe. Such modified models usually reduce to General Relativity in the strong-field regime, for example via a Vainshtein like mechanism~\cite{Vainshtein:1972sx,Deffayet:2001uk,Babichev:2013usa} in a static spherically-symmetric context. Extending this mechanism to highly-dynamical strong-field scenarios has not been fully worked out yet~\cite{deRham:2012fg,deRham:2012fw}. Gravitational wave tests of General Relativity, however, are concerned with modified theories that predict deviations in the strong-field, precisely where cosmological modified models do not. Clearly, property (4) is not necessary for a theory to be a valid description of nature. This is because a theory might be identical to General Relativity in the weak and strong fields, yet different at the Planck scale, where it would be unified with quantum mechanics. However, property (4) is a desirable feature if one is to test this theory with gravitational wave observations.

%------------------------------------------------------------------------------------------------------------------------------
\subsection{Well-posedness and effective theories}
\label{well-posed}

Property (3) does not only require the existence of an initial value
formulation, but also that it be well-posed, which is not necessarily
guaranteed. For example, the Cauchy--Kowalewski theorem states that a
system of $n$ partial differential equations for $n$ unknown functions
$\phi_{i}$ of the form $\phi_{i,tt} =
F_{i}(x^{\mu};\phi_{j,\mu};\phi_{j,ti};\phi_{j,ik})$, with $F_{i}$
analytic functions has an initial value formulation (see
e.g.~\cite{Wald:1984rg}). This theorem, however, does not guarantee
continuity or the causal conditions described above. For this, one has
to rely on more general energy arguments, for example constructing a
suitable energy measure that obeys the dominant energy condition and
using it to show well-posedness (see
e.g.~\cite{Hawking:1973uf,Wald:1984rg}). One can show that
second-order, hyperbolic partial differential equations, i.e., equations of the form
\be
\nabla^{\mu} \nabla_{\mu}\phi + A^{\mu} \nabla_{\mu} \phi + B \phi + C = 0\,,
\ee
where $A^{\mu}$ is an arbitrary vector field and $(B,C)$ are smooth functions, have a well-posed initial value formulation. Moreover, the Leray theorem proves that any quasilinear, diagonal, second-order hyperbolic system also has a well-posed initial value formulation~\cite{Wald:1984rg}. 

Proving the well-posedness of an initial value formulation for systems of higher-than-second-order, partial differential equations is much more difficult. In fact, to our knowledge, no general theorems exist of the type described above that apply to third, fourth or higher-order, partial, non-linear and coupled differential equations. Usually, one resorts to the Ostrogradski theorem~\cite{Ostro} to rule out (or at the very least cast serious doubt on) theories that lead to such higher-order field equations.  Ostrogradski's theorem states that Lagrangians that contain terms with higher than first time-derivatives possess a linear instability in the Hamiltonian (see e.g.~\cite{Woodard:2006nt} for a nice review)\epubtkFootnote{Stability and well-posedness are not the same concepts and they do not necessarily imply each other. For example, a well-posed theory might have stable and unstable solutions. For ill-posed theories, it does not make sense to talk about stability of solutions.}. As an example, consider the Lagrangian density
\be
{\cal{L}} = \frac{m}{2} \dot{q}^{2} - \frac{m \omega^{2}}{2} q^{2} - \frac{g m}{2 \omega^{2}} \ddot{q}^{2},
\label{eq:L}
\ee
whose equations of motion
\be
\ddot{q} + \omega^{2} q = - \frac{g}{\omega^{2}} \ddddot{q}\,,
\label{eq:sho-ho}
\ee
obviously contains higher derivatives. The exact solution to this differential equation is
\be
q = A_{1} \cos{k_{1} t} + B_{1} \sin{k_{1} t} + A_{2} \cos{k_{2} t} + B_{2} \sin{k_{2} t}\,, 
\label{eq:exact-sol}
\ee
where $(A_{i},B_{i})$ are constants and $k_{1,2}^{2}/\omega^{2} = (1 \mp \sqrt{1 - 4 g})/(2 g)$. The on-shell Hamiltonian is then
\be
H = \frac{m}{2} \sqrt{1 - 4 g} k_{1}^{2} \left(A_{1}^{2} + B_{1}^{2}\right) - \frac{m}{2} \sqrt{1 - 4 g} k_{2}^{2} \left(A_{2}^{2} + B_{2}^{2}\right)\,,
\ee
from which it is clear that mode $1$ carries positive energy, while mode $2$ carries negative energy and forces the Hamiltonian to be unbounded from below. The latter implies that dynamical degrees of freedom can reach arbitrarily negative energy states. If interactions are present, then an ``empty'' state would instantaneously decay into a collection of positive and negative energy particles, which cannot describe the Universe we live in~\cite{Woodard:2006nt}.  

The Ostrogradski theorem~\cite{Ostro}, however, can be evaded if the Lagrangian in Eq.~\eqref{eq:L} describes an {\emph{effective theory}}, i.e.~a theory that is a truncation of a more general or complete theory. Let us reconsider the particular example above, assuming now that the coupling constant $g$ is an effective theory parameter and Eq.~\eqref{eq:L} is only valid to linear order in $g$. One approach is to search for perturbative solutions of the form $q_{\rm pert} = x_{0} + g x_{1} + \ldots$, which leads to the system of differential equations
\be
\ddot{x}_{n} + \omega^{2} x_{n} = -\frac{1}{\omega^{2}} \ddddot{x}_{n-1}\,,
\ee
with $x_{-1} = 0$. Solving this set of $n$ differential equations and resumming, one finds
\be
q_{\rm pert} = A_{1} \cos{k_{1} t} + B_{1} \sin{k_{1} t}\,. 
\ee
Notice that $q_{\rm pert}$ contains only the positive (well-behaved) energy solution of Eq.~\eqref{eq:exact-sol}, ie.~perturbation theory acts to retain only the well-behaved, stable solution of the full theory in the $g\to0$ limit. One can also think of the perturbative theory as the full theory with additional constraints, ie.~the removal of unstable modes, which is why such an analysis is sometimes called {\emph{perturbative constraints}}~\cite{Cooney:2008wk,Cooney:2009rr,Yunes:2009hc}. 

Another way to approach effective field theories that lead to equations of motion with higher-order derivatives is to apply the method of {\emph{order-reduction}}. In this method, one substitutes the low-order derivatives of the field equations into the high-order derivative part, thus rendering the resulting new theory usually well-posed. One can think of this as a series resummation, where one changes the non-linear behavior of a function by adding uncontrolled, higher-order terms. Let us provide an explicit example by reconsidering the theory in Eq.~\eqref{eq:L}. To lowest order in $g$, the equation of motion is that of a simple harmonic oscillator, 
\be
\ddot{q} + \omega^{2} q = {\cal{O}}(g)\,,
\label{eq:sho-lo}
\ee
which is obviously well-posed. One can then order-reduce the full equation of motion, Eq.~\eqref{eq:sho-ho}, by substituting Eq.~\eqref{eq:sho-lo} into the right hand side of Eq.~\eqref{eq:sho-ho}. Doing so, one obtains the order-reduced equation of motion 
\be
\ddot{q} + \omega^{2} q =  g \ddot{q} + {\cal{O}}(g^{2})\,,
\ee
which clearly now has no high-order derivatives and is well posed provided $g \ll 1$. The solution to this order-reduced differential equation is $q_{\rm pert}$ once more, but with $k_{1}$ linearized in $g \ll 1$. Therefore, the solutions obtained with a perturbative decomposition and with the order-reduced equation of motion are the same to linear order  in $g$. Of course, since an effective field theory is only defined to a certain order in its perturbative parameter, both treatments are equally valid, with the unstable mode effectively removed in both cases.

Such a perturbative analysis, however, can say nothing about the well-posedness of the full theory from which the effective theory derives, or of the effective theory if treated as an exact one (ie.~not as a perturbative expansion). In fact, a well-posed full theory may have both stable and unstable solutions. The arguments presented above only discuss the stability of solutions in an effective theory, and thus, they are self-consistent only within their perturbative scheme. A full theory may have non-perturbative instabilities, but these can only be studied once one has a full (non-truncated in $g$) theory, from which Eq.~\eqref{eq:L} derives as a truncated expansion. Lacking a full quantum theory of Nature, quantum gravitational models are usually studied in a truncated low-energy expansion, where the leading-order piece is General Relativity and higher order pieces are multiplied by a small coupling constant. One can perturbatively explore the well-behaved sector of the truncated theory about solutions to the leading-order theory. Such an analysis, however, is incapable of answering questions about well-posedness or non-linear stability of the full theory.

%------------------------------------------------------------------------------------------------------------------------------
\subsection{Explored theories}

In this subsection we briefly describe the theories that have so far been studied in some depth as far as gravitational waves are concerned. In particular, we focus only on those theories that have been sufficiently studied so that predictions of the expected gravitational waveforms (the observables of gravitational wave detectors) have been obtained for at least a typical source, such as the quasi-circular inspiral of a compact binary.

%------------------------------------------------------------------------------------------------------------------------------
\subsubsection{Scalar-tensor theories}
\label{sec:ST}

Scalar-tensor theories in the Einstein frame~\cite{PhysRev.124.925,Damour:1992we,Faraoni:1998qx,Faraoni:1999hp,Fujii:2003pa,Goenner:2012cq} are defined by the action (where we will restore Newton's gravitational constant $G$ in this section)
\begin{equation}
S_{\rm ST}^{\rm (E)} = \frac{1}{16 \pi G} \int d^{4}x \sqrt{-g} \left[ R - 2 g^{\mu \nu} \left(\partial_{\mu}\varphi\right) \left(\partial_{\nu} \varphi\right) - V(\varphi)\right]
+ S_{\rm mat}[\psi_{\rm mat},A^{2}(\varphi) g_{\mu \nu}],
\label{ST-gen-action}
\end{equation}
where $\varphi$ is a scalar field, $A(\varphi)$ is a coupling function, $V(\varphi)$ is a potential function, $\psi_{\rm mat}$ represents matter degrees of freedom and $G$ is Newton's constant in the Einstein frame. For more details on this theory, we refer the interested reader to the reviews~\cite{lrr-2006-3,Will:1993ns}. One can of course consider more complicated scalar-tensor theories, for example by including multiple scalar fields, but we will ignore such generalizations here.

The Einstein frame is not the frame where the metric governs clocks and rods, and thus, it is convenient to recast the theory in the Jordan frame through the conformal transformation $\tilde{g}_{\mu \nu} = A^{2}(\varphi) g_{\mu \nu}$:
\begin{equation}
S_{\rm ST}^{\rm (J)} = \frac{1}{16 \pi G} \int d^{4}x \sqrt{-\tilde{g}} \left[ \phi \; \tilde{R} - \frac{\omega(\phi)}{\phi} \tilde{g}^{\mu \nu} \left(\partial_{\mu} \phi\right) \left(\partial_{\nu} \phi\right) - \phi^{2} V \right]
+ S_{\rm mat}[\psi_{\rm mat},\tilde{g}_{\mu \nu}],
\label{ST-action}
\end{equation}
where $\tilde{g}_{\mu \nu}$ is the physical metric, the new scalar field $\phi$ is defined via $\phi \equiv A^{-2}$, the coupling field is $\omega(\phi) \equiv (\alpha^{-2} - 3)/2$ and $\alpha \equiv A_{,\varphi}/A$. When cast in the Jordan frame, it is clear that scalar-tensor theories are metric theories (see~\cite{lrr-2006-3} for a definition), since the matter sector depends only on matter degrees of freedom and the physical metric (without a direct coupling of the scalar field). When the coupling $\omega(\phi) = \omega_{\rm BD}$ is constant, then Eq.~\eqref{ST-action} reduces to the massless version of Jordan--Fierz--Brans--Dicke theory~\cite{PhysRev.124.925}.  

The modified field equations in the Einstein frame are
\begin{align}
\square \varphi &= \frac{1}{4} \frac{dV}{d\varphi} - 4 \pi G  \frac{\delta S_{\rm mat}}{\delta \varphi}\,, 
\nonumber \\
G_{\mu \nu} &= 8 \pi G \left( T_{\mu \nu}^{\rm mat} + T_{\mu \nu}^{(\varphi)}\right)\,,
\end{align}
where
\begin{equation}
T_{\mu \nu}^{(\varphi)} = \frac{1}{4 \pi} \left[\varphi_{,\mu} \varphi_{,\nu} - \frac{1}{2} g_{\mu \nu} \varphi_{,\delta} \varphi^{,\delta} - \frac{1}{4} g_{\mu \nu} V(\varphi)\right]
\end{equation}
is a stress-energy tensor for the scalar field. The matter stress--energy tensor is not constructed from the Einstein-frame metric alone, but by the combination $A(\varphi)^{2} g_{\mu \nu}$. In the Jordan frame and neglecting the potential, the modified field equations are~\cite{Will:1993ns}
\begin{align}
\tilde{\square} \phi &= \frac{1}{3 + 2 \omega(\phi)} \left( 8 \pi T^{\rm mat} - \frac{d \omega}{d\phi} \tilde{g}^{\mu \nu} \phi_{,\mu} \phi_{,\nu} \right),
\nonumber \\
\tilde{G}_{\mu \nu} &= \frac{8 \pi G}{\phi} T_{\mu \nu}^{\rm mat} + \frac{\omega}{\phi^{2}} \left(\phi_{,\mu} \phi_{,\nu} 
- \frac{1}{2} \tilde{g}_{\mu \nu} \tilde{g}^{\sigma \rho} \phi_{,\sigma} \phi_{,\rho} \right)
+ \frac{1}{\phi} \left( \phi_{,\mu \nu} - \tilde{g}_{\mu \nu} \tilde{\square} \phi \right),
\end{align}
where $T^{\rm mat}$ is the trace of the matter stress-energy tensor $T_{\mu \nu}^{\rm mat}$ constructed from the physical metric $\tilde{g}_{\mu \nu}$. The form of the modified field equations in Jordan frame suggest that in the weak-field limit one may consider scalar-tensor theories as modifying Newton's gravitational constant via $G \to G(\phi) = G/\phi$. 

Using the decompositions of Eqs.~\eqref{eq:hab-def}-\eqref{eq:psi-def}, the field equations of massless Jordan--Fierz--Brans--Dicke theory can be linearized in the Jordan frame to find (see e.g.~\cite{Will:1989sk})
\begin{equation}
\square_{\eta} \theta^{\mu \nu} = -16 \pi \tau^{\mu \nu}\,,
\qquad
\square_{\eta} \psi = -16 \pi S\,,
\label{eq:ST-EOM}
\end{equation}
where $\square_{\eta}$ is the D'Alembertian operator of flat spacetime, we have defined a new metric perturbation
\begin{equation}
\theta^{\mu \nu} = h^{\mu \nu} - \frac{1}{2} \eta^{\mu \nu} h - \frac{\psi}{\phi_{0}} \eta^{\mu \nu}\,,
\end{equation}
ie.~the metric perturbation in the Einstein frame, 
with $h$ the trace of the metric perturbation and
\begin{align}
\tau^{\mu \nu} &= \phi_{0}^{-1} T^{\mu \nu}_{\rm mat} + t^{\mu \nu}\,,
\\
S &= - \frac{1}{6 + 4 \omega_{BD}} \left(T^{\rm mat} - 3 \phi  \frac{\partial T^{\rm mat}}{\partial \phi}\right) \left(1 - \frac{\theta}{2} - \frac{\psi}{\phi_{0}} \right) - \frac{1}{16 \pi} \left(\psi_{,\mu \nu} \theta^{\mu \nu} + \frac{1}{\phi_{0}} \phi_{,\mu} \psi^{,\mu} \right)\,,
\end{align}
with cubic remainders in either the metric perturbation or the scalar perturbation. The quantity $\partial T^{\rm mat} / \partial \phi$ arises in an effective point-particle theory, where the matter action is a functional of both the Jordan-frame metric and the scalar field. The quantity $t^{\mu \nu}$ is a function of quadratic or higher order in $\theta^{\mu \nu}$ or $\psi$. These equations can now be solved given a particular physical system, as done for quasi-circular binaries in~\cite{Will:1989sk,Saijo:1997wu,Ohashi:1996uz}. Given the above evolution equations, Jordan--Fierz--Brans--Dicke theory possesses a scalar (spin-0) mode, in addition to the two transverse-traceless (spin-2) modes of General Relativity, ie.~Jordan--Fierz--Brans--Dicke theory is of Type $N_{3}$ in the $E(2)$ classification~\cite{Eardleyprd,lrr-2006-3}. 

Let us now discuss whether scalar-tensor theories satisfy the properties discussed in Section~\ref{sec:properties}. Massless Jordan--Fierz--Brans--Dicke theory agrees with all known experimental tests provided $\omega_{\rm BD} > 4 \times 10^{4}$, a bound imposed by the tracking of the Cassini spacecraft through observations of the Shapiro time delay~\cite{Bertotti:2003rm}. Massive Jordan--Fierz--Brans--Dicke theory has been recently constrained to $\omega_{\rm BD} > 4 \times 10^{4}$ and $m_{\rm s} < 2.5 \times 10^{-20} \; {\textrm{eV}}$, with $m_{\rm s}$ the mass of the scalar field~\cite{Perivolaropoulos:2009ak,Alsing:2011er}. Of course, these bounds are not independent, as when $m_{\rm s} \to 0$ one recovers the standard massless constraint, while when $m_{\rm s} \to \infty$, $\omega_{\rm BD}$ cannot be bounded as the scalar becomes non-dynamical. Observations of the Nordtvedt effect with Lunar Laser Ranging observations, as well as observations of the orbital period derivative of white-dwarf/neutron-star binaries, yield similar constraints~\cite{Damour:1996ke,Damour:1998jk,Alsing:2011er,Freire:2012mg}. Neglecting any homogeneous, cosmological solutions to the scalar-field evolution equation, it is clear that in the limit $\omega \to \infty$ one recovers General Relativity, ie.~scalar-tensor theories have a continuous limit to Einstein's theory, but see~\cite{Faraoni:1999yp} for caveats for certain spacetimes. Moreover, Refs.~\cite{Salgado:2008xh,2007CQGra..24.5667L,Wald:1984rg} have verified that scalar-tensor theories with minimal or non-minimal coupling in the Jordan frame can be cast in a strongly-hyperbolic form, and thus, they possess a well-posed initial value formulation. Therefore, scalar-tensor theories possess both properties (1) and (3). 

Scalar-tensor theories also possess property (2), since they can be derived from the low-energy limit of certain string theories. The integration of string quantum fluctuations leads to a higher-dimensional string theoretical action that reduces locally to a field theory similar to a scalar-tensor one~\cite{Garay:1992ej,1985NuPhB.261....1F}, the mapping being $\phi = e^{-2 \psi}$, with $\psi$ one of the string moduli fields~\cite{Damour:1994zq,Damour:1994ya}. Moreover, scalar-tensor theories can be mapped to $f(R)$ theories, where one replaces the Ricci scalar by some functional of $R$. In particular, one can show that $f(R)$ theories are equivalent to Brans--Dicke theory with $\omega_{\rm BD} = 0$, via the mapping $\phi = df(R)/dR$ and $V(\phi) = R \; df(R)/dR - f(R)$~\cite{Chiba:2003ir,Sotiriou:2006hs}. For a recent Living Reviews article on this topic, see~\cite{lrr-2010-3}.

Black holes and stars continue to exist in scalar-tensor theories. Stellar configurations are modified from their General Relativity profile~\cite{Will:1989sk,Damour:1996ke,Harada:1997mr,Harada:1998ge,Tsuchida:1998jw,Damour:1998jk,Sotani:2004rq,DeDeo:2003ju,Sotani:2012eb,Horbatsch:2010hj}, while black holes are not, provided one neglects homogeneous, cosmological solutions to the scalar field evolution equation. Indeed, Hawking~\cite{Hawking:1972qk,Dykla:1972zz,hawking-uniqueness0,carter-uniqueness,israel2,robinson} has proved that Brans--Dicke black holes that are stationary and the endpoint of gravitational collapse are identical to those of General Relativity. This proof has recently been extended to a general class of scalar-tensor models~\cite{Sotiriou:2011dz}. That is, stationary black holes radiate any excess ``hair'', ie.~additional degrees of freedom, after gravitational collapse, a result sometimes referred to as the {\emph{no-hair}} theorem for black holes in scalar-tensor theories. This result has recently been extended even further to allow for quasi-stationary scenarios in generic scalar-tensor theories through the study of extreme-mass ratio inspirals~\cite{Yunes:2011aa} (small black hole in orbit around a much larger one), post-Newtonian comparable-mass inspirals~\cite{Mirshekari:2013vb} and numerical simulations of comparable-mass black hole mergers~\cite{Healy:2011ef,Berti:2013gfa}.

Damour and Esposito-Farese~\cite{Damour:1992we,Damour:1993hw} proposed a different type of scalar-tensor theory, one that can be defined by the action in Eq.~\eqref{ST-action} but with the conformal factor $A(\varphi) = e^{\alpha \varphi + \beta \varphi^{2}/2}$ or the coupling function $\omega(\phi) = -3/2 - 2\pi G/(\beta \log{\phi})$, where $\alpha$ and $\beta$ are constants. When $\beta = 0$ one recovers standard Brans-Dicke theory. When $\beta \lesssim -4$, non-perturbative effects that develop if the gravitational energy is large enough can force neutron stars to spontaneously acquire a non-trivial scalar field profile, to {\emph{spontaneously scalarize}}. Through this process, binary neutron stars that initially had no scalar hair in their early inspiral would acquire it before they merge, when their binding energy exceeds some threshold~\cite{Barausse:2012da}. Binary pulsar observations have constrained this theory in the $(\alpha,\beta)$ space; very roughly speaking $\beta > -4$ and $\alpha < 10^{-2}$~\cite{Damour:1996ke,Damour:1998jk,Freire:2012mg}

As for property (4), scalar tensor theories are not built with the aim to introduce strong-field corrections to General Relativity\epubtkFootnote{The process of spontaneous scalarization in a particular type of scalar-tensor theory~\cite{Damour:1992we,Damour:1993hw} does introduce strong-field modifications because it induces non-perturbative corrections that can affect the structure of neutron stars. These subclass of scalar-tensor theories would satisfy property (4).}. Instead, they naturally lead to modifications of Einstein's theory in the weak-field, modifications that dominate in scenarios with sufficiently weak gravitational interactions. Although this might seem strange, it is natural if one considers, for example, one of the key modifications introduced by scalar-tensor theories: the emission of dipolar gravitational radiation. Such dipolar emission dominates over the General Relativistic quadrupolar emission for systems in the weak to intermediate field regime, such as in binary pulsars or in the very early inspiral of compact binaries. Therefore, one would expect scalar-tensor theories to be best constrained by experiments or observations of weakly-gravitating systems, as it has recently been explicitly shown in~\cite{Yunes:2011aa}.  

%------------------------------------------------------------------------------------------------------------------------------
\subsubsection{Massive graviton theories and Lorentz violation}
\label{sec:MG-LV}

Massive graviton theories are those in which the gravitational
interaction is propagated by a massive gauge boson, i.e., a graviton with mass $m_{g} \neq 0$ or Compton wavelength $\lambda_{g} \equiv h/(m_{g} c) < \infty$. Einstein's theory predicts massless gravitons and thus gravitational propagation at light speed, but if this were not the case, then a certain delay would develop between electromagnetic and gravitational signals emitted simultaneously at the source. Fierz and Pauli~\cite{Fierz:1939ix} were the first to write down an action for a free massive graviton, and ever since then, much work has gone into the construction of such models. For a detailed review, see e.g.~\cite{Hinterbichler:2011tt}.

Gravitational theories with massive gravitons are somewhat well-motivated from a fundamental physics perspective, and thus, one can say they possess property (2). Indeed, in loop quantum cosmology~\cite{Ashtekar:2003hd,Bojowald:2006da}, the cosmological extension to loop quantum gravity, the graviton dispersion relation acquires holonomy corrections during loop quantization that endow the graviton with a mass~\cite{Bojowald:2007cd} $m_{g} = \Delta^{-1/2} \gamma^{-1} (\rho/\rho_{c})$, with $\gamma$ the Barbero--Immirzi parameter, $\Delta$ the area operator, and $\rho$ and $\rho_{c}$ the total and critical energy density respectively. In string theory-inspired effective theories, such as Dvali's compact, extra-dimensional theory~\cite{Dvali:2000hr} such massive modes also arise. 

Massive graviton modes also occur in many other modified gravity models. In Rosen's bimetric theory~\cite{Rosen:1974ua}, for example, photons and gravitons follow null geodesics of different metrics~\cite{lrr-2006-3,Will:1993ns}. In Visser's massive graviton theory~\cite{Visser:1997hd}, the graviton is given a mass at the level of the action through an effective perturbative description of gravity, at the cost of introducing a non-dynamical background metric, {\emph{i.e.~}}a prior geometry. A recent re-incarnation of this model goes by the name of Bigravity, where again two metric tensors are introduced~\cite{Pilo:2011zz,Paulos:2012xe,Hassan:2011zd,Hassan:2011ea}. In Bekenstein's Tensor-Vector-Scalar (TeVeS) theory~\cite{Bekenstein:2004ne}, the existence of a scalar and a vector field lead to subluminal GW propagation. 

Massive graviton theories have a theoretical issue, the van Dam--Veltman--Zakharov (vDVZ) discontinuity~\cite{vanDam:1970vg,Zakharov:1970cc}, which is associated with property 1.a, ie. a General Relativity limit.  The problem is that certain predictions of massive graviton theories do not reduce to those of General relativity in the $m_{g} \to 0$ limit. This can be understood qualitatively by studying how the $5$ spin states of the graviton behave in this limit. Two of them become the two General Relativity helicity states of the massless graviton. Another two become helicity states of a massless vector that decouples from the tensor perturbations in the $m_{g} \to 0$ limit. The last state, the scalar mode, however, retains a finite coupling to the trace of the stress-energy tensor in this limit. Therefore, massive graviton theories in the $m_{g} \to 0$ limit do not reduce to General Relativity, since the scalar mode does not decouple. 

The vDVZ discontinuity, however, can be evaded, for example by carefully including non-linearities. Vainshtein~\cite{Vainshtein:1972sx,Kogan:2000uy,Deffayet:2001uk,Babichev:2013usa} showed that around any spherically-symmetric source of mass $M$, there exists a certain radius $r < r_{V} \equiv (r_{S} \lambda_{g}^{4})^{1/5}$, with $r_{S}$ the Schwarzschild radius, where linear theory cannot be trusted. Since $r_{V} \to \infty$ as $m_{g} \to 0$, this implies that there is no radius at which the linear approximation (and thus vDVZ discontinuity) can be trusted. Of course, to determine then whether massive graviton theories have a continuous limit to General Relativity, one must include non-linear corrections to the action (see also an argument by~\cite{ArkaniHamed:2002sp}), which are more difficult to uniquely predict from fundamental theory. Recently, there has been much activity in the development of new, non-linear massive gravity theories~\cite{Bergshoeff:2009zz,deRham:2010kj,Gumrukcuoglu:2012wt,Bergshoeff:2013hr,deRham:2012fg,deRham:2012fw}. 

Lacking a particular action for massive graviton theories that modifies the strong-field regime and is free of non-linear and radiatively-induced ghosts, it is difficult to ascertain many of its properties, but this does not prevent us from considering certain {\emph{phenomenological}} effects. If the graviton is truly massive, whatever the action may be, two main modifications to Einstein's theory will be introduced:
\begin{itemize}
\item[(i)] Modification to Newton's laws;
\item[(ii)] Modification to gravitational wave propagation.  
\end{itemize}
Modifications of class (i) correspond to the replacement of the Newtonian potential by a Yukawa type potential (in the non-radiative, near-zone of any body of mass $M$): $V = (M/r) \to (M/r) \exp(-r/\lambda_{g})$, where $r$ is the distance to the body~\cite{Will:1997bb}. Tests of such a Yukawa interaction have been proposed through observations of bound clusters, tidal interactions between galaxies~\cite{Goldhaber:1974wg} and weak gravitational lensing~\cite{Choudhury:2002pu}, but such tests are model-dependent. 

Modifications of class (ii) are in the form of a non-zero graviton mass that induces a modified gravitational wave dispersion relation. Such a modification to the dispersion relation was originally parameterized via~\cite{Will:1997bb}
\begin{equation}
\frac{v_{g}^{2}}{c^{2}} = 1 - \frac{m_{g}^{2} c^{4}}{E^{2}}\,,
\label{eq:vg-standard}
\end{equation}
where $v_{g}$ and $m_{g}$ are the speed and mass of the graviton, while $E$ is its energy, usually associated to its frequency via the quantum mechanical relation $E = h f$. This modified dispersion relation is inspired by special relativity,  a more general version of which, inspired by quantum gravitational theories, is~\cite{Mirshekari:2011yq}
\begin{equation}
\frac{v_{g}^{2}}{c^{2}} = 1 - \lambda^{\alpha}\,,
\label{eq:vg-LV}
\end{equation}
where $\alpha$ is now a parameter that depends on the theory and $\lambda$ represents deviations from light speed propagation. For example, in Rosen's bimetric theory~\cite{Rosen:1974ua}, the graviton does not travel at the speed of light, but at some other speed partially determined by the prior geometry. In metric theories of gravity, $\lambda = A  m_{g}^{2} c^{4}/E^{2}$, where $A$ is some amplitude that depends on the metric theory (see discussion in~\cite{Mirshekari:2011yq}). Either modification to the dispersion relation has the net effect of slowing gravitons down, such that there is a difference in the time of arrival of photons and gravitons. Moreover, such an energy-dependent dispersion relation would also affect the accumulated gravitational wave phase observed at gravitational wave detectors, as we will discuss in Section~\ref{section:binary-sys-tests}. Given these modifications to the dispersion relation, one would expect the generation of gravitational waves to also be greatly affected in such theories, but again, lacking a particular healthy action to consider, this topic remains today mostly unexplored. 

From the structure of the above phenomenological modifications, it is clear that General Relativity can be recovered in the $m_{g} \to 0$ limit, avoiding the vDVZ issue altogether by construction. Such phenomenological modifications have been constrained by several types of experiments and observations. Using the modification to Newton's third law and precise observations of the motion of the inner planets of the Solar System together with Kepler's third law,~\cite{Will:1997bb} found a bound of $\lambda_{g} > 2.8 \times 10^{12} \; {\textrm{km}}$. Such a constraint is purely static, as it does not sample the radiative sector of the theory. Dynamical constraints, however, do exist: through observations of the decay of the orbital period of binary pulsars,~\cite{Finn:2001qi} found a bound of $\lambda_{g} > 1.6 \times 10^{10} \; {\textrm{km}}$\epubtkFootnote{The model considered by~\cite{Finn:2001qi} is not phenomenological, but it contains a ghost mode.}; by investigating the stability of Schwarzschild and Kerr black holes, Ref.~\cite{Brito:2013wya} placed the constraint $\lambda_{g} > 2.4 \times 10^{13} \; {\textrm{km}}$ in Fierz-Pauli theory~\cite{Fierz:1939ix}. New constraints that use gravitational waves have been proposed, including measuring a difference in time of arrival of electromagnetic and gravitational waves~\cite{Cutler:2002ef,2008ApJ...684..870K}, as well as direct observation of gravitational waves emitted by binary pulsars (see Section~\ref{section:binary-sys-tests}).  

Although massive gravity theories unavoidably lead to a modification to the graviton dispersion relation, the converse is not necessarily true. A modification of the dispersion relation is usually accompanied by a modification to either the Lorentz group or its action in real or momentum space. Such Lorentz-violating effects are commonly found in quantum gravitational theories, including loop quantum gravity~\cite{2008PhRvD..77b3508B} and string theory~\cite{2005hep.th....8119C,2010GReGr..42....1S}, as well as other effective models~\cite{Berezhiani:2007zf,Berezhiani:2008nr}.  In Doubly Special Relativity~\cite{2001PhLB..510..255A,2002PhRvL..88s0403M,AmelinoCamelia:2002wr,2010arXiv1003.3942A}, the graviton dispersion relation is modified at high energies by modifying the law of transformation of inertial observers. Modified graviton dispersion relations have also been shown to arise in generic extra-dimensional models~\cite{2011PhLB..696..119S}, in Ho\v{r}ava--Lifshitz theory~\cite{Horava:2008ih,Horava:2009uw,2010arXiv1010.5457V,Blas:2011zd} and in theories with non-commutative geometries~\cite{2011arXiv1102.0117G,Garattini:2011kp,Garattini:2011hy}. None of these theories necessarily requires a massive graviton, but rather the modification to the dispersion relation is introduced due to Lorentz-violating effects. 

One might be concerned that the mass of the graviton and subsequent modifications to the graviton dispersion relation should be suppressed by the Planck scale. However, Collins, et~al.~\cite{2004PhRvL..93s1301C,2006hep.th....3002C} have suggested that Lorentz violations in perturbative quantum field theories could be dramatically enhanced when one regularizes and renormalizes them. This is because terms that vanish upon renormalization due to Lorentz invariance do not vanish in Lorentz-violating theories, thus leading to an enhancement~\cite{2011arXiv1106.1417G}. Whether such an enhancement is truly present cannot be currently ascertained.

%------------------------------------------------------------------------------------------------------------------------------
\subsubsection{Modified quadratic gravity}
\label{subsec:MQG}

Modified quadratic gravity is a family of models first discussed in the context of black holes and gravitational waves in~\cite{Yunes:2011we,Yagi:2011xp}. The 4-dimensional action is given by  
\begin{align} 
S &\equiv \int d^4x \sqrt{-g} \left\{ \kappa R + \alpha_{1}
f_{1}(\vartheta) R^{2} + \alpha_{2} f_{2}(\vartheta) R_{\mu \nu} R^{\mu \nu}
+ \alpha_{3} f_{3}(\vartheta) R_{\mu \nu \delta \sigma}
R^{\mu \nu \delta \sigma} 
\right. \nonumber \\
&+ \left. \alpha_{4} f_{4}(\vartheta) 
R_{\mu \nu \delta \sigma} \!{}^{*}R^{\mu \nu \delta \sigma}
-  \frac{\beta}{2} \left[\nabla_{\mu}
\vartheta \nabla^{\mu} \vartheta + 2 V(\vartheta) \right] +
\mathcal{L}_{\rm mat} \right\}\,.
\label{exactaction} 
\end{align}
The quantity $^{*}R^{\mu}{}_{\nu\delta\sigma} = (1/2) \epsilon_{\delta \sigma}{}^{\alpha \beta} R^{\mu}{}_{\nu \alpha \beta}$ is the dual to the Riemann tensor. The quantity $\mathcal{L}_{\rm mat}$ is the external matter Lagrangian, while $f_{i}(\cdot)$ are functionals of the field $\vartheta$, with $(\alpha_{i},\beta)$ coupling constants and $\kappa = (16 \pi G)^{-1}$. Clearly, the two terms second to last in Eq.~\eqref{exactaction} represent a canonical kinetic energy term and a potential. At this stage, one might be tempted to set $\beta=1$ or the $\alpha_{i} = 1$ via a rescaling of the scalar field functional, but we shall not do so here.

The action in Eq.~\eqref{exactaction} is well-motivated from
fundamental theories, as it contains all possible quadratic, algebraic
curvature scalars with running (i.e., non-constant) couplings. The only restriction here is that all quadratic terms are assumed to couple to the {\emph{same}} field, which need not be the case. For example, in string theory some terms might couple to the dilaton (a scalar field), while other couple to the axion (a pseudo scalar field). Nevertheless, one can recover well-known and motivated modified gravity theories in simple cases. For example, Dynamical Chern--Simons modified gravity~\cite{Alexander:2009tp} is recovered when $\alpha_{4} =-\alpha_{\rm CS}/4$ and all other $\alpha_{i} = 0$. Einstein-Dilaton-Gauss--Bonnet gravity~\cite{Pani:2009wy} is obtained when $\alpha_{4} = 0$ and $(\alpha_1,\alpha_2,\alpha_3)=(1,-4,1)\alpha_{\rm EDGB}$\epubtkFootnote{Technically, Einstein-Dilaton-Gauss--Bonnet gravity has a very particular set of coupling functions $f_{1}(\vartheta) = f_{2}(\vartheta) = f_{3}(\vartheta) \propto e^{\gamma \vartheta}$, where $\gamma$ is a constant. In most cases, however, one can expand about $\gamma \vartheta \ll 1$, so that the functions become linear in the scalar field}. Both theories unavoidably arise as low-energy expansions of heterotic string theory~\cite{Green:1987sp,Green:1987mn,Alexander:2004xd,lrr-2004-5}. As such, modified quadratic gravity theories should be treated as a class of effective field theories. Moreover, Dynamical Chern--Simons Gravity also arises in loop quantum gravity~\cite{Ashtekar:2004eh,Rovelli:2004tv} when the Barbero--Immirzi parameter is promoted to a field in the presence of fermions~\cite{Ashtekar:1988sw,Alexander:2008wi,Taveras:2008yf,Mercuri:2009zt,Gates:2009pt}.

One should make a clean and clear distinction between the theory defined by the action of Eq.~\eqref{exactaction} and that of $f(R)$ theories. The latter are defined as functionals of the Ricci scalar only, while Eq.~\eqref{exactaction} contains terms proportional to the Ricci tensor and Riemann tensor squared. One could think of the subclass of $f(R)$ theories with $f(R) = R^{2}$ as the limit of modified quadratic gravity with only $\alpha_{1} \neq 0$ and $f_{1}(\vartheta) = 1$. In that very special case, one can map quadratic gravity theories and $f(R)$ gravity to a scalar-tensor theory. Another important distinction is that $f(R)$ theories are usually treated as exact, while the action presented above is to be interpreted as an {\emph{effective theory}}~\cite{lrr-2004-5} truncated to quadratic order in the curvature in a low-energy expansion of a more fundamental theory. This implies that there are cubic, quartic, etc.~terms in the Riemann tensor that are not included in Eq.~\eqref{exactaction} and that presumably depend on higher powers of $\alpha_{i}$. Thus, when studying such an effective theory one should also order-reduce the field equations and treat all quantities that depend on $\alpha_{i}$ perturbatively, the so-called {\emph{{small-coupling approximation}}. One can show that such an order reduction removes any additional polarization modes in propagating metric perturbations~\cite{Sopuerta:2009iy,Stein:2010pn} that naturally arise in $f(R)$ theories. In analogy to the treatment of the Ostrogradski instability in Section~\ref{sec:properties}, one would also expect that order-reduction would lead to a theory with a well-posed initial value formulation. 

This family of theories is usually simplified by making the assumption that coupling functions $f_{i}(\cdot)$ admit a Taylor expansion: $f_{i}(\vartheta) = f_{i}(0)+ f_{i}'(0)\vartheta + \mathcal{O}(\vartheta^{2})$ for small $\vartheta$, where $f_{i}(0)$ and $f_{i}'(0)$ are constants and $\vartheta$ is assumed to vanish at asymptotic spatial infinity. Reabsorbing $f_{i}(0)$ into the coupling constants $\alpha_{i}^{(0)} \equiv \alpha_{i} f_{i}(0)$ and $f_{i}^{\prime}(0)$ into the constants $\alpha_{i}^{(1)} \equiv \alpha_{i} f_{i}^{\prime}(0)$, Eq.~\eqref{exactaction} becomes $S = S_{\rm GR} + S_{0} + S_{1}$ with
\begin{subequations}
\label{eq:quad-action-simped}
\begin{align}
S_{\rm GR} &\equiv \int d^4x \sqrt{-g} \left\{ \kappa R + \mathcal{L}_{\rm mat} \right\}\,,
\\
S_{0} &\equiv  \int d^4x \sqrt{-g} \left\{ 
\alpha_{1}^{(0)} R^{2} 
+ \alpha_{2}^{(0)} R_{\mu \nu} R^{\mu \nu}
+ \alpha_{3}^{(0)} R_{\mu \nu \delta \sigma} R^{\mu \nu \delta \sigma} 
\right\}\,,
\\
S_{1} &\equiv  \int d^4x \sqrt{-g} \left\{ 
\alpha_{1}^{(1)} \vartheta R^{2} + \alpha_{2}^{(1)} \vartheta R_{\mu \nu} R^{\mu \nu}
+ \alpha_{3}^{(1)} \vartheta R_{\mu \nu \delta \sigma} R^{\mu \nu \delta \sigma} 
\right. \nonumber \\*
&+ \left. \alpha_{4}^{(1)} \vartheta  \; R_{\mu \nu \delta \sigma} \; \!{}^{*}R^{\mu \nu \delta \sigma}
- \frac{\beta}{2} \left[\nabla_{\mu}
\vartheta \nabla^{\mu} \vartheta + 2 V(\vartheta) \right] \right\}\,.
\end{align}
\label{action}
\end{subequations}
Here, $S_{\rm GR}$ is the Einstein--Hilbert plus matter action, while $S_{0}$ and $S_{1}$ are corrections. The former is decoupled from $\vartheta$, where the omitted term proportional to $\alpha_4^{(0)}$ does not affect the classical field equations since it is topological, i.e. it can be rewritten as the total $4$-divergence of some $4$-current. Similarly, if the $\alpha_{i}^{(0)}$ were chosen to reconstruct the Gauss--Bonnet invariant, $(\alpha_1^{(0)},\alpha_2^{(0)},\alpha_3^{(0)})=(1,-4,1)\alpha_{\mathrm{GB}}$, then this combination would also be topological and not affect the classical field equations. On the other hand, $S_{1}$ is a modification to GR with a direct (non-minimal) coupling to $\vartheta$, such that as the field goes to zero, the modified theory reduces to GR. 

Another restriction one usually makes to simplify modified gravity theories is to neglect the $\alpha_{i}^{(0)}$ terms and only consider the $S_{1}$ modification, the so-called {\emph{restricted}} modified quadratic gravity. The $\alpha_{i}^{(0)}$ terms represent corrections that are non-dynamical. The term proportional to $\alpha_{1}^{(0)}$ resembles a certain class of $f(R)$ theories. As such, it can be mapped to a scalar tensor theory with a complicated potential, which has been heavily constrained by torsion-balance E\"ot-Wash experiments to $\alpha_{1}^{(0)} < 2 \times 10^{-8} \; m^{2}$~\cite{Hoyle:2004cw,Kapner:2006si,Berry:2011pb}. Moreover, these theories have a fixed coupling constant that does not run with energy or scale. In restricted modified gravity, the scalar field is effectively forcing the running of the coupling.  

Let us then concentrate on restricted modified quadratic gravity and drop the superscript in $\alpha_i^{(1)}$. 
The modified field equations are
\begin{align}
G_{\mu \nu} &+ \frac{\alpha_1 \vartheta}{\kappa} \mathcal{H}_{\mu \nu}^{(0)} +
\frac{\alpha_2 \vartheta }{\kappa} \mathcal{I}_{\mu \nu}^{(0)} 
+ \frac{\alpha_3 \vartheta}{\kappa} \mathcal{J}_{\mu \nu}^{(0)} 
+ \frac{\alpha_1}{\kappa}
\mathcal{H}_{\mu \nu}^{(1)} + 
\frac{\alpha_2}{\kappa} \mathcal{I}_{\mu \nu}^{(1)} 
+ \frac{\alpha_3}{\kappa} \mathcal{J}_{\mu \nu}^{(1)} +
\frac{\alpha_4}{\kappa} \mathcal{K}_{\mu \nu}^{(1)} 
= \frac{1}{2\kappa} \left( T_{\mu \nu}^{\rm mat} + T_{\mu \nu}^{(\vartheta)} \right)\,,
\label{FEs} 
\end{align}
where we have defined 
\begin{subequations}
\allowdisplaybreaks[1]
\begin{align}
\mathcal{H}_{\mu \nu}^{(0)} \equiv & 2 R R_{\mu \nu}  -
\frac{1}{2} g_{\mu \nu} R^{2} - 2 \nabla_{\mu \nu} R + 2 g_{\mu \nu} \square R\,,\\
\mathcal{I}_{\mu \nu}^{(0)} \equiv & \square R_{\mu \nu} + 2
R_{\mu \delta \nu \sigma} R^{\delta \sigma} - \frac{1}{2} g_{\mu \nu} 
R^{\delta \sigma} R_{\delta \sigma}
+ \frac{1}{2} g_{\mu \nu} \square R - \nabla_{\mu \nu} R\,,\\
\mathcal{J}_{\mu \nu}^{(0)}
\equiv & 8 R^{\delta \sigma} R_{\mu \delta \nu \sigma} - 
2 g_{\mu \nu} R^{\delta \sigma} R_{\delta \sigma} + 4 \square R_{\mu \nu}
- 2 R \, R_{\mu \nu} + \frac{1}{2} g_{\mu \nu} R^{2} - 2 \nabla_{\mu \nu} R\,,\\
\mathcal{H}_{\mu \nu}^{(1)} \equiv & -4
(\nabla_{(\mu} \vartheta) \nabla_{\nu)} R - 2 R \nabla_{\mu\nu} \vartheta 
+ g_{\mu \nu} \left[2 R \square \vartheta + 4 (\nabla^{\delta}\vartheta) \nabla_{\delta}R \right]\,,\\
\mathcal{I}_{\mu \nu}^{(1)} \equiv & -(\nabla_{(\mu}\vartheta) \nabla_{\nu)} R - 2\nabla^\delta\vartheta
  \nabla_{(\mu} R_{\nu)\delta} 
+ 2 \nabla^\delta\vartheta \nabla_{\delta} R_{\mu \nu}
+R_{\mu\nu}\square \vartheta
\nonumber \\*
&- 2 R_{\delta(\mu}\nabla^{\delta} \nabla_{\nu)}\vartheta + g_{\mu \nu} \left( \nabla^\delta \vartheta
  \nabla_{\delta} R + R^{\delta \sigma} \nabla_{\delta\sigma} \vartheta \right)\,,\\
\mathcal{J}_{\mu \nu}^{(1)}
\equiv & - 8 \left(\nabla^\delta \vartheta \right) \left( \nabla_{(\mu} R_{\nu)\delta} - \nabla_{\delta} R_{\mu \nu}\right) + 4
R_{\mu \delta \nu \sigma} \nabla^{\delta\sigma} \vartheta\,, \\
\mathcal{K}_{\mu \nu}^{(1)}
\equiv & -4 \left( \nabla^\delta\vartheta \right) \epsilon_{\delta \sigma \chi(\mu} \nabla^{\chi} R_{\nu)}^{~\sigma} + 4
(\nabla_{\delta\sigma} \vartheta) {}^{*}\!R_{(\mu}{}^{\delta}{}_{\nu)}{}^{\sigma}\,.
\end{align}
\end{subequations}
The $\vartheta$ stress-energy tensor is
\be
T_{\mu \nu}^{(\vartheta)} = \beta \left[(\nabla_{\mu}\vartheta) (\nabla_{\nu}\vartheta)
 - \frac{1}{2}g_{\mu \nu} \left(\nabla_{\delta}\vartheta \nabla^{\delta}\vartheta - 2
V(\vartheta) \right) \right]\,.
\label{theta-Tab}
\ee
The field equations for the scalar field are
\begin{align}
\beta \square \vartheta - \beta \frac{dV}{d\vartheta}
=&\, -\alpha_1 R^{2} - \alpha_2 R_{\mu \nu} R^{\mu \nu} 
-\alpha_3 R_{\mu \nu \delta \sigma} R^{\mu \nu \delta \sigma} 
- \alpha_4 R_{\mu \nu \delta \sigma} \!{}^{*}R^{\mu \nu \delta \sigma}\,.
\label{EOM} 
\end{align}
Notice that unlike traditional scalar-tensor theories, the scalar field is here sourced by the geometry and not by the matter distribution. This directly implies that black holes in such theories are likely to be hairy. 

From the structure of the above equations, it should be clear that the dynamics of $\vartheta$ guarantee that the modified field equations are covariantly conserved exactly. That is, one can easily verify that the covariant divergence of Eq.~\eqref{FEs} identically vanishes upon imposition of Eq.~\eqref{EOM}. Such a result had to be so, as the action is diffeomorphism invariant. If one neglected the kinetic and potential energies of $\vartheta$ in the action, as was originally done in~\cite{Jackiw:2003pm}, the theory would possess preferred-frame effects and would not be covariantly conserved. Moreover, such a theory requires an additional constraint, ie.~the right-hand side of \eqref{EOM} would have to vanish, which is an unphysical consequence of treating $\vartheta$ as prior structure~\cite{Yunes:2007ss,Grumiller:2007rv}.

One last simplification that is usually made when studying modified quadratic gravity theories is to ignore the potential $V(\vartheta)$, ie.~set $V(\vartheta) = 0$. This potential can in principle be non-zero, for example if one wishes to endow $\vartheta$ with a mass or if one wishes to introduce a cosine driving term, like that for axions in field and string theory. However, reasons exist to restrict the functional form of such a potential. First, a mass for $\vartheta$ will modify the evolution of any gravitational degree of freedom only if this mass is comparable to the inverse length scale of the problem under consideration (such as a binary system). This could be possible if there is an incredibly large number of fields with different masses in the theory, such as perhaps in the string axiverse picture~\cite{PhysRevD.83.044026,Kodama:2011zc,PhysRevD.85.103514}. In that picture, however, the moduli fields are endowed with a mass due to shift-symmetry breaking by non-perturbative effects; such masses are not expected to be comparable to the inverse length scale of binary systems. Second, no mass term may appear in a theory with a shift symmetry, ie.~invariance under the transformation $\vartheta \to \vartheta + {\rm{const}}$. Such symmetries are common in four-dimensional, low-energy, effective string theories~\cite{Boulware:1985wk,Green:1987mn,Green:1987sp,1992PhLB..285..199C,lrr-2004-5}, such as dynamical Chern--Simons and Einstein-Dilaton-Gauss--Bonnet theory. Similar considerations apply to other more complicated potentials, such as a cosine term. 

Given these field equations, one can linearize them about Minkowski space to find evolution equations for the perturbation in the small-coupling approximation. Doing so, one finds~\cite{Yagi:2011xp}
\begin{align}
\square_{\eta} \vartheta =&
- \frac{\alpha_1}{\beta} \left( \frac{1}{2\kappa} \right)^{2} T_{\mathrm{mat}}^2
- \frac{\alpha_2}{\beta} \left( \frac{1}{2\kappa} \right)^{2} T_{\mathrm{mat}}^{\mu\nu} T^{\mathrm{mat}}_{\mu\nu}
\nonumber \\
& -  \frac{2\alpha_3}{\beta} (h_{\alpha \beta ,\mu \nu} h^{\alpha [\beta ,\mu] \nu} + h_{\alpha \beta ,\mu \nu} h^{\mu [\nu ,\alpha] \beta} ) 
\nonumber \\
& -  \frac{2 \alpha_4}{\beta} \bar{\epsilon}^{\alpha \beta \mu \nu} h_{\alpha \delta,\gamma \beta} h_{\nu}{}^{[\gamma,\delta]}{}_{\mu}\,,
\end{align}
where we have order-reduced the theory where possible and used the harmonic gauge condition (which is preserved in this class of theories~\cite{Sopuerta:2009iy,Stein:2010pn}). The corresponding equation for the metric perturbation is rather lengthy and can be found in Eqs.~$(17)$-$(24)$ in~\cite{Yagi:2011xp}. Since these theories are to be considered effective, working always to leading order in $\alpha_{i}$, one can show that they are perturbatively of type $N_{2}$ in the $E(2)$ classification~\cite{Eardleyprd}, ie.~in the far zone, the only propagating modes that survive are the two transverse-traceless (spin-2) metric perturbations~\cite{Sopuerta:2009iy}. In the strong-field region, however, it is possible that additional modes are excited, although they decay rapidly as they propagate to future null infinity. 

Lastly, let us discuss what is known about whether modified quadratic gravities satisfy the requirements discussed in Section~\ref{sec:properties}. As it should be clear from the action itself, this modified gravity theory satisfies the fundamental requirement, ie.~passing all precision tests, provided the couplings $\alpha_{i}$ are sufficiently small. This is because such theories have a continuous limit to General Relativity as $\alpha_{i} \to 0$\epubtkFootnote{Formally, as $\alpha_{i} \to 0$, one recovers General Relativity with a dynamical scalar field. The latter, however, does not couple to the metric or the matter sector, so it does not lead to any observable effects that distinguish it from General Relativity.}. Dynamical Chern--Simons gravity is constrained only weakly at the moment, $\xi_{4}^{1/4} < 10^{8} \; {\rm{km}}$, where $\xi_{4} \equiv \alpha_{4}^{2}/(\beta \kappa)$, only through observations of Lense--Thirring precession in the Solar System~\cite{AliHaimoud:2011fw}. The Einstein-Dilaton-Gauss--Bonnet gravity coupling constant $\xi_{3} \equiv \alpha_{3}^{2}/(\beta \kappa)$, on the other hand, has been constrained by several experiments: Solar System observations of the Shapiro time delay with the Cassini spacecraft placed the bound $\xi_{3}^{1/4} < 1.3 \times 10^{7} \; {\rm{km}}$~\cite{Bertotti:2003rm,Amendola:2007ni}; the requirement that neutron stars still exist in this theory placed the constraint $\xi_{3}^{1/4} \lesssim 26 {\rm{km}}$~\cite{Pani:2011xm}, with the details depending somewhat on the central density of the neutron star; 
% Berti's constraint is \alpha \beta = M^{2}, where the coupling function is f = \alpha \beta \Phi, so that \alpha \beta = \alpha_{4}/\beta^{1/2}
% This then means that a constraint \alpha beta < 100 Msun^{2} is really \alpha_{4}/\beta^{1/2} < 100 Msun^{2}, 
% converting to \xi = \alpha^{2}/(\beta \kappa) you get \alpha_{4}^{2}/(\kappa \beta)  < 10^{4} Msun^{4}/\kappa = 50 10^{4} Msun^{4}
% so then \xi^{1/4} = \alpha^{1/2}/(\beta \kappa)^{1/4} <  26 Msun
observations of the rate of change of the orbital period in the low-mass X-ray binary $A0620-00$~\cite{Psaltis:2005ai,Johannsen:2008tm} has led to the constraint $\xi_{3}^{1/4} < 1.9 \; {\rm{km}}$~\cite{Yagi:2012gp}. 

Not all sub-properties of the fundamental requirement, however, are known to be satisfied. One can show that certain members of modified quadratic gravity possess known solutions and these are stable, at least in the small-coupling approximation. For example, in Dynamical Chern--Simons gravity, spherically symmetric vacuum solutions are given by the Schwarzschild metric with constant $\vartheta$ to all orders in $\alpha_{i}$~\cite{Jackiw:2003pm,Yunes:2007ss}.  Such a solution is stable to small perturbations~\cite{Molina:2010fb,Garfinkle:2010zx}, as also are non-spinning black holes and branes in anti De~Sitter space~\cite{Delsate:2011qp}. On the other hand, spinning solutions continue to be elusive, with approximate solutions in the slow-rotation/small-coupling limit known both for black holes~\cite{Yunes:2009hc,Konno:2009kg,Pani:2011gy,Yagi:2012ya} and stars~\cite{AliHaimoud:2011fw,Pani:2011xm}; nothing is currently known about the stability of these spinning solutions. In Einstein-Dilaton-Gauss--Bonnet theory even spherically symmetric solutions are modified~\cite{Yunes:2011we,Pani:2011gy} and these are stable to axial perturbations~\cite{Pani:2009wy}.

The study of modified quadratic gravity theories as effective theories is valid provided one is sufficiently far from its cut-off scale, ie.~the scale beyond which higher-order curvature terms cannot be neglected anymore. One can estimate the magnitude of this scale by studying the size of loop corrections to the quadratic curvature terms in the action due to $n$-point interactions~\cite{Yagi:2012ya}. Simple counting requires that the number of scalar and graviton propagators, $P_{s}$ and $P_{g}$, satisfy the following relation in terms of the number of vertices $V$:
\begin{equation}
P_s = \frac{V}{2}, \quad P_g = (n-1)\frac{V}{2}\,.
\end{equation}
Loop corrections are thus suppressed by factors of $\alpha_{i}^{V} M_\mathrm{pl}^{(2-n)V} \Lambda^{nV}$, with $M_\mathrm{pl}$ the Planck mass and $\Lambda$ the energy scale introduced by dimensional arguments. The cut-off scale above which the theory cannot be treated as an effective one can be approximated as the value of $\Lambda$ at which the suppression factor becomes equal to unity:
\begin{equation}
\Lambda_c \equiv M_\mathrm{pl}^{1-2/n} \alpha_{i}^{1/n}\,,
\label{lambdac}
\end{equation}
This cut-off scale automatically places a constraint on the magnitude of $\alpha_{i}$ above which higher-curvature corrections must be included. Setting the largest value of $\Lambda_{c}$ to be equal to $\mathcal{O}(10\mu$m), thus saturating bounds from table-top experiments~\cite{Kapner:2006si}, and solving for $\alpha_{i}$, we find
\begin{equation}
\alpha_{i}^{1/2} < \mathcal{O}(10^8 \mathrm{km}).
\label{alpha-ineq}
\end{equation}
Current Solar System bounds on $\alpha_{i}$ already require the coupling constant to be smaller than $10^{8} \; {\rm{km}}$, thus justifying the treatment of these theories as effective models.

As for the other requirements discussed in Section~\ref{sec:properties}, it is clear that modified quadratic gravity is well-motivated from fundamental theory, but it is not clear at all whether it has a well-posed initial value formulation. From an effective point of view, a perturbative treatment in $\alpha_{i}$ naturally leads to stable solutions and a well-posed initial value problem, but this is probably not the case when it is treated as an exact theory. In fact, if one were to treat such a theory as exact (to all orders in $\alpha_{i}$), then the evolution system would likely not be hyperbolic, as higher than second time derivatives now drive the evolution. Although no proof exists, it is likely that such an exact theory is not well-posed as an initial value problem. Notice, however, that this says nothing about the fundamental theories that modified quadratic gravity derives from. This is because even if the truncated theory were ill-posed, higher order corrections that are neglected in the truncated version could restore well-posedness. 

As for the last requirement (that the theory modifies the strong-field), modified quadratic theories are ideal in this respect. This is because they introduce corrections to the action that depend on higher powers of the curvature. In the strong-field, such higher powers could potentially become non-negligible relative to the Einstein--Hilbert action. Moreover, since the curvature scales inversely with the mass of the objects under consideration, one expects the largest deviations in systems with small total mass, such as stellar-mass black hole mergers. On the other hand, deviations from General Relativity should be small for small compact objects spiraling into a supermassive black hole, since here the spacetime curvature is dominated by the large object, and thus it is small, as discussed in~\cite{Sopuerta:2009iy}. 

%------------------------------------------------------------------------------------------------------------------------------
\subsubsection{Variable \textit{G} theories and large extra dimensions}
\label{sec:Variable-G}

Variable $G$ theories are defined as those where Newton's gravitational constant is promoted to a spacetime function. Such a modification breaks the principle of equivalence (see~\cite{lrr-2006-3}) because the laws of physics now become local position dependent. In turn, this implies that experimental results now depend on the spacetime position of the laboratory frame at the time of the experiment. 

Many known alternative theories that violate the principle of equivalence, and in particular, the strong equivalence principle, predict a varying gravitational constant. A classic example is scalar-tensor theory~\cite{Will:1993ns}, which, as explained in Section~\ref{sec:ST}, modifies the gravitational sector of the action by multiplying the Ricci scalar by a scalar field (in the Jordan frame). In such theories, one can effectively think of the scalar as promoting the coupling between gravity and matter to a field-dependent quantity $G \to G(\phi)$, thus violating local position invariance when $\phi$ varies. Another example are bimetric theories, such as that of Lightman-Lee~\cite{1973PhRvD...8.3293L}, where the gravitational constant becomes time-dependent even in the absence of matter, due to possibly time-dependent cosmological evolution of the prior geometry. A final example are higher-dimensional, brane-world scenarios, where enhanced Hawking radiation inexorably leads to a time-varying effective 4D gravitational constant~\cite{Deffayet:2007kf}, whose rate of change depends on the curvature radius of extra dimensions~\cite{Johannsen:2008tm}. 

One can also construct $f(R)$-type actions that introduce variability to Newton's constant. For example, consider the $f(R)$ model~\cite{Frolov:2011ys}
\be
S = \int d^{4}x \sqrt{-g} \; \kappa R \left[1 + \alpha_{0} \ln \left(\frac{R}{R_{0}}\right) \right] + S_{\rm mat}\,,
\ee
where $\kappa = (16 \pi G)^{-1}$, $\alpha_{0}$ is a coupling constant and $R_{0}$ is a curvature scale. This action is motivated by certain renormalization group flow arguments~\cite{Frolov:2011ys}. The field equations are
\begin{align}
G_{\mu \nu} = \frac{1}{2 \bar{\kappa}} T_{\mu \nu}^{\rm mat} - \frac{\alpha_{0}}{\bar{\kappa}} R_{\mu \nu} - 2 \frac{\kappa}{\bar{\kappa}} \frac{\alpha_{0}}{R^{2}} \nabla_{(\mu} R \nabla_{\nu)} R - \frac{1}{2} \frac{\alpha_{0} \kappa}{\bar{\kappa}} g_{\mu \nu} \square R\,,
\end{align}
where we have defined the new constant
\begin{equation}
\bar{\kappa} := \kappa \left[1 + \frac{\alpha_{0}}{\kappa} \ln \left(\frac{R}{R_{0}}\right)\right]\,.
\end{equation}
Clearly, the new coupling constant $\bar{\kappa}$ depends on the curvature scale involved in the problem, and thus, on the geometry, forcing $G$ to run to zero in the ultraviolet limit. 

An important point to address is whether variable $G$ theories can lead to modifications to a {\emph{vacuum}} spacetime, such as a black hole binary inspiral. In Einstein's theory, $G$ appears as the coupling constant between geometry, encoded by the Einstein tensor $G_{\mu\nu}$, and matter, encoded by the stress energy tensor $T_{\mu\nu}^{\rm mat}$. When considering vacuum spacetimes, $T_{\mu \nu}^{\rm mat} = 0$ and one might naively conclude that a variable $G$ would not introduce any modification to such spacetimes. In fact, this is the case in scalar-tensor theories (without homogeneous, cosmological solutions to the scalar field equation), where the no-hair theorem establishes that black hole solutions are not modified. On the other hand, scalar-tensor theories with a cosmological, homogeneous scalar field solution can violate the no-hair theorem, endowing black holes with time-dependent hair, which in turn would introduce variability into $G$ even in vacuum spacetimes~\cite{Jacobson:1999vr,Horbatsch:2011ye,Berti:2013gfa}. 

In general, Newton's constant plays a much more fundamental role than merely a coupling constant: it defines the relationship between energy and length. For example, for the \emph{vacuum} Schwarzschild solution, $G$ establishes the relationship between the radius $R$ of the black hole and the rest-mass energy $E$ of the spacetime via $R=2 G E/ c^4$. Similarly, in a binary black hole spacetime, each black hole introduces an energy scale into the problem that is quantified by a specification of Newton's constant. Therefore, one can treat variable $G$ modifications as induced by some effective theory that modifies the mapping between the curvature scale and the energy scale of the problem, as is done for example in theories with extra dimensions. 

An explicit example of this idea is realized in braneworld models. Superstring theory suggests that physics should be described by 4 large dimensions, plus another 6 that are compactified and very small~\cite{1998stth.book.....P,Polchinski:1998rr}. The size of these extra dimensions is greatly constrained by particle theory experiments. Braneworld models, where a certain higher-dimensional membrane is embedded in a higher dimensional bulk spacetime, can however evade this constraint as only gravitons can interact with the bulk. The ADD model~\cite{ArkaniHamed:1998rs,ArkaniHamed:1998nn} is a particular example of such a braneworld, where the bulk is flat and compact and the brane is tensionless with ordinary fields localized on it. Since gravitational wave experiments have not yet constrained deviations from Einstein's theory in the strong-field, the size of these extra dimensions is constrained to micrometer scales only by table-top experiments~\cite{Kapner:2006si,Adelberger:2006dh}. 

What is relevant to gravitational wave experiments is that in many of these braneworld models black holes may not remain static~\cite{Emparan:2002px,Tanaka:2002rb}. The argument goes roughly as follows: a five-dimensional black hole is dual to a four-dimensional one with conformal fields on it by the ADS/CFT conjecture~\cite{Maldacena:1997re,Aharony:1999ti}, but since the latter must evolve via Hawking radiation, the black hole must be losing mass. The Hawking mass loss rate is here enhanced by the large number of degrees of freedom in the conformal field theory, leading to an effective modification to Newton's laws and to the emission of gravitational radiation. Effectively, one can think of the black hole mass loss as due to the black hole being stretched away from the brane into the bulk due to an universal acceleration, that essentially reduces the size of the brane-localized black hole. For black-hole binaries, one can then draw an analogy between this induced time-dependence in the black hole mass and a variable $G$ theory, where Newton's constant decays due to the presence of black holes. Of course, this is only analogy, since large extra dimensions would not predict a time-evolving mass in neutron star binaries.

Recently, however, Figueras et al.~\cite{Figueras:2011va,Figueras:2011gd,Figueras:2013jja} numerically found stable solutions that do not require a radiation component. If such solutions were the ones realized in Nature as a result of gravitational collapse on the brane, then the black hole mass would be time-independent, up to quantum correction due to Hawking evaporation, a negligible effect for realistic astrophysical systems.  Unfortunately, we currently lack numerical simulations of the dynamics of gravitational collapse in such scenarios.  

Many experiments have been carried out to measure possible deviations from a constant $G$ value, and they can broadly be classified into two groups: (a) those that search for the present or nearly present rate of variation (at redshifts close to zero); (b) those that search for {\emph{secular}} variations over long time periods (at very large redshifts). Examples of experiments or observations of the first class include planetary radar-ranging~\cite{2005AstL...31..340P}, surface temperature observations of low-redshift millisecond pulsars~\cite{Jofre:2006ug,Reisenegger:2009cq}, lunar ranging observations~\cite{Williams:2004qba} and pulsar timing observations~\cite{Kaspi:1994hp,Deller:2008jx}, the latter two being the most stringent. Examples of experiments of the second class include the evolution of the Sun~\cite{1998ApJ...498..871G} and Big Bang Nucleosynthesis (BBN) calculations~\cite{Copi:2003xd,Bambi:2005fi}, again with the latter being more stringent. For either class, the strongest constraint are about $\dot{G}/G \lesssim 10^{-13} \; {\rm yr}^{-1}$, varying somewhat from experiment to experiment.

Lacking a particularly compelling action to describe variable $G$ theories, one is usually left with a phenomenological model of how such a modification to Einstein's theory would impact gravitational waves. Given that the part of the waveform that detectors are most sensitive to is the gravitational wave phase, one can model the effect of variable $G$ theories by studying how the rate of change of its frequency would be modified. Assuming a Taylor expansion for Newton's constant one can derive the modification to the evolution equation for the gravitational wave frequency, given whichever physical scenario one is considering. Solving such an evolution equation then leads to a modification in the accumulated gravitational wave phase observed at detectors on Earth. In Section~\ref{section:binary-sys-tests} we will provide an explicit example of this for a compact binary system. 

Let us now discuss whether such theories satisfy the criteria defined in Section~\ref{sec:properties}. The fundamental property can be satisfied if the rate of change of Newton's constant is small enough, as variable $G$ theories usually have a continuous limit to General Relativity (as all derivatives of $G$ go to zero). Whether variable $G$ theories are well-motivated from fundamental physics (property 2) depends somewhat on the particular effective model or action that one considers. But in general, property 2 is usually satisfied, considering that such variability naturally arises in theories with extra dimensions, and the latter are also natural in all string theories. Variable $G$ theories, however, usually fail at introducing modifications in the strong-field region. Usually, such variability is parameterized as a Taylor expansion about some initial point with constant coefficients. That is, the variability of $G$ is not usually constructed so as to become stronger closer to merger. 

The well-posed property and the sub-properties of the fundamental property depend somewhat on the particular effective theory used to describe varying $G$ modifications. In the $f(R)$ case, one can impose restrictions on the functional form $f(\cdot)$ such that no ghosts ($f' >0$) or instabilities ($f'' > 0$) arise~\cite{Frolov:2011ys}. This, of course, does not guarantee that this (or any other such) theory is well-posed. A much more detailed analysis would be required to prove well-posedness of the class of theories that lead to a variable Newton's constant, but such is currently lacking.  

%------------------------------------------------------------------------------------------------------------------------------
\subsubsection{Non-commutative geometry}
\label{sec:NCG}

Non-commutative geometry is a gravitational theory that generalizes the continuum Riemannian manifold of Einstein's theory with the product of it with a tiny, discrete, finite non-commutative space, composed of only two points. Although the non-commutative space has zero spacetime dimension, as the product manifold remains four dimensional, its internal dimensions are $6$ to account for Weyl and chiral fermions. This space is discrete to avoid the infinite tower of massive particles that would otherwise be generated, as in string theory. Through this construction, one can recover the Standard Model of elementary particles, while accounting for all (elementary particle) experimental data to date. Of course, the simple non-commutative space described above is expected to be replaced by a more complex model at Planckian energies. Thus, one is expected to treat such non-commutative geometry models as effective theories. Essentially nothing is currently known about the full non-commutative theory of which the theories described in this section are an effective low-energy limit.

Before proceeding with an action-principle description of non-commutative geometry theories, we must distinguish between the spectral geometry approach championed by Connes~\cite{Connes:1996gi}, and Moyal-type non-commutative geometries~\cite{Snyder:1946qz,Groenewold:1946kp,Moyal:1949sk}. In the former, the manifold is promoted to a non-commutative object through the product of a Riemann manifold with a non-commutative space. In the latter, instead, a non-trivial set of commutation relations is imposed between operators corresponding to position. These two theories are in principle unrelated. In this review, we will concentrate only on the former, as it is the only type of non-commutative General Relativity extension that has been studied in the context of gravitational wave theory.  

The effective action for spectral non-commutative geometry theories (henceforth, non-commutative geometries for short) is
\begin{align}
S = \int d^{4}x \sqrt{-g} \left( \kappa R + \alpha_{0} C_{\mu \nu \delta \sigma} C^{\mu \nu \delta \sigma} + \tau_{0} R^{*} R^{*} - \xi_{0} R \; |H|^{2}\right) + S_{\rm mat}\,,
\label{eq:NCG-action}
\end{align}
where $H$ is related to the Higgs field, $C_{\mu \nu \delta \sigma}$ is the Weyl tensor, $(\alpha_{0}, \tau_{0}, \xi_{0})$ are couplings constants and we have defined the quantity
\begin{equation}
R^{*} R^{*} :=\frac{1}{4} \epsilon^{\mu \nu \rho \sigma} \epsilon_{\alpha \beta \gamma \delta} R_{\mu \nu}{}^{\alpha \beta} R_{\rho \sigma}{}^{\gamma \delta}\,.
\end{equation}
Notice that this term integrates to the Euler characteristic, and since $\tau_{0}$ is a constant, it is topological and does not affect the classical field equations. The last term of Eq.~\eqref{eq:NCG-action} is usually ignored as $H$ is assumed to be relevant only in the early universe. Finally, the second term can be rewritten in terms of the Riemann and Ricci tensors as
\begin{equation}
C_{\mu \nu \delta \sigma} C^{\mu \nu \delta \sigma} = \frac{1}{3} R^{2} -2 R_{\mu \nu} R^{\mu \nu} + R_{\mu \nu \delta \sigma} R^{\mu \nu \delta \sigma}\,.
\end{equation}
Notice that this corresponds to the modified quadratic gravity action of Eq.~\eqref{action} with all $\alpha_{i}^{(1)} = 0$ and $(\alpha_{1}^{(0)},\alpha_{2}^{(0)},\alpha_{3}^{(0)}) = (1/3,-2,1)$, which is not the Gauss--Bonnet invariant. Notice also that this model is not usually studied in modified quadratic gravity theory, as one usually concentrates on the terms that have an explicit scalar field coupling. 

The field equations of this theory can be read directly from Eq.~\eqref{FEs}, but we repeat them here for completeness: 
\begin{align}
G_{\mu \nu} - \frac{2 \alpha_{0}}{\kappa} \left[2 \nabla^{\kappa \lambda } + R^{\lambda \kappa} \right] C_{\mu \lambda \nu \kappa} = \frac{1}{2 \kappa} T_{\mu \nu}^{\rm mat}\,.
\label{eq:NCG-FEs}
\end{align}
One could in principle rewrite this in terms of the Riemann and Ricci tensors, but the expressions become quite complicated, as calculated explicitly in Eqs.~(2) and (3) of~\cite{Yunes:2011we}. Due to the absence of a dynamical degree of freedom coupling to the modifications to the Einstein--Hilbert action, this theory is not covariantly conserved in vacuum. By this we mean that the covariant divergence of Eq.~\eqref{eq:NCG-FEs} does not vanish in vacuum, thus violating the weak-equivalence principle and leading to additional equations that might over-constrain the system. In the presence of matter, the equations of motion will not be given by the vanishing of the covariant divergence of the matter stress-energy alone, but now there will be additional geometric terms. 

Given these field equations, one can linearize them about a flat background to find the evolution equations for the metric perturbation~\cite{Nelson:2010rt,Nelson:2010ru}
\begin{equation}
\left(1 - \beta^{-2} \square_{\eta}\right) \square_{\eta} h_{\mu \nu} = -16 \pi T_{\mu \nu}^{\rm mat}\,,
\label{eq:NCG-h-eq}
\end{equation}
where the term proportional to $\beta^{2} = (-32 \pi \alpha_{0})^{-1}$ acts like a mass term. Here, one has imposed the transverse-traceless gauge (a refinement of Lorenz gauge), which can be shown to exist~\cite{Nelson:2010rt,Nelson:2010ru}. Clearly, even though the full non-linear equations are not covariantly conserved, its linearized version is, as one can easily show that the divergence of the left-hand side of Eq.~\eqref{eq:NCG-h-eq} vanishes. Because of these features, if one works perturbatively in $\beta^{-1}$, then such a theory will only possess the two usual transverse-traceless (spin-2) polarization modes, ie.~it is perturbatively of type $N_{2}$ in the $E(2)$ classification~\cite{Eardleyprd}.  

Let us now discuss whether such a theory satisfies the properties discussed in Section~\ref{sec:properties}. Non-commutative geometry theories clearly possess the fundamental property, as one can always take $\alpha_{0} \to 0$ (or equivalently $\beta^{-2} \to 0$) to recover General Relativity. Therefore, there must exist a sufficiently small $\alpha_{0}$ such that all precision tests carried out to date are satisfied. As for the existence and stability of known solutions, Refs.~\cite{Nelson:2010rt,Nelson:2010ru} have shown that Minkowski spacetime is stable only for $\alpha_{0} < 0$, as otherwise a tachyonic term appears in the evolution of the metric perturbation, as can be seen from Eq.~\eqref{eq:NCG-h-eq}. This then automatically implies that $\beta$ must be real.

Current constraints on Weyl terms of this form come mostly from Solar System experiments. Ni~\cite{Ni:2012sa} recently studied an action of the form of Eq.~\eqref{eq:NCG-action} minimally coupled to matter in light of Solar System experiments. He calculated the relativistic Shapiro time-delay and light deflection about a massive body in this theory and found that observations of the Cassini satellite place constraints on $|\alpha_{0}|^{1/2} < 5.7 \; {\rm{km}}$~\cite{Ni:2012sa}. This is currently the strongest bound we are aware of on $\alpha_{0}$. 

Many solutions of General Relativity are preserved in non-commutative geometries. Regarding black holes, all solutions that are Ricci flat (vacuum solutions of the Einstein equations) are also solutions to Eq.~\eqref{eq:NCG-FEs}. This is because by the second Bianchi identity, one can show that
\begin{equation}
\nabla^{\kappa \lambda} R_{\mu \lambda \nu \kappa} = \nabla^{\kappa}{}_{\nu}R_{\mu \kappa} - \square R_{\mu \nu}\,,
\end{equation}
and the right-hand side vanishes in vacuum, forcing the entire left-hand side of Eq.~\eqref{eq:NCG-FEs} to vanish. This is not so for neutron stars, however, where the equations of motion are likely to be modified, unless they are static~\cite{2010PhRvD..82j4026N}. Moreover, as of now there has been no stability analysis of black hole or stellar solutions and no study of whether the theory is well-posed as an initial value problem, even as an effective theory. Thus, except for the fundamental property, it is not clear that non-commutative geometries satisfy any of the other criteria listed in Section~\ref{sec:properties}.

%------------------------------------------------------------------------------------------------------------------------------
\subsubsection{Gravitational parity violation}
\label{sec:GPV}

Parity, the symmetry transformation that flips the sign of the spatial triad, has been found to be broken in the Standard Model of elementary interactions. Only the combination of a parity transformation, time inversion and charge conjugation (CPT) remains still a true symmetry of the Standard Model. Experimentally, it is curious that the weak interaction exhibits maximal parity violation, while other fundamental forces seem to not exhibit any. Theoretically, parity violation unavoidably arises in the Standard Model~\cite{Bell:1969ts,1969PhRv..177.2426A,AlvarezGaume:1983ig}, as there exist one-loop chiral anomalies that give rise to parity violating terms coupled to lepton number~\cite{Weinberg:1996kr}. In certain sectors of string theory, such as in heterotic and in Type~I superstring theories, parity violation terms are also generated through the Green--Schwarz gauge anomaly-canceling mechanism~\cite{Green:1987mn,Polchinski:1998rr,Alexander:2004xd}. Finally, in loop quantum gravity~\cite{Ashtekar:1988sw}, the scalarization of the Barbero--Immirzi parameter coupled to fermions leads to an effective action that contains parity-violating terms~\cite{Taveras:2008yf,Calcagni:2009xz,Mercuri:2009zt,Gates:2009pt}. Even without a particular theoretical model, one can generically show that effective field theories of inflation generically contain non-vanishing, second-order, parity violating curvature corrections to the Einstein--Hilbert action~\cite{Weinberg:2008hq}. Alternatively, phenomenological parity-violating extensions of General Relativity have been proposed through a scalarization of the fundamental constants of Nature~\cite{Contaldi:2008yz}.

One is then naturally led to ask whether the gravitational interaction is parity invariant in the strong field. A violation of parity invariance would occur if the Einstein--Hilbert action were modified through a term that involved a Levi-Civita tensor and parity invariant tensors or scalars. Let us try to construct such terms with only single powers of the Riemann tensor and a single scalar field $\vartheta$:
\begin{enumerate}
\item[(ia)] $R_{\alpha \beta \gamma \delta} \; \epsilon^{\alpha \beta \gamma \delta}$,
\qquad (ib) $R_{\alpha \beta \gamma \mu} \; \epsilon^{\alpha \beta \gamma \nu} \; \nabla^{\mu}{}_{\nu} \vartheta$,
\item[(ic)] $R_{\alpha \beta \gamma \mu} \; \epsilon^{\alpha \beta \delta \nu} \; \nabla^{\mu\gamma}{}_{\nu\delta} \vartheta$\,,
\qquad (id) $R_{\alpha \zeta \gamma \mu} \; \epsilon^{\alpha \beta \delta \nu} \; \nabla^{\mu\gamma}{}_{\beta \nu\delta}{}^{\zeta} \vartheta$\,.
\end{enumerate}
Option (ia) and (ib) vanish by the Bianchi identities. Options (ic) and (id) include the commutator of covariant derivatives, which can be rewritten in terms of a Riemann tensor, and thus it leads to terms that are at least quadratic in the Riemann tensor. Therefore, no scalar can be constructed that includes contractions with the Levi-Civita tensor from a single Riemann curvature tensor and a single field. One can try to construct a scalar from the Ricci tensor 
\begin{enumerate}
\item[(iia)] $R_{\alpha \beta} \; \epsilon^{\alpha \beta \gamma \delta} \nabla_{\gamma \delta} \vartheta$,
\qquad (iib) $R_{\alpha \beta} \; \epsilon^{\alpha \mu \gamma \delta} \nabla_{\gamma \delta \mu}{}^{\beta} \vartheta$,
\end{enumerate}
but again (iia) vanishes by the symmetries of the Ricci tensor, while (iib) involves the commutator of covariant derivatives, which introduces another power of the curvature tensor. Obviously, the only term one can write with the Ricci scalar would lead to a double commutator of covariant derivatives, leading to extra factors of the curvature tensor. 

One is then forced to consider either theories with two mutually independent fields or theories with quadratic curvature tensors. Of the latter, the only combination that can be constructed and that does not vanish by the Bianchi identities is the so called Pontryagin density, ie.~$R{}^{*}R$, and therefore, the action~\cite{Jackiw:2003pm,Alexander:2009tp}
\begin{equation}
S = \int d^{4}x \sqrt{-g} \left(\kappa \; R + \frac{\alpha}{4} \; \vartheta \; R{}^{*}R\right)\,,
\label{CS-action}
\end{equation}
is the most general, quadratic action with a single scalar field that violates parity invariance, where we have rescaled the $\alpha$ prefactor to follow historical conventions. This action defines non-dynamical Chern--Simons modified gravity, initially proposed by Jackiw and Pi~\cite{Jackiw:2003pm,Alexander:2009tp}. Notice that this is the same as the term proportional to $\alpha_{4}$ in the quadratic gravity action of Eq.~\eqref{eq:quad-action-simped}, except that here $\vartheta$ is prior geometry, ie.~it does not possess self-consistent dynamics or an evolution equation. Such a term violates parity invariance because the Pontryagin density is a pseudo-scalar, while $\vartheta$ is assumed to be a scalar. 

The field equations for this theory are\epubtkFootnote{The tensor ${\cal{K}}^{(1)}_{\mu \nu}$ is sometimes written as $C_{\mu \nu}$ and referred to as the C-tensor.}
\begin{equation}
G_{\mu \nu} + \frac{\alpha}{4 \kappa} {\cal{K}}^{(1)}_{\mu \nu} = \frac{1}{2 \kappa} T_{\mu \nu}^{\rm mat}\,,
\end{equation}
which is simply Eq.~\eqref{FEs} with $(\alpha_{1},\alpha_{2},\alpha_{3})$ set to zero and no stress-energy for $\vartheta$. Clearly, these field equations are not covariantly conserved in vacuum, ie.~taking the covariant divergence one finds the constraint
\begin{equation}
\alpha R{}^{*}R = 0\,.
\label{eq:RR}
\end{equation}
This constraint restricts the space of allowed solutions, for example disallowing the Kerr metric~\cite{Grumiller:2007rv}. Therefore, it might seem that the evolution equations for the metric are now overconstrained, given that the field equations provide 10 differential conditions for the 10 independent components of the metric tensor, while the constraint adds one additional, independent differential condition. Moreover, unless the Pontryagin constraint, Eq.~\eqref{eq:RR}, is satisfied, matter fields will not evolve according to $\nabla^{\mu} T_{\mu \nu}^{\rm mat} = 0$, thus violating the equivalence principle. 

From the field equations, we can derive an evolution equation for the metric perturbation when linearizing about a flat background, namely
\begin{equation}
\square_{\eta} h_{\mu \nu} 
+ \frac{\alpha}{\kappa}  \left(\vartheta_{,\gamma} \; \epsilon_{(\mu}{}^{\gamma \delta \chi} \square_{\eta} h_{\nu) \delta,\chi}
-  \vartheta_{,\gamma}{}^{\zeta} \; \epsilon_{(\mu}{}^{\gamma \delta \chi} h_{|\delta \zeta|,\nu)\chi}
+ \vartheta_{,\gamma}{}^{\zeta} \;  \epsilon_{(\mu}{}^{\gamma \delta \chi} h_{\nu) \delta,\chi \zeta} \right) = -\frac{2}{\kappa} T_{\mu \nu}^{\rm mat}\,.
\label{eq:h-EOM-NDCS}
\end{equation}
in a transverse-traceless gauge, which can be shown to exist in this theory~\cite{Alexander:2007kv,Yunes:2008bu}. The constraint of Eq.~\eqref{eq:RR} is here identically satisfied to second order in the metric perturbation. Without further information about $\vartheta$, however, one cannot proceed any  further, except for a few general observations. As is clear from Eq.~\eqref{eq:h-EOM-NDCS}, the evolution equation for the metric perturbation can contain third time derivatives, which generically will lead to instabilities. In fact, as shown in~\cite{Alexander:2004wk} the general solution to these equations will contain exponentially growing and decaying modes. The theory defined by Eq.~\eqref{CS-action}, however, is an effective theory, and thus, there can be higher order operators not included in this action that may stabilize the solution. Regardless, when studying this theory order-reduction is necessary if one is to consider it an effective model.

Let us now discuss the properties of such an effective theory. Because of the structure of the modification to the field equations, one can always choose a sufficiently small value for $\alpha$ such that all Solar System tests are satisfied. In fact, one can see from the equations in this section that in the limit $\alpha \to 0$, one recovers General Relativity. Non-dynamical Chern--Simons gravity leads to modifications to the non-radiative (near-zone) metric in the so-called gravitomagnetic sector, leading to corrections to Lense--Thirring precession~\cite{Alexander:2007zg,Alexander:2007vt}. This fact has been used to constrain the theory through observations of the orbital motion of the LAGEOS satellites~\cite{Smith:2007jm} to $(\alpha/\kappa) \dot{\vartheta} < 2 \times 10^{4} \; {\textrm{km}}$, or equivalently $(\kappa/\alpha) \dot{\vartheta}^{-1} \gtrsim 10^{-14} \; {\textrm{eV}}$. 
% Note: To convert between our conventions and Smith, et al, notice that -4 l_{smith} = \alpha$ and \kappa = 1/(2 k^{2}) 
% and thus, l = -\alpha/4, while k^{2} = 1/(2 \kappa). Therefore, 
% m_{cs, smith} := -3/(l k^{2} \dot{\theta}) = -3/(-alpha/4 1/(2 kappa) \dot{\theta}) = 24 kappa/(alpha \dot{\theta})  .
% Then, since the Smith bound is m = 24/kappa/(alpha \dot{\theta}) > 10^{-3} km^{-1} --> (\alpha/kappa) \dot\theta < 24 10^{3} km ~ 2e4 km.  
Much better constraints, however, can be placed through observations of the double binary pulsar~\cite{Yunes:2008ua,AliHaimoud:2011bk}: $(\alpha_{4}/\kappa) \dot{\vartheta} < 0.8 \; {\textrm{km}}$. 
% Yacine's conventions. kcs = (8 pi l^{2} \dot\theta)^{-1}, and l^{2} = alpha, and 1/(2 kappa) = 8 pi (with kappa = 1/(16 \pi)),  
% so then kcs = (alpha/(2 kappa) \dot{\theta})^{-1} = 2 kappa/(alpha \dot{\theta}). The constraint he gets is k^{-1} < 0.4 km, 
% so then (alpha/kappa) \dot{\theta} < 0.8 km. 

Some of the sub-properties of the fundamental requirement are satisfied in non-Dynamical Chern--Simons Gravity. On the one hand, all spherically symmetric metrics that are solutions to the Einstein equations are also solutions in this theory for a ``canonical'' scalar field ($\theta \propto t$)~\cite{Grumiller:2007rv}. On the other hand, axisymmetric solutions to the Einstein equations are generically not solutions in this theory. Moreover, although spherically symmetric solutions are preserved, perturbations of such spacetimes that are solutions to the Einstein equations are not generically solutions to the modified theory~\cite{Yunes:2007ss}. What is perhaps worse, the evolution of perturbations to non-spinning black holes have been found to be generically overconstrained~\cite{Yunes:2007ss}. This is a consequence of the lack of scalar field dynamics in the modified theory, which via Eq.~\eqref{eq:RR} tends to overconstrain it. Such a conclusion also suggests that this theory does not posses a well-posed initial value problem. 

One can argue that Non-Dynamical Chern--Simons Gravity is well-motivated from fundamental theories~\cite{Alexander:2009tp}, except that in the latter, the scalar field is always dynamical, instead of having to be prescribed {\emph{a priori}}. Thus, perhaps the strongest motivation for such a model is as a phenomenological proxy to test whether the gravitational interaction remains parity invariant in the strong field, a test that is uniquely suited to this modified model.

%------------------------------------------------------------------------------------------------------------------------------
\subsection{Currently unexplored theories in the gravitational-wave sector}

The list of theories we have here described is by no means exhaustive. In fact, there are many fascinating theories that we have chosen to leave out  because they have not yet been analyzed in the gravitational wave context in detail. Examples of these include the following:
\begin{itemize}
\item Einstein-Aether Theory~\cite{Jacobson:2008aj} and Horava--Lifshitz Theory~\cite{Horava:2009uw};
\item Standard Model Extension~\cite{Colladay:1998fq};
\item Eddington-inspired Born--Infeld gravity\cite{Banados:2010ix};
\item New Massive Gravity\cite{Bergshoeff:2009zz,deRham:2010kj} and Bi-Gravity Theories\cite{Pilo:2011zz,Paulos:2012xe,Hassan:2011zd,Hassan:2011ea};
\end{itemize}
We will update this review with a description of these theories, once a detailed gravitational wave study for compact binaries or supernovae sources is carried out and the predictions for the gravitational waveform observables are made for any physical system plausibly detectable by current or near future gravitational wave experiments.

%% file: detectors.tex
% Xavi write this
%
%   a-GW interferometers
%   b-GW bars
%   c-Pulsar Timing Arrays
%   d-CMB Detectors (Plank and WMAP)
%------------------------------------------------------------------------------------------------------------------------------

%------------------------------------------------------------------------------------------------------------------------------
\subsection{Gravitational-wave interferometers}

Kilometer scale gravitational-wave interferometers have been in operation for over a decade.  These type of
detectors use laser interferometry to monitor the locations of test masses at the ends of the arms with
exquisite precision. Gravitational waves change the relative length of 
the optical cavities in the interferometer (or equivalently, the proper travel time of photons) resulting in a strain
$$
h=\frac{\Delta L}{L},
$$
where $\Delta L$ is the path length difference between the two arms of the interferometer.

Fractional changes in the difference in path lengths along the two arms can be 
monitored to better than 1 part in $10^{20}$.  It is not hard to understand how such 
precision can be achieved. For a simple Michelson interferometer,
a difference in path length of order the size of a fringe can easily be detected. 
For the typically-used, infrared lasers of wavelength $\lambda \sim 1\,\mu$m, 
and interferometer arms of length $L=4$~km, the minimum 
detectable strain is 
$$
h \sim \frac{\lambda}{L} \sim 3 \times 10^{-10}.
$$

This is still far off the $10^{-20}$ mark. In principle, however, changes in the length of the cavities corresponding to 
fractions of a single fringe can also be measured provided we have a sensitive photodiode 
at the dark port of the interferometer, and enough photons to perform the 
measurement. This way we can track changes in the amount of light incident on the photodiode
as the lengths of the arms change and we move over a fringe.  The rate at which photons arrive
at the photodiode is a Poisson process and the fluctuations in the number of photons is $\sim N^{1/2}$, 
where $N$ is the number of photons. Therefore we can track changes in the path length difference of order
$$
\Delta L \sim \frac{\lambda}{N^{1/2}}.
$$
The number of photons depends on the laser power $P$, and the amount of time available to 
perform the measurement. For a gravitational wave of frequency $f$, we can collect photons for a time 
$t \sim 1/f$, so the number of photons is
$$
N \sim \frac{P}{f h_p \nu},
$$
where $h_p$ is Planck's constant and $\nu=c/\lambda$ is the laser frequency. For a typical laser 
power $P \sim 1$~W, a gravitational wave frequency $f=100$~Hz, and $\lambda \sim 1\,\mu$m the number of photons
is
$$
N \sim 10^{16}, 
$$
so that the strain we are sensitive to becomes
$$
h \sim 10^{-18}.
$$

The sensitivity can be further improved by increasing the effective length of the arms. 
In the LIGO instruments, for example, each of the two arms forms a resonant Fabry-Perot cavity. For
gravitational wave frequencies smaller than the inverse of the light storage time, the light in the 
cavities makes many back and forth trips in the arms while the wave is traversing the instrument. 
For gravitational waves of frequencies around 
100~Hz and below, the light makes about 
a thousand back and forth trips while the gravitational wave is traversing the interferometer, which results in a 
3 order of magnitude improvement in sensitivity,
$$
h \sim 10^{-21}.
$$
For frequencies larger than 100~Hz the number of round trips the light makes in the Fabry-Perot cavities
while the gravitational wave is traversing the instrument is reduced and the sensitivity is degraded.

The proper light travel time of photons in interferometers is controlled by the metric perturbation, which can be expressed as a sum over polarization modes
\begin{equation}
h_{ij}(t,\vec{x}) = \sum_{A} h^A_{ij}(t,\vec{x}),
\end{equation}
where $A$, labels the six possible polarization modes in metric 
theories of gravity. The metric perturbation for each mode can be written in terms of a plane wave expansion,
\begin{equation}
 h^A_{ij}(t,\vec{x})=\int_{-\infty}^{\infty}df\, \int_{S^2}d\hat{\Omega}\, 
e^{i2\pi f(t-\hat{\Omega}\cdot\vec{x})}
\tilde h^A(f,\hat{\Omega})\epsilon_{ij}^A(\hat{\Omega}).
\label{pwexp2}
\end{equation}
Here $f$ is the frequency of the gravitational waves, 
$\vec k= 2 \pi f \hat \Omega$ is the wave vector, $\hat \Omega$ is a unit vector
that points in the direction of propagation of the gravitational waves, $e_{ij}^{A}$ is the $A$th polarization tensor,
with $i,j=x,y,z$ spatial indices. The
metric perturbation due to mode $A$ from the direction $\hat \Omega$  can be written by
integrating over all frequencies,
\begin{equation}
 h^A_{ij}(t -\hat \Omega \cdot \vec{x})=\int_{-\infty}^{\infty}df\
e^{i2\pi f(t-\hat{\Omega}\cdot\vec{x})}
\tilde h^A(f,\hat{\Omega}) \epsilon_{ij}^A(\hat{\Omega}).
\label{pwexp}
\end{equation}
By integrating Eq.~\eqref{pwexp2} over all frequencies we have an
expression for the metric perturbation from a particular direction
$\hat \Omega$, i.e. only a function of $t -\hat \Omega \cdot \vec{x}$.
The full metric perturbation due to a gravitational wave from a 
direction $\hat \Omega$ can be written as a sum over all polarization modes
\begin{equation}
h_{ij}(t -\hat \Omega \cdot \vec{x})=\sum_A h^A(t-\hat \Omega \cdot \vec{x}) \epsilon_{ij}^A (\hat \Omega).
\end{equation}

The response of an interferometer to gravitational waves is generally referred to as the 
antenna pattern response, and depends on the geometry of the detector and the
direction and polarization of the gravitational wave. To derive the
antenna pattern response 
of an interferometer for all
six polarization modes we follow the discussion in~\cite{Nishizawa:2009bf} closely. 
For a gravitational wave propagating in the $z$ direction, the
polarization tensors are as follows
\begin{eqnarray}
\epsilon_{ij}^{+}&=& 
\left(
\begin{array}{ccc} 
1 & 0 & 0  \\
0 & -1 & 0  \\
0 & 0 & 0 
\end{array}
\right), 
\epsilon_{ij}^{\times}= 
\left(
\begin{array}{ccc} 
0 & 1 & 0  \\
1 & 0 & 0  \\
0 & 0 & 0 
\end{array}
\right),\nonumber \\ 
\epsilon_{ij}^{x} &=&
\left(
\begin{array}{ccc} 
0 & 0 & 1  \\
0 & 0 & 0  \\
1 & 0 & 0 
\end{array}
\right), 
\epsilon_{ij}^{y}= 
\left(
\begin{array}{ccc} 
0 & 0 & 0  \\
0 & 0 & 1  \\
0 & 1 & 0 
\end{array}
\right),  \nonumber \\
\epsilon_{ij}^{b}&=& 
\left(
\begin{array}{ccc} 
1 & 0 & 0  \\
0 & 1 & 0  \\
0 & 0 & 0 
\end{array}
\right), 
\epsilon_{ij}^{\ell}=
\left(
\begin{array}{ccc} 
0 & 0 & 0  \\
0 & 0 & 0  \\
0 & 0 & 1 
\end{array}
\right), 
\label{polarizationsz} 
\end{eqnarray}
where the superscripts $+$, $\times$, $x$, $y$, $b$, and $\ell$
correspond to the plus, cross, vector-x, vector-y,  breathing, and
longitudinal modes.

%%%%%%%%% 
\epubtkImage{}{%
\begin{figure}[htbp]
\centerline{\includegraphics[width=7cm]{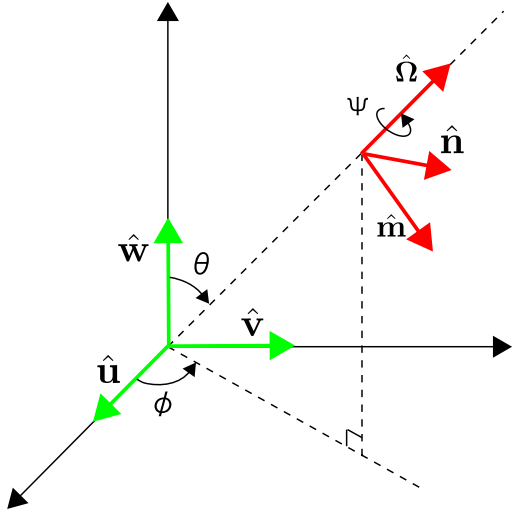}}
\caption{Detector coordinate system and gravitational wave coordinate
  system.}
\label{coodinatesystems}
\end{figure}}
%%%%%%%%%

Suppose that the coordinate system for the detector is $\hat x=
(1,0,0), \hat y = (0,1,0),
\hat z = (0,0,1)$, as in Fig.~\ref{coodinatesystems}. Relative to the detector, the gravitational wave coordinate system is 
rotated by angles $(\theta, \phi)$,
$\hat x ^{\prime} =(\cos \theta \cos \phi , \cos \theta \sin \phi , -\sin \theta)$, 
$\hat y^{\prime} = (- \sin \phi , \cos \phi , 0)$, and 
$\hat z^{\prime} = (\sin \theta \cos \phi , \sin \theta \sin \phi , \cos \theta)$.
We still have the freedom to perform a rotation about the gravitational wave propagation direction 
which introduces the polarization angle $\psi$, 
\begin{equation}
\begin{array}{lll} 
\displaystyle
\hat m= \hat x^{ \prime} \cos \psi + \hat y^{\prime} \sin \psi\,,\\ 
\displaystyle 
\hat n= - \hat x^{ \prime} \sin \psi + \hat y ^{\prime} \cos \psi\,,
\\ 
\displaystyle 
\hat \Omega = \hat z^{ \prime}\,.
\end{array}
\nonumber
\end{equation}
The coordinate systems $(\hat x,\hat y,\hat z)$ 
and  $(\hat m,\hat n,\hat \Omega)$ 
are also shown in Fig.~\ref{coodinatesystems}. 
To generalize the polarization tensors in Eq.~\eqref{polarizationsz} to a wave coming from a 
direction $\hat{{\Omega}}$, we use the unit vectors $\hat{{m}}$, $\hat{{n}}$, 
and $\hat{{\Omega}}$ as follows
\begin{eqnarray}
\epsilon^{+} &=& \hat{{m}} \otimes \hat{{m}} -\hat{{n}} \otimes \hat{{n}} , \nonumber \\
\epsilon^{\times} &=& \hat{{m}} \otimes \hat{{n}} +\hat{{n}} \otimes \hat{{m}} , \nonumber \\
\epsilon^{x} &=& \hat{{m}} \otimes \hat{{\Omega}} +\hat{{\Omega}} \otimes \hat{{m}} , \nonumber \\
\epsilon^{y} &=& \hat{{n}} \otimes \hat{{\Omega}}+\hat{{\Omega}} \otimes \hat{{n}}, \nonumber \\
\epsilon^{b} &=& \hat{{m}} \otimes \hat{{m}} + \hat{{n}} \otimes \hat{{n}} ,  \nonumber \\
\epsilon^{\ell} &=& \hat{{\Omega}} \otimes \hat{{\Omega}}
.\end{eqnarray}
For LIGO and VIRGO the arms are perpendicular
so that the antenna pattern response can be written as the difference of projection of the 
polarization tensor onto each of the interferometer arms,
$$
F^A (\hat \Omega, \psi) = \frac{1}{2} 
\left(\hat x^i \hat x^j - \hat y^i \hat y^j  \right)\epsilon^A_{ij} (\hat \Omega, \psi).
$$
This means that the strain measured by an interferometer due to a
gravitational wave from direction $\hat \Omega$ and polarization angle
$\psi$ takes the
form
\begin{equation}
h(t)=\sum_A h_A(t-\hat \Omega \cdot x) F^A (\hat \Omega, \psi).
\label{strain}
\end{equation}
Explicitly, the antenna pattern functions are,
\begin{eqnarray}
F^{+}(\theta, \phi, \psi) &=& \frac{1}{2} (1+ \cos ^2 \theta ) \cos 2\phi \cos 2 \psi
 - \cos \theta \sin 2\phi \sin 2 \psi, \nonumber \\
F^{\times}(\theta, \phi, \psi) &=& -\frac{1}{2} (1+ \cos ^2 \theta ) \cos 2\phi \sin 2 \psi 
 -  \cos \theta \sin 2\phi \cos 2 \psi,\nonumber \\
F^{x}(\theta, \phi, \psi) &=& \sin \theta \,(\cos \theta \cos 2 \phi
\cos \psi -\sin 2\phi \sin \psi) ,\nonumber \\
F^{y}(\theta, \phi, \psi) &=& - \sin \theta \,(\cos \theta \cos 2 \phi
\sin \psi +\sin 2\phi \cos \psi) ,\nonumber \\
F^{b}(\theta, \phi) &=& -\frac{1}{2} \sin^2 \theta \cos 2\phi,
\nonumber \\
F^{\ell}(\theta, \phi) &=& \frac{1}{2} \sin^2 \theta \cos 2\phi. 
\label{IFOresponse} 
\end{eqnarray}
The dependence on the polarization angles $\psi$, reveals that 
the $+$ and $\times$ polarizations are spin-2 tensor modes, 
the $x$ and $y$ polarizations are spin-1 vector modes, and the $b$ 
and $\ell$ polarizations are spin-0 scalar modes. Note that for interferometers
the antenna pattern responses of the scalar modes are degenerate. Figure
\ref{LIGOresponse} shows the antenna patterns for the various
polarizations given in Eq.~\eqref{IFOresponse} with $\psi=0$. The color indicates the
strength of the response with red being the strongest and blue being
the weakest.

\epubtkImage{}{%
\begin{figure}[htbp]
\centerline{\includegraphics[width=15cm]{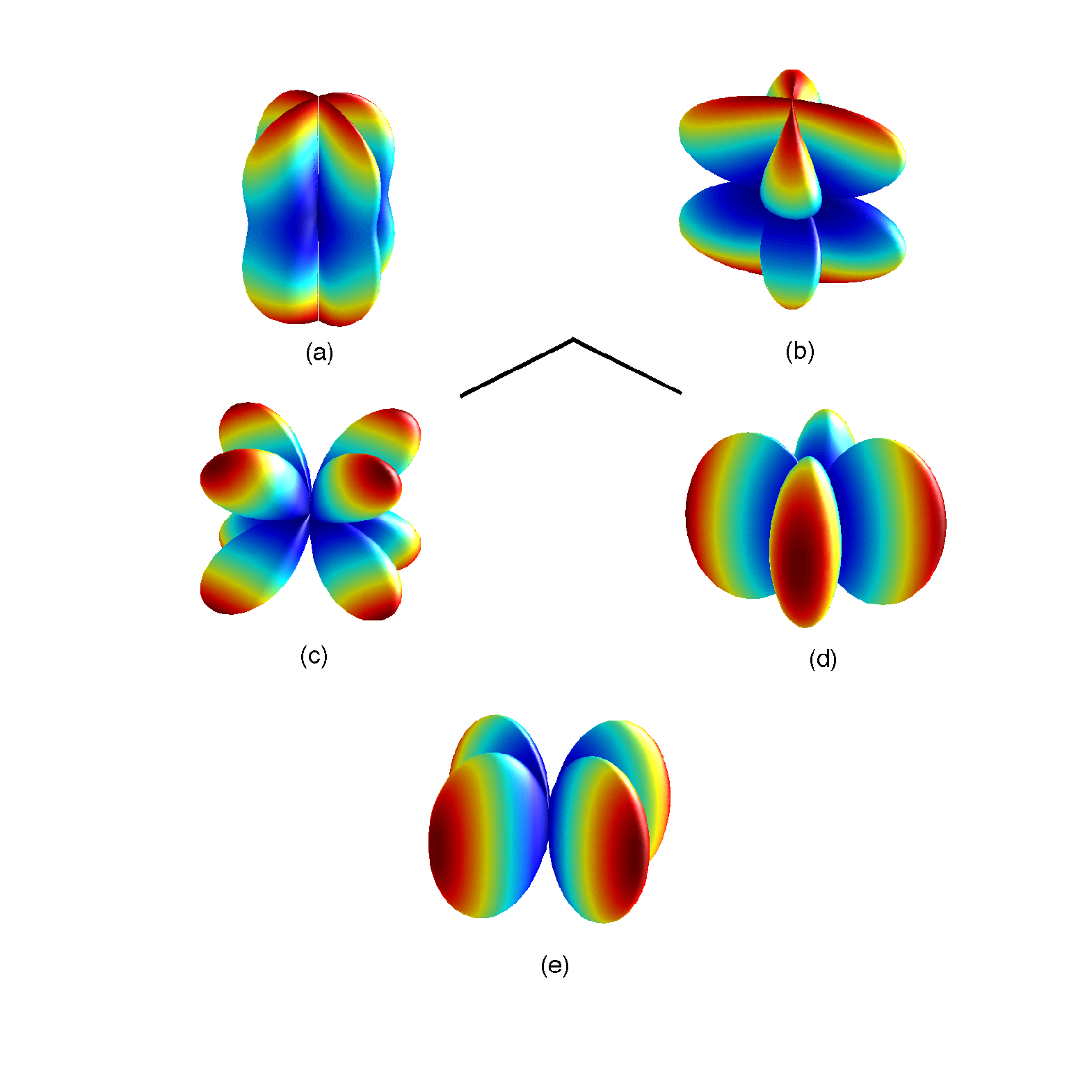}}
\caption{Antenna pattern response functions of an interferometer (see
  Eqs.~\eqref{IFOresponse}) for $\psi=0$. Panels (a) and (b) 
show the plus ($|F_+|$) and cross ($|F_{\times}|$) modes, panels (c) and (d) the vector x and vector y 
modes ($|F_{x}|$ and $|F_{y}|$), and panel (e) shows the scalar modes
(up to a sign, it is the same for both breathing and
longitudinal). Color  indicates the
strength of the response with red being the strongest and blue being
the weakest. The black lines near the center give the orientation of
the interferometer arms.}
\label{LIGOresponse}
\end{figure}}

%------------------------------------------------------------------------------------------------------------------------------
\subsection{Pulsar timing arrays}

Neutron stars can emit powerful beams of radio waves from their
magnetic poles. If the rotational and magnetic axes are not aligned,
the beams sweep through space like the beacon on a lighthouse.  
If the line of sight is aligned with the magnetic axis at any point 
during the neutron star's rotation the star is observed as a source of
periodic radio wave bursts.  Such a neutron star is referred to as a pulsar.
Due to their large moment of inertia pulsars are very stable rotators,
and their radio pulses arrive on Earth with extraordinary regularity. Pulsar timing experiments
exploit this regularity: gravitational waves are expected to cause
fluctuations in the time of arrival of radio pulses from
pulsars. 

The effect of a gravitational wave on the pulses propagating from a
pulsar to Earth was first computed in the late 1970s by Sazhin and Detweiler~\cite{saz78,det79}.
Gravitational waves induce a redshift in the pulse train 
\begin{equation}
z(t,\hat{\Omega}) = 
\frac{1}{2}
\frac{\hat{p}^i\hat{p}^j}{1+\hat{\Omega}\cdot\hat{p}}\Delta h_{ij},
\label{eqzsom}
\end{equation}
where $\hat p$ is a unit vector that points in the direction of the pulsar,
$\hat \Omega$ is the unit vector of gravitational wave propagation,
and 
\begin{equation}
\Delta h_{ij}
\equiv
h_{ij}(t_{\rm e},\hat{\Omega}) - 
h_{ij}(t_{\rm p},\hat{\Omega}),
\label{delhdef}
\end{equation}
is the difference in the metric perturbation at the pulsar when the
pulse was emitted and the metric perturbation on Earth when the pulse
was received. The inner product in Eq.~\eqref{eqzsom} is computed with 
the Euclidean metric. 

In pulsar timing experiments it is not the redshift, but rather the 
\emph{timing residual} that is measured.  The times of arrival of
pulses are measured and the timing residual is produced by subtracting
off a model that includes the rotational frequency of the pulsar, the
spin-down (frequency derivative), binary parameters if the pulsar is 
in a binary, sky location and proper motion, etc. The timing
residual induced by a gravitational wave, $R(t)$, is just the integral of the redshift
\begin{equation}
R(t)\equiv \int_0^t dt'\, z(t').
\end{equation}
Times of arrival (TOAs) are measured a few times a year over the
course of several years allowing for gravitational waves in the
nano-Hertz band to be probed.  Currently, the best timed pulsars
have residual RMSs of a few 10s of ns over a few years. 

The equations above can be used
to estimate the strain sensitivity of pulsar timing experiments.  
For gravitational waves of frequency $f$ the expected
induced residual is
$$
R \sim \frac{h}{f},
$$
so that for pulsars with RMS residuals $R \sim 100$~ns, and gravitational waves
of frequency $f \sim 10^{-8}$~Hz, gravitational waves with strains
$$
h \sim Rf \sim 10^{-15}
$$
would produce a measurable effect.

To find the antenna pattern response of the pulsar-Earth system, we are free to place the pulsar on the $z$-axis. The response to gravitational waves of
different polarizations can then be written as
$$
F^A (\hat \Omega, \psi) = \frac{1}{2} \frac{\hat z^i \hat z^j}{1+ \cos
 \theta} \epsilon^A_{ij} (\hat \Omega, \psi),
$$
which allows us to express the Fourier transform of \eqref{eqzsom} as
\begin{eqnarray}\label{FTantpatts}
	\tilde{z}(f, \hat{\Omega}) = \left(1-e^{-2 \pi i f L 
(1+\hat{\Omega} \cdot \hat{p})} \right) \sum_A \tilde h_A(f,\hat{\Omega})F^A(\hat{\Omega})
\end{eqnarray}
where the sum is over all possible GW polarizations: $A=+, 
\times, x, y, b,l$, and $L$ is the distance to the pulsar. 

Explicitly,
\begin{eqnarray}
F^{+}(\theta, \psi) &=& \sin ^2 \frac{\theta}{2}  \cos 2 \psi\\
F^{\times}(\theta,  \psi) &=&  -\sin ^2 \frac{\theta}{2}  \sin 2 \psi\\
F^{x}(\theta, \psi) &=&  -\frac{1}{2} \frac{\sin 2 \theta }{1+\cos \theta}  \cos \psi,\\
F^{y}(\theta, \psi) &=& \frac{1}{2} \frac{\sin 2 \theta }{1+\cos \theta}\sin \psi,\\
F^{b}(\theta) &=&  \sin ^2 \frac{\theta}{2}\\
F^{\ell}(\theta) &=&  \frac{1}{2} \frac{\cos ^2 \theta}{1+\cos \theta}.
\label{PulsarResponse} 
\end{eqnarray}
Just like for
the interferometer case, the dependence on the polarization angle $\psi$, reveals that 
the $+$ and $\times$ polarizations are spin-2 tensor modes, 
the $x$ and $y$ polarizations are spin-1 vector modes, and the $b$ 
and $\ell$ polarizations are spin-0 scalar modes. Unlike
interferometers, the antenna pattern responses of the pulsar-Earth system do not
depend on the azimuthal angle of the gravitational wave, and
the scalar modes are not degenerate. 

\epubtkImage{}{%
\begin{figure}[htbp]
\centerline{\includegraphics[width=15cm]{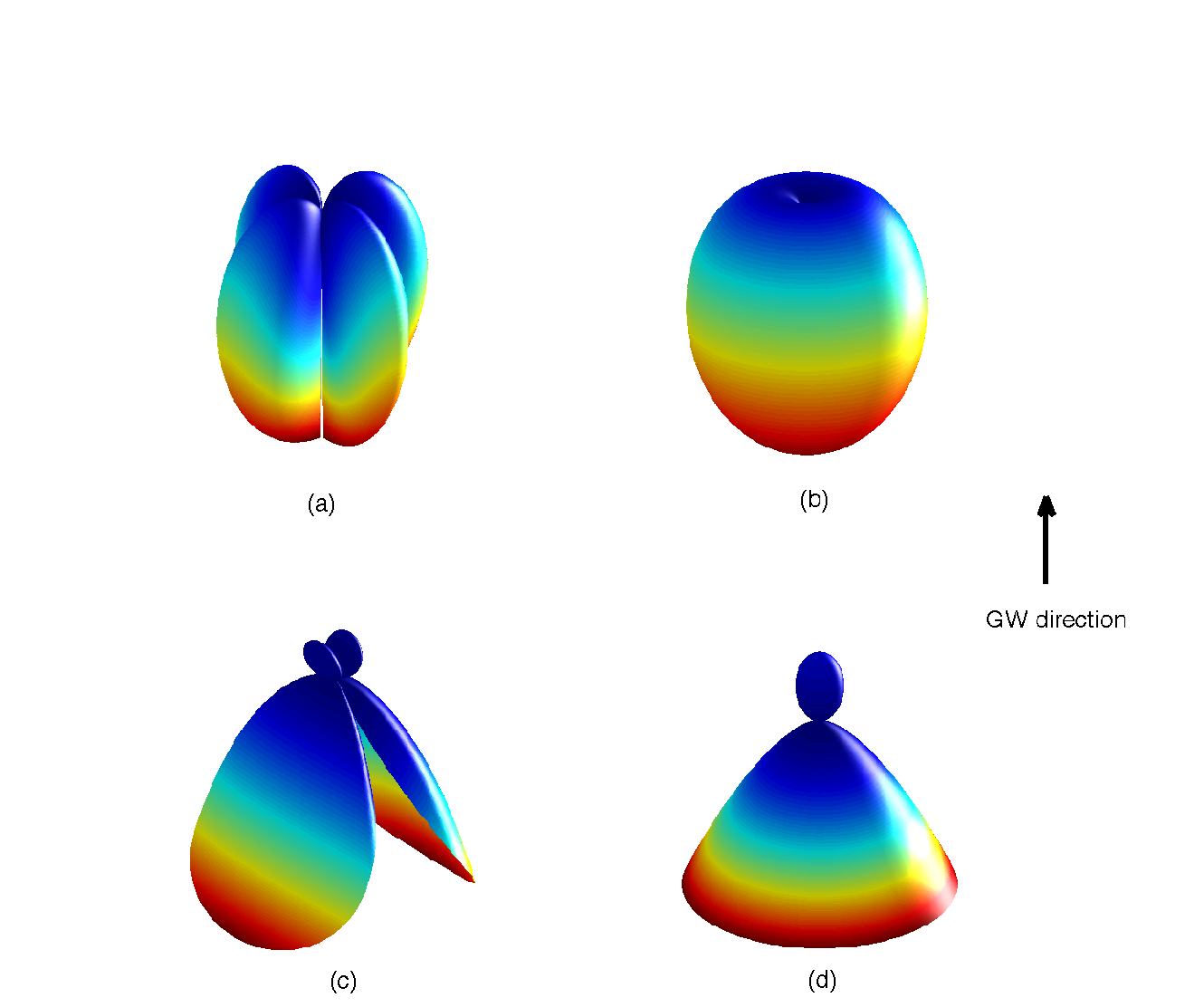}}
\caption{Antenna patterns for the pulsar-Earth system. The plus mode
  is shown in (a), breathing modes in (b), the vector-x mode in (c), and longitudinal 
	modes in (d), as computed from Eq.~\eqref{PulsarResponse2}. The
        cross mode and the vector-y mode are
	rotated versions of the plus mode and the vector-x mode, respectively,  so we did not include them here.
	The GW propagates in the positive z-direction with the Earth at the origin, 
	and the antenna pattern depends on the pulsar's location.
        The color indicates the strength of the response, red being
        the largest and blue the smallest.}
\label{PTAantenna}
\end{figure}}

In the literature, it is common to write the antenna pattern response
by fixing the GW direction and changing the location of the pulsar. In
this case the antenna pattern responses are~\cite{LeeJenetPrice,daSilvaAlves:2011fp,Chamberlin:2011ev},
\begin{eqnarray}
\tilde F^{+}(\theta_p, \phi_p) &=& \sin ^2 \frac{\theta_p}{2}  \cos 2
\phi_p \nonumber \\
\tilde F^{\times}(\theta_p,  \phi_p) &=& \sin ^2 \frac{\theta_p}{2}  \sin 2 \phi_p \nonumber \\
\tilde F^{x}(\theta_p, \phi_p) &=&  \frac{1}{2} \frac{\sin 2 \theta_p }{1+\cos \theta_p}  \cos \phi_p,\nonumber \\
\tilde F^{y}(\theta_p, \phi_p) &=& \frac{1}{2} \frac{\sin 2 \theta_p }{1+\cos \theta_p}\sin \phi_p,\nonumber \\
\tilde F^{b}(\theta_p) &=&  \sin ^2 \frac{\theta_p}{2}\nonumber \\
\tilde F^{\ell}(\theta_p) &=&  \frac{1}{2} \frac{\cos ^2 \theta_p}{1+\cos \theta_p},
\label{PulsarResponse2} 
\end{eqnarray} 
where $\theta_p$ and $\phi_p$ are the polar and azimuthal angles
of the vector pointing to the pulsar.  Up to signs, these expressions are the same
as Eq.~\eqref{PulsarResponse} taking $\theta \rightarrow \theta_p$ and
$\psi \rightarrow \phi_p$.
This is because fixing the GW propagation direction
while allowing the pulsar location to change 
is analogous to fixing the pulsar position while 
allowing the direction of GW propagation to change--there is degeneracy in the GW polarization 
angle and the pulsar's azimuthal angle $\phi_p$. For example, changing
the polarization angle of a
gravitational wave traveling in the $z$-direction is the same as
performing a rotation about the $z$-axis that changes the pulsar's
azimuthal angle. Antenna patterns for the pulsar-Earth system using
Eqs.~\eqref{PulsarResponse2} are shown
in Fig.~\ref{PTAantenna}.  The color indicates the strength of the response, red being
the largest and blue the smallest.

%% file: techniques.tex
% Xavi write this
%
%   a-Bust Analysis
%   b-Inspiral Analysis
%        b1-Matched Filtering and Fisher Analysis
%        b2-Bayesian Theory and Model Testing
%   c-Pulsar Analysis
%   d-Stochastic Analysis
%------------------------------------------------------------------------------------------------------------------------------

\newcommand{\Deqn}[1]{{Eq.~(\ref{#1})}}
\newcommand{\Deqns}[1]{{Eqs.~(\ref{#1})}}
\newcommand{\Ceqn}[1]{{Equation~(\ref{#1})}}
\newcommand{\Dfig}[1]{{Fig.~\ref{#1}}}
\newcommand{\beq}{\begin{equation}}
\newcommand{\eeq}{\end{equation}}
\newcommand{\bea}{\begin{eqnarray}}
\newcommand{\eea}{\end{eqnarray}}
\newcommand{\tr}{{\mathop{\mathrm{tr}}\nolimits}}

%------------------------------------------------------------------------------------------------------------------------------
\subsection{Coalescence analysis}

Gravitational waves emitted during the inspiral, merger and ringdown of compact binaries are the most studied in the context of data analysis and parameter estimation. In this section, we will review some of the main data analysis techniques employed in the context of parameter estimation and tests of General Relativity. We begin with a discussion of matched filtering and Fisher theory (for a detailed review, see~\cite{Finn:1992xs,Chernoff:1993th,Cutler:1994ys,Finn:2001qi,lrr-2012-4}). We then continue with a discussion of Bayesian parameter estimation and hypothesis testing (for a detailed review, see~\cite{Sivia-Skilling,Gregory,Cornish:2007if,Littenberg:2009bm}). 

%------------------------------------------------------------------------------------------------------------------------------
\subsubsection{Matched filtering and Fishers analysis}

When the detector noise $n(t)$ is Gaussian and stationary, and when the signal $s(t)$ is known very well, the optimal detection strategy is matched filtering. For any given realization, such noise can be characterized by its power spectral density $S_{n}(f)$, defined via 
\begin{equation}
\left< \tilde{n}(f) \, \tilde{n}^{*}(f')\right> = \frac{1}{2} S_{n}(f) \delta\left(f - f'\right)\,,
\label{eq:Sn(f)-def}
\end{equation}
where recall that the overhead tilde stands for the Fourier transform, the overhead star stands for complex conjugation and the brackets for the expectation value.

The detectability of a signal is determined by its {\emph{signal-to-noise ratio}} or SNR, which is defined via
\begin{equation}
\rho^{2} = \frac{\left(s|h\right)}{\sqrt{\left(h|n\right)\left(n|h\right)}}\,,
\end{equation}
where $h$ is a template with parameters $\lambda^{i}$ and we have defined the inner product
\begin{equation}
\left(A|B\right) \equiv 4 \Re \int_{0}^{\infty} \frac{\tilde{A}^{*} \tilde{B}}{S_{n}} df\,.
\label{eq:inner-prod-def}
\end{equation}
If the templates do not exactly match the signal, then the signal-to-noise ratio is reduced by a factor of ${\cal{M}}$, called the {\emph{match}}:
\begin{equation}
\bar{{\cal{M}}} \equiv \frac{\left(s|h\right)}{\sqrt{\left(s|s\right)\left(h|h\right)}}\,,
\end{equation}
where $1 - \bar{{\cal{M}}} = {\cal{MM}}$ is the mismatch.

For the noise assumptions made here, the probability of measuring $s(t)$ in the detector output, given a template $h$, is given by
\begin{equation}
p \propto e^{-\left(s - h|s -h\right)/2}\,,
\label{eq:likelihood}
\end{equation}
and thus the waveform $h$ that best fits the signal is that with best-fit parameters such that the argument of the exponential is minimized. For large signal-to-noise ratio, the best-fit parameters will have a multivariate Gaussian distribution
centered on the true values of the signal $\hat{\lambda}^{i}$, and thus, the waveform parameters that best fit the signal minimize the argument of the exponential. The parameter errors $\delta \lambda ^{i}$ will be distributed according to
\begin{equation}
p(\delta \lambda^{i}) \propto e^{-\frac{1}{2} \Gamma_{ij} \delta \lambda^{i} \delta \lambda^{j}}\,,
\end{equation}
where $\Gamma_{ij}$ is the {\emph{Fisher matrix}}
\begin{equation}
\Gamma_{ij} \equiv \left.\left(\frac{\partial h}{\partial \lambda^{i}}\right|\frac{\partial h}{\partial \lambda^{j}}\right)\,.
\end{equation}
The root-mean-squared ($1\sigma$) error on a given parameter $\lambda^{\bar{i}}$ is then
\begin{equation}
\label{Fisher-error}
\sqrt{\left<(\delta \lambda^{\bar{i}})^{2}\right>} = \sqrt{\Sigma^{\bar{i} \bar{i}}}\,,
\end{equation}
where $\Sigma^{ij} \equiv (\Gamma_{ij})^{-1}$ is the variance-covariance matrix and summation is not implied in Eq.~\eqref{Fisher-error} ($\lambda^{\bar{i}}$ denotes a particular element of the vector $\lambda^{i}$). This root-mean-squared error is sometimes referred to as the {\emph{statistical}} error in the measurement of $\lambda^{\bar{i}}$. One can use Eq.~\eqref{Fisher-error} to estimate how well modified gravity parameters can be measured. Put another way, if a gravitational wave were detected and found consistent with General Relativity, Eq.~\eqref{Fisher-error} would provide an estimate of how close to zero these modified gravity parameters would have to be. 

The Fisher method to estimate projected constraints on modified gravity theory parameters is as follows. First, one constructs a waveform model in the particular modified gravity theory one wishes to constrain. Usually, this waveform will be similar to the General Relativity one, but it will contain an additional parameter, $\kappa$, such that the template parameters are now $\lambda^{i}$ plus $\kappa$. Let us assume that as $\kappa \to 0$, the modified gravity waveform reduces to the General Relativity expectation. Then, the accuracy to which $\kappa$ can be measured, or the accuracy to which we can say $\kappa$ is zero, is approximately $(\Sigma^{\kappa \kappa})^{1/2}$, where the Fisher matrix associated with this variance-covariance matrix must be computed with the non-General Relativity model evaluated at the General Relativity limit ($\kappa \to 0$). Such a method for estimating how well modified gravity theories can be constrained was pioneered by Will in~\cite{Will:1994fb,Poisson:1995ef}, and since then, it has been widely employed as a first-cut estimate of the accuracy to which different theories can be constrained.  

The Fisher method described above can dangerously lead to incorrect results if abused~\cite{Vallisneri:2007ev,Vallisneri:2011ts}. One must understand that this method is suitable only if the noise is stationary and Gaussian and if the signal-to-noise ratio is sufficiently large. How large a signal-to-noise ratio is required for Fisher methods to work depends somewhat on the signals considered, but usually for applications concerning tests of General Relativity, one would be safe with $\rho \gtrsim 30$ or so. In real data analysis, the first two conditions are almost never satisfied. Moreover, the first detections that will be made will probably be of low signal-to-noise ratio, ie.~$\rho \sim 8$, for which again the Fisher method fails. In such cases, more sophisticated parameter estimation methods need to be employed. 

%------------------------------------------------------------------------------------------------------------------------------
\subsubsection{Bayesian theory and model testing}

Bayesian theory is ideal for parameter estimation and model selection. Let us then assume that a detection has been made and that the gravitational wave signal in the data can be described by some model ${\cal{M}}$, parameterized by the vector $\lambda^{i}$. Using Bayes' theorem, the posterior distribution function (PDF) or the probability density function for the model parameters, given data $d$ and model ${\cal{M}}$, is 
\begin{equation}
p(\lambda^{i}|d,{\cal{M}}) = \frac{p(d|\lambda^{i},{\cal{M}}) p(\lambda^{i}|{\cal{M}})}{p(d|{\cal{M}})}\,.
\label{eq:PDF}
\end{equation}
Obviously, the global maximum of the PDF in the parameter manifold gives the best fit parameters for that model. The prior probability density $p(\lambda^{i}|{\cal{M}})$ represents our prior beliefs of the parameter range in model ${\cal{M}}$. The marginalized likelihood or {\emph{evidence}}, is the normalization constant
\begin{equation}
p(d|{\cal{M}}) = \int d\lambda^{i} p(d|\lambda^{i},{\cal{M}}) \; p(\lambda^{i}|{\cal{M}})\,,
\label{eq:evidence}
\end{equation}
which clearly guarantees that the integral of Eq.~\eqref{eq:PDF} integrates to unity. 
The quantity $p(d|\lambda^{i},{\cal{M}})$ is the likelihood function, which is simply given by Eq.~\eqref{eq:likelihood}, with a given normalization. In that equation we used slightly different notation, with $s$ being the data $d$ and $h$ the template associated with model ${\cal{M}}$ and parameterized by $\lambda^{i}$.  The marginalized PDF, which represents the probability density function for a given parameter $\lambda^{\bar{i}}$ (recall that $\lambda^{\bar{i}}$ is a particular element of $\lambda^{i}$), after marginalizing over all other parameters, is given by
\begin{equation}
p(\lambda^{\bar{i}}|d,{\cal{M}}) = \int_{i \neq \bar{i}} d \lambda^{i} p(\lambda^{i}|{\cal{M}}) p(d|\lambda^{i},{\cal{M}})\,,
\label{eq:marginalizedPDF}
\end{equation}
where the integration is not to be carried out over $\bar{i}$. 
 
Let us now switch gears to model selection. In hypothesis testing, one wishes to determine whether the data is more consistent with {\emph{hypothesis A}} (eg.~that a General Relativity waveform correctly models the signal) or with {\emph{hypothesis B}} (eg.~that a non-General Relativity waveform correctly models the signal). Using Bayes' theorem, the PDF for model $A$ given the data is
\begin{equation}
p(A|d) = \frac{p(d|A) p(A)}{p(d)}\,,
\label{eq:PDF2}
\end{equation}
As before, $p(A)$ is the prior probability of hypothesis $A$, namely the strength of our prior belief that hypothesis $A$ is correct. The normalization constant $p(d)$ is given by
\be
p(d) = \int d{\cal{M}} \; p(d|{\cal{M}}) \; p({\cal{M}})\,.
\label{eq:evidence2}
\ee
where the integral is to be taken over all models. Thus, it is clear that this normalization constant does not depend on the model. Similar relations hold for hypothesis $B$ by replacing $A \to B$ in Eq.~\eqref{eq:PDF2}.

When hypothesis A and B refer to fundamental theories of nature we can take different viewpoints regarding the priors. If we argue that we know nothing about whether hypothesis A or B better describes nature, then we would assign equal priors to both hypotheses. If, on the other hand, we believe General Relativity is the correct theory of Nature, based on all previous experiments performed in the Solar System and with binary pulsars, then we would assign $p(A)>p(B)$. This assigning of priors necessarily biases the inferences derived from the calculated posteriors, which is sometimes heavily debated when comparing Bayesian theory to a frequentist approach. However, this ``biasing'' is really unavoidable and merely a reflection of our state of knowledge of nature (for a more detailed discussion on such issues, please refer to~\cite{Littenberg:2009bm}).      

The integral over all models in Eq.~\eqref{eq:evidence2} can never be calculated in practice, simply because we do not know all models. Thus, one is forced to investigate {\emph{relative}} probabilities between models, such that the normalization constant $p(d)$ cancels out. The so-called {\emph{odds-ratio}} is defined by
\begin{equation}
{\cal{O}}_{A,B} = \frac{p(A|d)}{p(B|d)} = \frac{p(A)}{p(B)} {\cal{B}}_{A,B}\,,
\end{equation}
where ${\cal{B}}_{A,B} \equiv p(d|A)/p(d|B)$ is the {\emph{Bayes Factor}} and the prefactor $p(A)/p(B)$ is the {\emph{prior odds}}. The odds-ratio is a convenient quantity to calculate because the evidence $p(d)$, which is difficult to compute, actually cancels out. Recently, Vallisneri~\cite{Vallisneri:2012qq} has investigated the possibility of calculating the odds-ratio using only frequentist tools and without having to compute full evidences. The odds-ratio should be interpreted as the betting-odds of model $A$ over model $B$. For example, an odds-ratio of unity means that both models are equally supported by the data, while an odds-ratio of $10^{2}$ means that there is a 100 to 1 odds that model $A$ better describes the data than model $B$. 

The main difficulty in Bayesian inference (both in parameter
estimation and model selection) is sampling the PDF sufficiently
accurately. Several methods have been developed for this purpose, but
currently the two main workhorses in gravitational wave data analysis
are Markov chain Monte Carlo and Nested Sampling. In the former, one
samples the likelihood through the Metropolis--Hastings
algorithm~\cite{Metropolis:1979dx,Hastings:1970,Cornish:2005qw,Rover:2006ni}. This
is computationally expensive in high-dimensional cases, and thus,
there are several techniques to improve the efficiency of the method,
e.g., parallel tempering~\cite{1986PhRvL..57.2607S}. Once the PDF has been sampled, one can then calculate the evidence integral, for example via thermodynamic integration~\cite{Veitch:2008wd,Feroz:2009de,vanderSluys:2008qx}. In Nested Sampling, the evidence is calculated directly by laying out a fixed number of points in the prior volume, which are then allowed to move and coalesce toward regions of high posterior probability. With the evidence in hand, one can then infer the PDF. As in the previous case, Nested Sampling can be computationally expensive in high-dimensional cases.  

Del Pozzo et al.~\cite{DelPozzo:2011pg} were the first to carry out a Bayesian implementation of model selection in the context of tests of General Relativity. Their analysis focused on tests of a particular massive graviton theory, using the gravitational wave signal from  quasi-circular inspiral of non-spinning black holes.  Cornish, et al.~\cite{Cornish:2011ys,Sampson:2013lpa} extended this analysis by considering model-independent deviations from General Relativity, using the ppE approach (Section~\ref{subsubsection:ppE})~\cite{Yunes:2009ke}. Recently, this was continued by Li et al.~\cite{Li:2011cg,Li:2011vx}, who carried out a similar analysis to that of Cornish et al.~\cite{Cornish:2011ys,Sampson:2013lpa} on a large statistical sample of advanced LIGO detections using simulated data and a restricted ppE model. All of these studies suggest that Bayesian tests of General Relativity are possible, given sufficiently high signal-to-noise ratio events. Of course, whether deviations from General Relativity are observable will depend on the strong-field character and strength of the deviation, as well as the availability of sufficiently accurate General Relativity waveforms.

%------------------------------------------------------------------------------------------------------------------------------
\subsubsection{Systematics in model selection}

The model selection techniques described above are affected by other systematics present in data analysis. In general, we can classify these into the following~\cite{Vallisneri:2013rc}: 
\begin{itemize}
\item {\bf{Mismodeling Systematic}}, caused by inaccurate models of the gravitational wave template.
\item {\bf{Instrumental Systematic}}, caused by inaccurate models of the gravitational wave response.
\item {\bf{Astrophysical Systematic}}, caused by inaccurate models of the astrophysical environment.
\end{itemize}
Mismodeling systematics are introduced due to the lack of an exact solution to the Einstein equations from which to extract an exact template, given a particular astrophysical scenario. Inspiral templates, for example, are approximated through post-Newtonian theory and become increasingly less accurate as the binary components approach each other. Cutler and Vallisneri~\cite{Cutler:2007mi} were the first carry out a formal and practical investigation of such a systematic in the context of parameter estimation from a frequentist approach. 

Mismodeling systematics will prevent us from testing General Relativity effectively with signals that we do not understand sufficiently well. For example, when considering signals from black hole coalescences, if the the total mass of the binary is sufficiently high, the binary will merge in band. The higher the total mass, the fewer the inspiral cycles that will be in band, until eventually only the merger is in band. Since the merger phase is the least understood phase, it stands to reason that our ability to test General Relativity will deteriorate as the total mass increases. Of course, we do understand the ringdown phase very well, and tests of the no-hair theorem would be allow during this phase, provided a sufficiently large signal-to-noise-ratio~\cite{Berti:2007zu}. On the other hand, for neutron star binaries or very low-mass black hole binaries, the merger phase is expected to be essentially out of band for Ad.~LIGO (above $1$kHz), and thus, the noise spectrum itself may shield us from our ignorance.   

Instrumental systematics are introduced by our ignorance of the transfer function, which connects the detector output to the incoming gravitational waves. Through sophisticated calibration studies with real data, one can approximate the transfer function very well~\cite{Accadia:2010aa,Abadie:2010px}. However, this function is not time-independent, because the noise in the instrument is not stationary or Gaussian. Thus, un-modeled drifts in the transfer function can introduce systematics in parameter estimation that are as large as $10\%$ in the amplitude and the phase~\cite{Accadia:2010aa}.  

Instrumental systematics can affect tests of General Relativity, if these are performed with a single instrument. However, one expects multiple detectors to be online in the future and for gravitational wave detections to be made in several of them simultaneously. Instrumental systematics should be present in all such detections, but since the noise will be mostly uncorrelated between different instruments, one should be able to ameliorate its effects through cross-correlating outputs from several instruments. 

Astrophysical systematics are induced by our lack of {\emph{a priori}} knowledge of the gravitational wave source. As explained above, matched filtering requires knowledge of a waveform template with which to filter the data. Usually, we assume the sources are in a perfect vacuum and isolated. For example, when considering inspiral signals, we ignore any third bodies, electric or magnetic fields, neutron star hydrodynamics, the expansion of the Universe, etc. Fortunately, however, most of these effects are expected to be small: the probability of finding third bodies sufficiently close to a binary system is very small~\cite{Yunes:2010sm}; for low redshift events, the expansion of the Universe induces an acceleration of the center of mass, which is also very small~\cite{Yunes:2009bv}; electromagnetic fields and neutron star hydrodynamic effects may affect the inspiral of black holes and neutron stars, but not until the very last stages, when most signals will be out of band anyways. For example, tidal deformation effects enter a neutron star binary inspiral waveform at 5 post-Newtonian order, which therefore affects the signal outside of the most sensitive part of the Adv.~LIGO sensitivity bucket. 

Perhaps the most dangerous source of astrophysical systematics is due to the assumptions made about the astrophysical systems we expect to observe. For example, when considering neutron star binary inspirals, one usually assumes the orbit will have circularized by the time it enters the sensitivity band. Moreover, one assumes that any residual spin angular momentum that the neutron stars may possess is very small and aligned or counter-aligned with the orbital angular momentum. These assumptions certainly simplify the construction of waveform templates, but if they happen to be wrong, they would introduce mismodeling systematics that could also affect parameter estimation and tests of General Relativity.

%------------------------------------------------------------------------------------------------------------------------------
\subsection{Burst analyses}

In alternative theories of gravity, gravitational-wave sources such as core collapse supernovae may result in the production of gravitational waves in more than just the plus- and cross-polarizations~\cite{Shibata:1994qd,Scheel:1994yr,Harada:1996wt,Novak:1999jg,Novak:1997hw,Ruiz:2012jt}. Indeed, the near-spherical geometry of the collapse can be a source of scalar breathing-mode gravitational waves.  The precise form of the waveform, however, is unknown because it is sensitive to the initial conditions. 

When searching for un-modeled bursts in alternative theories of gravity, a general approach involves optimized linear combinations of data streams from all available detectors to form maximum likelihood estimates of the waveforms in the various polarizations, and the use of {\emph{null streams}}.  In the context of ground-based detectors and General Relativity, these ideas were first explored by G\"ursel and Tinto~\cite{Guersel:1989th} and later by Chatterji et al.~\cite{Chatterji:2006nh} with the aim of separating false-alarm events from real detections. The main idea was to construct a linear combination of data streams received by a network of detectors, so that the combination contained only noise. In General Relativity, of course, one need only include $h_{+}$ and $h_{\times}$ polarizations, and thus a network of three detectors suffices.  This concept can be extended to develop null tests of General Relativity, as originally proposed by Chatziioannou, et al.~\cite{Chatziioannou:2012rf} and recently implemented by Hayama, et al.~\cite{Hayama:2012au}. 

%%%%%%%%%%%%%%%%%

Let us consider a network of $D \geq 6$ detectors with uncorrelated noise and a detection by all $D$ detectors.  For a source that emits gravitational waves in the direction $\hat \Omega$, a single data point (either in the time-domain, or a time-frequency pixel) from an array of $D$ detectors (either pulsars or interferometers) can be written as
\begin{equation}
{\boldsymbol{{d}}}={\boldsymbol{F}}{\boldsymbol{{h}}}+{\boldsymbol{{n}}}.
\end{equation}
Here
\begin{equation}
{\boldsymbol{{d}}}\equiv
\left[
        \begin{array}{c}
                {d}_{1}\\ {d}_{2}\\ \vdots\\ {d}_{D}
        \end{array} \right] \, , \quad
         {\boldsymbol{{h}}} \equiv  \left[
        \begin{array}{c}
                        {h}_+ \\ {h}_\times \\ {h}_x \\ {h}_y \\ {h}_b \\ {h}_\ell
        \end{array} \right] \, , \quad
        {\boldsymbol{{n}}} \equiv \left[
        \begin{array}{c}
                {n}_{1}\\ {n}_{1}\\ \vdots\\ {n}_{D}
        \end{array} \right],
\end{equation}
where $ {\boldsymbol{{n}}}$ is a vector with the noise.
The antenna pattern functions are given by the matrix,
\begin{equation}
\left[ \begin{array}{c c c c c c}
                {\boldsymbol{F^+}}& {\boldsymbol{F^\times}}& {\boldsymbol{F^x}}& {\boldsymbol{F^y}} & {\boldsymbol{F^b}}& {\boldsymbol{F^\ell}}
        \end{array} \right]
        \equiv \left[
        \begin{array}{c c c c c c}
                F^+_{1} & F^\times_{1}  & F^x_{1} & F^y_{1}  & F^b_{1} & F^\ell_{1} \\
                F^+_{2} & F^\times_{2}  & F^x_{2} & F^y_{2}  & F^b_{2} & F^\ell_{2}\\
                \vdots & \vdots &\vdots & \vdots& \vdots & \vdots\\
                F^+_{D} & F^\times_{D}  & F^x_{D} & F^y_{D}  & F^b_{D} & F^\ell_{D}
        \end{array} \right].
\end{equation}
For simplicity we have suppressed the sky-location dependence of the antenna pattern functions.  These can be either the interferometric antenna pattern functions in Eqs.~\eqref{IFOresponse}, or the pulsar response functions in Eqs.~\eqref{PulsarResponse}. For interferometers, since the breathing and longitudinal antenna pattern response functions are degenerate,  and even though ${\boldsymbol{F}}$ is a $6\times D$ matrix, there are only $5$ linearly independent 
vectors~\cite{Boyle:2010gc,Boyle:2010gn,Chatziioannou:2012rf,Hayama:2012au}.
 
If we do not know the form of the signal present in our data, we can obtain maximum likelihood estimators for it. For simplicity, let us assume the data are Gaussian and of unit variance (the latter can be achieved by whitening the data). Just as we did in Eq.~\eqref{eq:likelihood}, we can write the probability of obtaining datum ${\boldsymbol{d}}$, in the presence of a gravitational wave  ${\boldsymbol{h}}$ as
\begin{equation}
P({\boldsymbol{d}} | {\boldsymbol{h}} ) =
                \frac{1}{(2\pi)^{D/2}}
                \exp\left[-\frac12\left|{\boldsymbol{d}} -{\boldsymbol{F}} {\boldsymbol{h}} \right|^2\right].
\end{equation}
The logarithm of the likelihood ratio, ie. the logarithm of the ratio of the likelihood when a signal is present to that when a signal is absent, can then be written as
\begin{equation}
L\equiv\ln\frac{P( {\boldsymbol{d}} | {\boldsymbol{h}} )}{P({\boldsymbol{d}} | 0 )} = 
                \frac{1}{2} \left[ \left|{\boldsymbol{d}}\right|^2 - \left|{\boldsymbol{d}}-{\boldsymbol{F}}{\boldsymbol{h}}\right|^2 \right] \,.
\end{equation}
If we treat the waveform values for each datum as free parameters, we can maximize the likelihood ratio
\begin{equation}
0 = \left.\frac{\partial L}{\partial {\boldsymbol{h}}}\right|_{{\boldsymbol{h}}={\boldsymbol{h}}_{\rm MAX}} \,,       
\end{equation}
and obtain maximum likelihood estimators for the gravitational wave,
\begin{equation}
{\boldsymbol{h}}_{\rm MAX}= ({\boldsymbol{F}}^T {\boldsymbol{F}})^{-1} {\boldsymbol{F}}^T \, {\boldsymbol{d}}.
\end{equation}
We can further substitute this solution into the likelihood, to obtain the value of the likelihood at the maximum,
\begin{equation}
E_{\mathrm{SL}} \equiv 2 L({\boldsymbol{h}}_{\rm MAX}) 
        = {\boldsymbol{d}}^T \boldsymbol{P}^{\text{GW}} {\boldsymbol{d}},
\end{equation}
where
\begin{equation}
\boldsymbol{P}^{\text{GW}} \equiv {\boldsymbol{F}} \, ({\boldsymbol{F}}^T {\boldsymbol{F}})^{-1} {\boldsymbol{F}}^T.
\end{equation}
The maximized likelihood can be thought of as the power in the signal, and can be used as a detection statistic. $\boldsymbol{P}^{\text{GW}} $ is a projection operator that projects the data into the subspace spanned by ${\boldsymbol{F}} $.  An orthogonal projector can also be constructed,
\begin{equation}
\boldsymbol{P}^{\text{null}}\equiv(\boldsymbol{I}-\boldsymbol{P}^{\text{GW}}),
\end{equation}
which projects the data onto a sub-space orthogonal to ${\boldsymbol{F}} $. Thus one can construct a certain linear combination of data streams that has no component of a certain polarization by projecting them to a direction orthogonal to the direction defined by the beam pattern functions of this polarization mode
\begin{equation}
\boldsymbol{d}^{\text{null}}=\boldsymbol{P}^{\text{null}}\boldsymbol{d}.
\end{equation}
This is called a \textit{null stream} and, in the context of General Relativity, it was introduced as a means of separating false-alarm events due, say, to instrumental glitches from real detections~\cite{Guersel:1989th,Chatterji:2006nh}. 

With just three independent detectors, we can choose to eliminate the two tensor modes (the plus- and cross-polarizations) and construct a \emph{GR null stream}: a linear combination of data streams that contains no signal consistent within General Relativity, but could contain a signal in another gravitational theory, as illustrated in Fig.~\ref{figurenull}. With such a GR null stream, one can carry out null tests of General Relativity and study whether such a stream contains any statistically significant deviations from noise. Notice that this approach does not require a template; if one were parametrically constructed, such as in~\cite{Chatziioannou:2012rf}, more powerful null tests could be applied. In the future, we expect several gravitational wave detectors to be online: the two Ad.~LIGO ones in the United States, Ad.~VIRGO in Italy, LIGO-India in India, and KAGRA in Japan. Given a gravitational wave observation that is detected by all five detectors, one can then construct three GR null streams, each with power in a signal direction. 

For pulsar timing experiments where one is dealing with the data streams of about a few tens of pulsars, waveform reconstruction for all polarization states, as well as numerous null streams, can be constructed.

\epubtkImage{}{%
\begin{figure}[htbp]
\centerline{\includegraphics[height=6cm,clip=true]{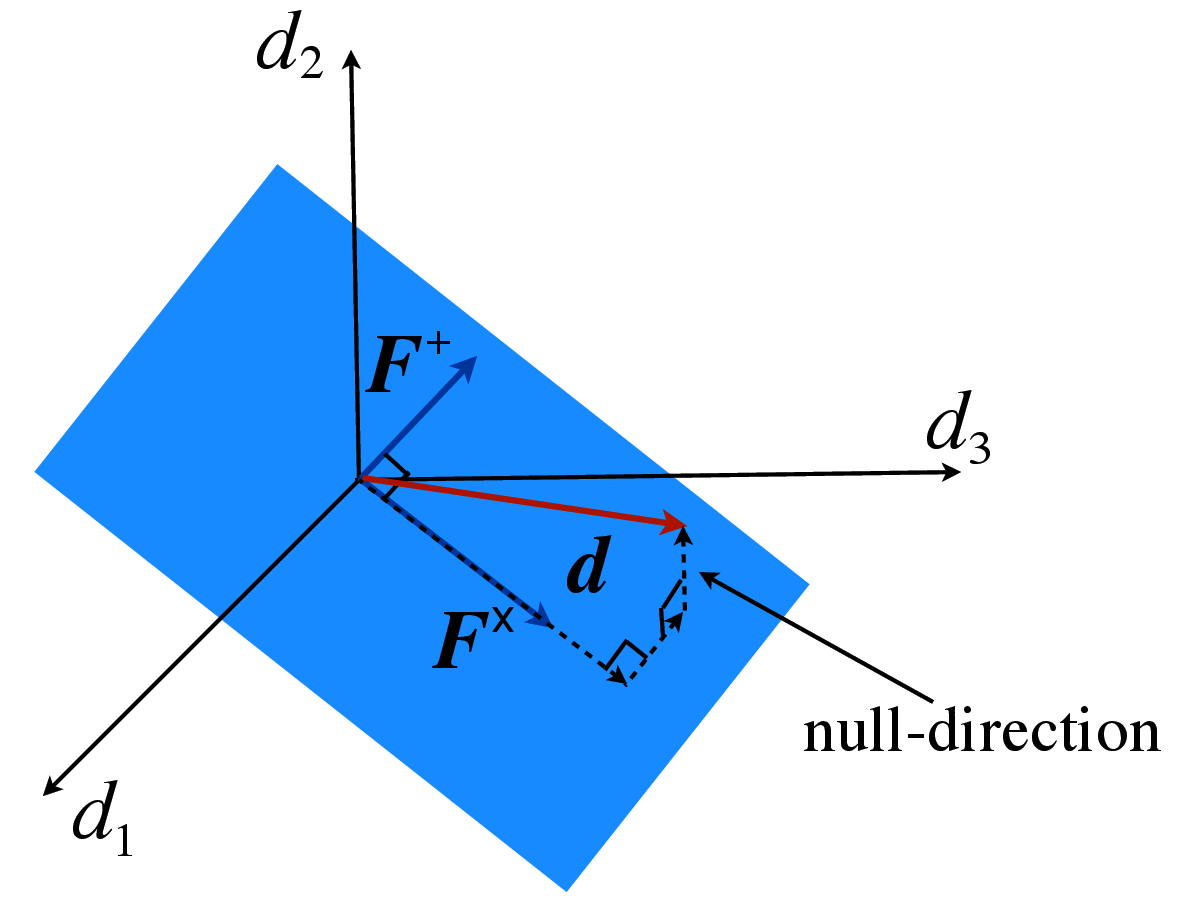}}
\caption{Schematic diagram of the projection of the data stream
  ${\boldsymbol{d}} $ orthogonal to the GR subspace spanned by $F^+$
  and $F^\times$, along with a perpendicular subspace, for 3 detectors
  to build the GR null stream.}
\label{figurenull}
\end{figure}}

%------------------------------------------------------------------------------------------------------------------------------
\subsection{Stochastic background searches}
\label{sec:Stoch-Anal}

Much work has been done on the response of ground-based interferometers to non-tensorial polarization modes, stochastic background detection prospects, and data analysis techniques~\cite{Maggiore:1999wm,Nakao:2000ug,Gasperini:2001mw,Nishizawa:2009bf,Corda:2010zza}. In the context of pulsar timing, the first work to deal with the detection of such backgrounds in the context of alternative theories of gravity is due to Lee et al.~\cite{LeeJenetPrice}, who used a coherence statistic approach to the detection of non-Einsteinian polarizations. They studied the number of pulsars required to detect the various extra polarization modes, and found that pulsar timing arrays are especially sensitive to the longitudinal mode. Alves and Tinto~\cite{daSilvaAlves:2011fp} also found enhanced sensitivity to longitudinal and vector modes. Here we follow the work in~\cite{Nishizawa:2009bf,Chamberlin:2011ev} that deals with the LIGO and pulsar timing cases using the optimal statistic, a cross-correlation that maximizes the signal-to-noise ratio. 

In the context of the optimal statistic, the derivations of the effect of extra polarization states for ground-based instruments and pulsar timing are very similar.  We begin with the metric perturbation written in terms of a plane wave expansion, as in Eq.~\eqref{pwexp2}. If we assume that the background is unpolarized, isotropic, and stationary, we have that
\beq
\langle \tilde h^*_A(f,\hat{\Omega}) \tilde h_{A'}(f',\hat{\Omega}')\rangle = 
\delta^2(\hat{\Omega},\hat{\Omega}')\delta_{AA'}
\delta(f-f')H_A(f),
\label{hev}
\eeq
where $H_A(f)$ is the gravitational wave power spectrum for polarization $A$.  $H_A(f)$ is related to the energy density in 
gravitational waves per logarithmic frequency interval for that polarization through
\beq
 \Omega_{A}(f)\equiv \frac{1}{\rho_{\rm crit}}\frac{d\rho_A}{d\ln f},
\label{omdef}
\eeq
where $\rho_{\rm crit}=3H_0^2/8\pi$ is the closure density of the universe, and
\beq
\rho_A = \frac{1}{32\pi}\langle \dot{h}_{A\,ij}(t,\vec{x})
\dot{h}_A^{ij}(t,\vec{x})\rangle,
\label{eqrho}
\eeq
is the energy density in gravitational waves for polarization $A$.  It follows from the plane
wave expansion in \Deqn{pwexp}, along with \Deqns{hev} and~(\ref{omdef}) in 
\Deqn{eqrho}, that
\beq
H_A(f)=\frac{3H_0^2}{16\pi^3}|f|^{-3}\Omega_A(|f|),
\eeq
and therefore
\bea
\langle \tilde h_A^*(f,\hat{\Omega}) \tilde h_{A'}(f',\hat{\Omega}')\rangle =
\frac{3H_0^2}{16\pi^3}\delta^2(\hat{\Omega},\hat{\Omega}')\delta_{AA'}
\delta(f-f')|f|^{-3}\Omega_A(|f|).
\label{hevom}
\eea

For both ground-based interferometers and pulsar timing experiments, an isotropic stochastic background of GWs appears in the data as correlated noise between measurements from different instruments. The data set from the $a^{\rm th}$ instrument is of the form 
\begin{eqnarray}
	d_a(t) &=& s_a(t) + n_a(t)\,,
\end{eqnarray}
where $s_a(t)$ corresponds to the GW signal and $n_a(t)$ to noise. The noise is assumed in this case to be stationary and Gaussian, and uncorrelated between different detectors,
\begin{eqnarray}
	\langle n_a(t) \rangle &=& 0\,, \\
	\langle n_a(t) n_b(t) \rangle &=& 0,
\end{eqnarray}
for $a \neq b$.

Since the GW signal is correlated, we can use cross-correlations to search for it. The cross-correlation statistic is defined as 
\begin{eqnarray}\label{ccstatQ}
	S_{ab} = \int_{-T/2}^{T/2} dt \int_{-T/2}^{T/2} dt' d_a(t) d_b(t') Q_{ab}(t-t')\,,
\end{eqnarray}
where $Q_{ab}(t-t')$ is a filter function to be determined. Henceforth, no summation is implied on the detector indices $(a,b,\ldots)$.  At this stage it is not clear why  $Q_{ab}(t-t')$ depends on the pair of data sets being correlated. We will show how this comes about later.
 The optimal filter is determined by maximizing the expected signal-to-noise ratio
\begin{eqnarray}\label{snr}
	{\rm{SNR}} = \frac{\mu_{ab}}{\sigma_{ab}}.
\end{eqnarray}
Here $\mu_{ab}$ is the mean $\langle S_{ab} \rangle$ and $\sigma_{ab}$ is the square root of the variance $\sigma_{ab}^2=\langle S_{ab}^2 \rangle - \langle S_{ab} \rangle^2$. 

The expressions for the mean and variance of the cross-correlation statistic, $\mu_{ab}$ and $\sigma^2_{ab}$ respectively, take the same form for both pulsar timing and ground-based instruments. In the frequency domain, Eq.~\eqref{ccstatQ} becomes
\begin{eqnarray}
	S_{ab} = \int_{-\infty}^{\infty} df \int_{-\infty}^{\infty} df' \delta_T(f-f') \tilde{d}^*_a(f) \tilde{d}_b(f') \tilde{Q}_{ab}(f'),
\label{ccstatQ2}
\end{eqnarray}
by the convolution theorem, and the mean $\mu$ is then
\begin{eqnarray}
	\mu_{ab} \equiv \langle S_{ab} \rangle = \int_{-\infty}^\infty df \int_{-\infty}^\infty df' \, \delta_T(f-f') \langle \tilde{s}_a^{*}(f) \tilde{s}_b(f') \rangle \tilde{Q}_{ab}(f')\,,
\label{mueq}
\end{eqnarray}
where $\delta_T$ is the finite time approximation to the delta function, $\delta_T(f)={\sin{\pi f t}}/({\pi f})$. With this in hand, the mean of the cross-correlation statistic is
\begin{eqnarray}
\mu_{ab} = \frac{3H_0^2}{16\pi^3} T \sum_{A} \int_{-\infty}^{\infty} df |f|^{-3} \tilde{Q}_{ab}(f) \Omega_A (f) \Gamma^A_{ab} (f),
\end{eqnarray}
and the variance in the weak signal limit  is
\bea
\sigma_{ab}^2
&\equiv&
\langle S_{ab}^2\rangle-\langle S_{ab}\rangle^2\approx \langle S_{ab}^2\rangle
\nonumber
\\
&\approx& \frac{T}{4}\int_{-\infty}^{\infty} df\, P_a(|f|)P_b(|f|)
\left|\tilde{Q}_{ab}(f)\right|^2,
\label{eqsig}
\eea
where the one-sided power spectra of the noise are defined by
\beq\label{e:psd}
\langle \tilde{n}_a^*(f)\tilde{n}_a(f')\rangle = \frac{1}{2}\delta(f-f') P_a(|f|)\,,
\eeq
in analogy to Eq.~\eqref{eq:Sn(f)-def}, where $P_{a}$ plays here the role of $S_{n}(f)$.

The mean and variance can be rewritten more compactly if we define a positive-definite inner 
product using the noise power spectra of the two data streams
\beq
(A,B)_{ab}\equiv\int_{-\infty}^{\infty}df\, A^*(f)B(f)P_a(|f|)P_b(|f|)\,,
\eeq
again in analogy to the inner product in Eq.~\eqref{eq:inner-prod-def}, when considering inspirals.
Using this definition
\bea
\mu_{ab} &=& \frac{3 H_0^2}{16\pi^3}T\,\left(\tilde{Q}_{ab}, 
\frac{\sum_A \Omega_A(|f|)\Gamma^A_{ab}(|f|)}{|f|^3P_a(|f|)P_b(|f|)}
\right)_{ab},\label{optfiltmu}\\
\sigma_{ab}^2 &\approx& \frac{T}{4}\left(\tilde{Q},\tilde{Q}\right)_{ab},
\label{sigip}
\eea
where we recall that the capital Latin indices $(A,B,\ldots)$ stand for the polarization content. From the definition of the signal-to-noise ratio and the Schwartz's inequality, it follows that the optimal filter is given by
\beq
\tilde{Q}_{ab}(f)=N\frac{\sum_A\Omega_A(|f|)\Gamma^A_{ab}(|f|)}{|f|^3P_a(|f|)P_b(|f|)},
\label{optfilt}
\eeq
where $N$ is an arbitrary normalization constant, normally chosen so that the mean of the statistic gives the amplitude of the stochastic background.  

The differences in the optimal filter between interferometers and pulsars arise only from differences in the so-called {\emph{overlap reduction functions}}, $\Gamma^A_{ab} (f)$. For ground based instruments, the signal data $s_a$ are the strains given by Eq.~\eqref{strain}. The overlap reduction functions are then given by
\begin{eqnarray}
\Gamma^A_{ab} (f) = \int_{S^2}  d \hat{\Omega} F_a^A(\hat{\Omega}) F_b^A(\hat{\Omega}) e^{2\pi if \hat{\Omega}\cdot (\vec{x}_a-\vec{x}_b) },
\label{ligoorf}
\end{eqnarray}
where $\vec{x}_a$ and $\vec{x}_b$ are the locations of the two interferometers. The integrals in this case all have solutions in terms of spherical Bessel functions ~\cite{Nishizawa:2009bf}, which we do not summarize here for brevity.

For pulsar timing arrays, the signal data $s_a$ are the redshifts $z_a$, given by Eq.~\eqref{FTantpatts}. The overlap reduction functions are then given by
\bea
\Gamma^A_{ab}(f) = \frac{3}{4\pi} \int_{S^2}d\hat{\Omega}\,
\left(e^{i2\pi f L_a(1+\hat{\Omega}\cdot\hat{p}_a)}-1\right)
\left(e^{-i2\pi f L_b(1+\hat{\Omega}\cdot\hat{p}_b)}-1\right)
F_a^A(\hat{\Omega})F_b^{A}(\hat{\Omega}),
\label{pulsarorf}
\eea
where $L_a$ and $L_b$ are the distances to the two pulsars. For all transverse modes
pulsar timing experiments are in a regime where the exponential
factors in \Deqn{pulsarorf} can be neglected~\cite{Anholm:2008wy,Chamberlin:2011ev}, and the overlap reduction functions 
effectively become frequency independent.  For the $+$ and $\times$ mode the
overlap reduction function becomes
\bea
\Gamma^+_{ab} = 3 \left\{ \frac{1}{3} + \frac{1-\cos\xi_{ab}}{2}\left[ \ln\left(
    \frac{1-\cos\xi_{ab}}{2}\right)  -\frac{1}{6}\right]\right\},
\label{orfapprox}
\eea
where $\xi_{ab}=\cos^{-1}(\hat p_a \cdot \hat p_b)$ is the angle between the two 
pulsars.  This quantity is proportional to the Hellings and Downs 
curve~\cite{hd83}. For the breathing mode, the overlap reduction function takes the closed form expression~\cite{LeeJenetPrice}:
\begin{eqnarray}
	\Gamma^b _{ab} = \frac{1}{4} \left( 3+ \cos{\xi_{ab}} \right).
\end{eqnarray} 
For the vector and and longitudinal modes the overlap reduction functions remain frequency dependent and there are no general analytic solutions.

The result for the combination of cross-correlation pairs to form an optimal network statistic is also the same in both ground-based interferometer and pulsar timing cases: a sum of the cross-correlations of all detector pairs weighted by their variances. The detetctor network optimal statistic is,
\beq
S_{\rm opt} = \frac{\sum_{ab} \sigma^{-2} _{ab} S_{ab} }{\sum_{ab} \sigma^{-2} _{ab}},
\label{optnetwork}
\eeq
 where $\sum_{ab} $ is a sum over all detector pairs.  

In order to perform a search for a given polarization mode one first needs to compute the overlap reduction functions (using either Eq.~\eqref{ligoorf} or \eqref{pulsarorf}) for that mode. With that in hand and a form for the stochastic background spectrum $\Omega_A(f)$, one can construct optimal filters for all pairs in the detector network using Eq.~\eqref{optfilt}, and perform the cross-correlations using either Eq.~\eqref{ccstatQ} (or equivalently Eq.~\eqref{ccstatQ2}). Finally, we can calculate the overall network statistic Eq.~\eqref{optnetwork}, by first finding the variances using Eq.~\eqref{eqsig}.  

It is important to point out that the procedure outlined above is straightforward for ground-based interferometers. Pulsar timing data, however, are irregularly sampled, and have a pulsar timing model subtracted out. This needs to be accounted for, and generally, a time-domain approach is more appropriate for these data sets.  The procedure is similar to what we have outlined above, but power spectra and gravitational wave spectra in the frequency domain need to be replaced by auto-covariance and cross-covariance matrices in the time domain that account for the model fitting (for an example of how to do this see~\cite{Ellis:2013nrb}).

Interestingly, Nishizawa et al.~\cite{Nishizawa:2009bf} show that with three spatially separated detectors the tensor, vector, and  scalar contributions to the energy density in GWs can be measured independently. Lee et al.~\cite{LeeJenetPrice} and Alves and Tinto~\cite{daSilvaAlves:2011fp}  showed that pulsar timing experiments are especially sensitive to the longitudinal mode, and to a lesser extent the vector modes. Chamberlin and Siemens~\cite{Chamberlin:2011ev} showed that the sensitivity of the cross-correlation to the longitudinal mode using nearby pulsar pairs can be enhanced significantly compared to that of the transverse modes. For example, for the NANOGrav pulsar timing array, two pulsar pairs separated by $3^\circ$ result in an enhancement of 4 orders of magnitude in sensitivity to the longitudinal mode relative to the transverse modes. The main contribution to this effect is due to GWs that are coming from roughly the same direction as the pulses from the pulsars. In this case, the induced redshift for any GW polarization mode is proportional to $fL$, the product of the GW frequency and the distance to the pulsar, which can be large. When the GWs and the pulse direction are exactly parallel, the redshift for the transverse and vector modes vanishes, but it is proportional to $fL$ for the scalar-longitudinal mode. 

Lee et al.~\cite{2010ApJ...722.1589L} studied the detectability of massive gravitons in pulsar timing arrays through stochastic background searches. They introduced a modification to Eq.~\eqref{eqzsom} to account for graviton dispersion, and found the modified overlap reduction functions (i.e., modifications to the Hellings--Downs curves Eq.~\eqref{orfapprox}) for various values of the graviton mass. They conclude that a large number of stable pulsars ($\geq 60$) are required to distinguish between the massive and massless cases, and that future pulsar timing experiments could be sensitive to graviton masses of about $10^{-22}$~eV ($\sim 10^{13}$~km). This method is competitive with some of the compact binary tests described later in Section~\ref{generic-tests:MG-LV} (see Table~\ref{table:comparison-MG}). In addition, since the method of Lee et al.~\cite{2010ApJ...722.1589L} only depends on the form of the overlap reduction functions, it is advantageous in that it does not involve matched filtering (and therefore prior knowledge of the waveforms), and generally makes few assumptions about the GW source properties.

%% file: binary-sys-tests.tex
% Nico writes this
%   5.1 Direct Tests versus Generic Tests
%   5.2 Direct Tests
%       5.2.1 Jordan--Fierz--Brans--Dicke Theory
%       5.2.2 Modified Quadratic Gravity
%       5.2.3 Non-Commutative Geometry
%   5.3 Generic Tests
%       5.3.1 Massive Graviton Theories and Lorentz Violation
%       5.3.2 Variable G Theories and Large Extra-dimensions
%       5.3.3 Parity Violation
%       5.3.4 Parametrized post-Einsteinian Framework
%   5.4 Search for Exotica
%       5.4.1 Bumpy Black Holes
%       5.4.2 Horizon-less Objects
%       5.4.3 Other Strange Objects
%------------------------------------------------------------------------------------------------------------------------------
In this section, we discuss gravitational wave tests of General Relativity with signals emitted by compact binary systems. We begin by explaining the difference between direct and generic tests. We then proceed to describe the many direct or top-down tests and generic or bottom-up tests that have been proposed once gravitational waves are detected, including tests of the no-hair theorems. We concentrate here only on binaries composed of compact objects, such as neutron stars, black holes or other compact exotica. We will not discuss tests one could carry out with electromagnetic information from binary (or double binary) pulsars, as these are already described in~\cite{lrr-2006-3}. We will also not review tests of General Relativity with accretion disk observations, for which we refer the interested reader to~\cite{lrr-2008-9}.

%------------------------------------------------------------------------------------------------------------------------------
\subsection{Direct and generic tests}

Gravitational wave tests of Einstein's theory can be classified in two distinct subgroups: direct tests
and generic tests. Direct tests employ a top-down approach, where one starts from a particular modified 
gravity theory with a known action, derives the modified field equations and solves them for a particular
gravitational wave-emitting system. On the other hand, generic tests adopt a bottom-up approach, where
one takes a particular feature of General Relativity and asks what type of signature its absence would leave 
on the gravitational wave observable; one then asks whether the data presents a statistically significant anomaly 
pointing to that particular signature. 

Direct tests have by far been the traditional approach to testing General Relativity with gravitational waves. 
The prototypical examples here are tests of Jordan--Fierz--Brans--Dicke theory. As described in Section~\ref{section:alt-theories}, one can solve the modified field equations for a binary system in the post-Newtonian approximation to find a prediction for the gravitational wave observable, as we will see in more detail later in this section. Other examples of direct tests include those concerning modified quadratic gravity models and non-commutative geometry theories. 

The main advantage of such direct tests is also its main disadvantage: one has to pick a particular modified gravity
theory. Because of this, one has a well-defined set of field equations that one can solve, but at the same time, 
one can only make predictions about that particular modified gravity model. Unfortunately, we currently lack a particular
modified gravity theory that is particularly compelling; many modified gravity theories exist,
but none possess all the criteria described in Section~\ref{section:alt-theories}, except perhaps for the subclass
of scalar-tensor theories with spontaneous scalarization. Lacking a clear alternative to 
General Relativity, it is not obvious which theory one should pick. Given that the full development (from the action 
to the gravitational wave observable) of any particular theory can be incredibly difficult, time and computationally consuming, 
carrying out direct tests of all possible modified gravity models once gravitational waves are detected is clearly
unfeasible. 

Given this, one is led to generic tests of General Relativity, where one asks how the absence of specific features
contained in General Relativity could impact the gravitational wave observable. For example, one can ask how such
observable would be modified if the graviton had a mass, if the gravitational interaction were Lorentz or parity violating, or if
there existed large extra dimensions. From these general considerations, one can then construct a ``meta''-observable, 
ie.~one that does not belong to a particular theory, but that interpolates over all known possibilities in a well-defined way.
This model has become to be known as the parameterized post-Einsteinian framework, in analogy to the parameterized
post-Newtonian scheme used to test General Relativity in the Solar System~\cite{lrr-2006-3}. Given such a construction, one can 
then ask whether the data points to a statistically significant deviation from General Relativity. 

The main advantage of generic tests is precisely that one does not have to specify a particular model, but instead
one lets the data select whether it contains any statistically significant deviations from our canonical beliefs. 
Such an approach is, of course, not new to physics, having most recently been successfully employed by the WMAP 
team~\cite{Bennett:2010jb}. The intrinsic disadvantage of this method is that, if a deviation is found, there is no
one-to-one mapping between it and a particular action, but instead one has to point to a class of possible models. 
Of course, such a disadvantage is not that limiting, since it would provide strong hints as to what type of symmetries
or properties of General Relativity would have to be violated in an ultra-violet completion of Einstein's theory.   

%------------------------------------------------------------------------------------------------------------------------------
\subsection{Direct tests}
\label{sec:Direct-tests}

%------------------------------------------------------------------------------------------------------------------------------
\subsubsection{Scalar-tensor theories}
\label{sec:direct-test-BD}

Let us first concentrate on Jordan--Fierz--Brans--Dicke theory, where black holes and neutron stars have been shown to exist. In this theory, the gravitational mass depends on the value of the scalar field, as Newton's constant is effectively promoted to a function, thus leading to violations of the weak-equivalence principle~\cite{1975ApJ...196L..59E,1977ApJ...214..826W,Will:1989sk}. The usual prescription for the modeling of binary systems in this theory is due to Eardley~\cite{1975ApJ...196L..59E}\footnote{A modern interpretation in terms of effective field theory can be found in~\cite{Goldberger:2004jt,Goldberger:2006bd}.}. He showed that such a scalar-field effect can be captured by replacing the constant inertial mass by a function of the scalar field in the distributional stress-energy tensor and then Taylor expanding about the cosmological constant value of the scalar field at spatial infinity, ie.
\begin{equation}
m_{a} \to m_{a}(\phi) = m_{a}(\phi_{0}) \left\{1 + s_{a} \frac{\psi}{\phi_{0}} - \frac{1}{2} \left(s_{a}' - s_{a}^{2} + s_{a} \right) \left(\frac{\psi}{\phi_{0}}\right)^{2} + {\cal{O}}\left[\left(\frac{\psi}{\phi_{0}}\right)^{3}\right] \right\}\,, 
\end{equation}
where the subscript $a$ stands for $a$ different sources, while $\psi \equiv \phi - \phi_{0} \ll 1$ and the sensitivities $s_{a}$ and $s_{a}'$ are defined by 
\begin{equation}
s_{a} \equiv - \left[\frac{\partial \left(\ln m_{a}\right)}{\partial \left(\ln G\right)}\right]_{0}\,,
\label{eq:sensitivity-def}
\qquad
s_{a}' \equiv - \left[\frac{\partial^{2} \left(\ln m_{a}\right)}{\partial \left(\ln G\right)^{2}}\right]_{0}\,,
\end{equation}
where we remind the reader that $G = 1/\phi$, the derivatives are to be taken with the baryon number held fixed and evaluated at $\phi = \phi_{0}$. These sensitivities encode how the gravitational mass changes due to a non-constant scalar field; one can think of them as measuring the gravitational binding energy per unit mass. The internal gravitational field of each body leads to a non-trivial variation of the scalar field, which then leads to modifications to the gravitational binding energies of the bodies.  In carrying out this expansion, one assumes that the scalar field takes on a constant value at spatial infinity $\phi \to \phi_{0}$, disallowing any homogeneous, cosmological solution to the scalar field evolution equation [Eq.~\eqref{eq:ST-EOM}]. 

With this at hand, one can solve the massless Jordan--Fierz--Brans--Dicke modified field equations [Eq.~\eqref{eq:ST-EOM}] for the non-dynamical, near-zone field of $N$ compact objects to obtain~\cite{Will:1989sk}
\begin{align}
\frac{\psi}{\phi_{0}} &= \frac{1}{2 + \omega_{\rm BD}} \sum_{a} \left(1 - 2 s_{a} \right) \frac{m_{a}}{r_{a}} + \ldots\,,
\label{eq:psi-sol-BD}
\\
g_{00} &=-1 + \sum_{a} \left( 1 - \frac{s_{a}}{2 + \omega_{\rm BD}} \right) \frac{2 m_{a}}{r_{a}} + \ldots\,,
\\
g_{0i} &= -2 \left(1 + \gamma \right)\sum_{a} \frac{m_{a}}{r_{a}} v_{a}^{i} + \ldots\,,
\\
g_{ij} &= \delta_{ij} \left[1 + 2 \gamma \sum_{a} \left(1 + \frac{s_{a}}{1 + \omega_{\rm BD}} \right) \frac{m_{a}}{r_{a}} + \ldots \right]\,,
\end{align}
where $a$ runs from $1$ to $N$, we have defined the spatial field point distance $r_{a} \equiv |x^{i} - x_{a}^{i}|$, the parameterized post-Newtonian quantity $\gamma = (1 + \omega_{\rm BD}) (2 + \omega_{\rm BD})^{-1}$ and we have chosen units in which $G = c = 1$. This solution is obtained in a post-Newtonian expansion~\cite{Blanchet:2006zz}, where the ellipses represent higher order terms in $v_{a}/c$ and $m_{a}/r_{a}$. From such an analysis, one can also show that compact objects follow geodesics of such a spacetime, to leading order in the post-Newtonian approximation~\cite{1975ApJ...196L..59E}, except that Newton's constant in the coupling between matter and gravity is replaced by $G \to {\cal{G}}_{12} = 1 - (s_{1} + s_{2} - 2 s_{1} s_{2}) (2 + \omega_{\rm BD})^{-1}$, in geometric units.

As is clear from the above analysis, black hole and neutron star solutions in this theory generically depend on the quantities $\omega_{\rm BD}$ and $s_{a}$. The former determines the strength of the correction, with the theory reducing to General Relativity in the $\omega_{\rm BD} \to \infty$ limit~\cite{Faraoni:1999yp}. The latter depends on the compact object that is being studied. For neutron stars, this quantity can be computed as follows. First, neglecting scalar corrections to neutron star structure and using the Tolman--Oppenheimer--Volkoff equation, one notes that the mass $m \propto N \propto G^{-3/2}$, for a fixed equation of state and central density, with $N$ the total baryon number. Thus, using Eq.~\eqref{eq:sensitivity-def}, one has that 
\begin{equation}
s_{a} \equiv \frac{3}{2} \left[1 - \frac{\partial \left(\ln m_{a}\right)}{\partial \left(\ln N\right)}_{G} \right]\,,
\end{equation}
where the derivative is to be taken holding $G$ fixed. In this way, given an equation of state and central density, one can compute the gravitational mass as a function of baryon number, and from this, obtain the neutron star sensitivities. Eardley~\cite{1975ApJ...196L..59E}, Will and Zaglauer~\cite{Will:1989sk}, and Zaglauer~\cite{Zaglauer:1992bp} have shown that these sensitivities are always in the range $s_{a} \in (0.19,0.3)$ for a soft equation of state and $s_{a} \in (0.1,0.14)$ for a stiff one, in both cases monotonically increasing with mass in $m_{a} \in (1.1,1.5) M_{\odot}$. Recently, Gralla~\cite{Gralla:2010cd} has found a more general method to compute sensitivities is generic modified gravity theories.

What is the sensitivity of black holes in generic scalar-tensor theories? Will and Zaglauer~\cite{Zaglauer:1992bp} have argued that the no-hair theorems require $s_{a} = 1/2$ for all black holes, no matter what their mass or spin is. As already explained in Section~\ref{section:alt-theories}, stationary black holes that are the byproduct of gravitational collapse (ie.~with matter that satisfies the energy conditions) in a general class of scalar-tensor theories are identical to their General Relativity counterparts~\cite{Hawking:1972qk,1971ApJ...166L..35T,Dykla:1972zz,Sotiriou:2011dz}\epubtkFootnote{One should note in passing that more general black hole solutions in scalar-tensor theories have been found~\cite{Kim:1998hc,Campanelli:1993sm}. These, however, usually violate the weak energy condition, and sometimes they require unreasonably small values of $\omega_{\rm BD}$ that have already been ruled out by observation.}. This is because the scalar field satisfies a free wave equation in vacuum, which forces the scalar field to be constant in the exterior of a stationary, asymptotically flat spacetime, provided one neglects a homogeneous, cosmological solution. If the scalar field is to be constant, then by Eq.~\eqref{eq:psi-sol-BD}, $s_{a} = 1/2$ for a single black hole spacetime. 

Such an argument formally applies only to stationary scenarios, so one might wonder whether a similar argument holds for binary systems that are in a quasi-stationary arrangement. Will and Zaglauer~\cite{Zaglauer:1992bp} and Mirshekari and Will~\cite{Mirshekari:2013vb} extended this discussion to quasi-stationary spacetimes describing binary black holes to higher post-Newtonian order. They argued that the only possible deviations from $\psi = 0$ are due to tidal deformations of the horizon due to the companion, which are known to arise at very high order in post-Newtonian theory, $\psi = {\cal{O}}[(m_{a}/r_{a})^{5}]$. Recently, Yunes, et al.~\cite{Yunes:2011aa} extended this argument further by showing that to all orders in post-Newtonian theory, but in the extreme mass-ratio limit, black holes cannot have scalar hair in generic scalar tensor theories. Finally, Healy, et al.~\cite{Healy:2011ef} have carried out a full numerical simulations of the non-linear field equations confirming this argument in the full non-linear regime. 

The activation of dynamics in the scalar field for a vacuum spacetime requires either a non-constant distribution of initial scalar field (violating the constant cosmological scalar field condition at spatial infinity) or a pure geometrical source to the scalar field evolution equation. The latter would lead to the quadratic modified gravity theories discussed in Section~\ref{subsec:MQG}. As for the former, Horbatsch and Burgess~\cite{Horbatsch:2010hj} have argued that if, for example, one lets $\psi = \mu t$, which clearly satisfies $\square \psi = 0$ in a Minkowski background~\epubtkFootnote{The scalar field of Horbatsch and Burgess satisfies $\square \psi = \mu g^{\mu \nu} \Gamma^{t}_{\mu \nu}$, and thus $\square \psi = 0$ for stationary and axisymmetric spacetimes, since the metric is independent of time an azimuthal coordinate. Notice, however that is not necessarily needed for Jacobson's construction~\cite{Jacobson:1999vr} to be possible.}, then a Schwarzschild black hole will acquire modifications that are proportional to $\mu$. Alternatively, scalar hair could also be induced by spatial gradients in the scalar field~\cite{Berti:2013gfa}, possibly anchored in matter at galactic scales. Such cosmological hair, however, is likely to be suppressed by a long time-scale; in the example above $\mu$ must have units of inverse time, and if it is to be associated with the expansion of the Universe, then it would be natural to assume $\mu = {\cal{O}}(H)$, where $H$ is the Hubble parameter. Therefore, although such cosmological hair might have an effect on black holes in the early Universe, it should not affect black hole observations at moderate to low redshifts. 

Scalar field dynamics can be activated in non-vacuum spacetimes, even if initially the stars are not scalarized provided one considers a more general scalar tensor theory, like the one introduced by Damour and Esposito-Farese~\cite{Damour:1992we,Damour:1993hw}. As discussed in Section~\ref{sec:ST}, when the conformal factor takes on a particular functional form, non-linear effects induced when the gravitational energy exceeds a certain threshold can spontaneously scalarize merging neutron stars, as demonstrated recently by Barausse, {\emph{et al}}~\cite{Barausse:2012da}. Therefore, neutron stars in binaries are likely to have hair in generic scalar-tensor theories, even if they start their inspiral unscalarized.

What do gravitational waves look like in Jordan--Fierz--Brans--Dicke theory? As described in Section~\ref{sec:ST}, both the scalar field perturbation $\psi$ and the new metric perturbation $\theta^{\mu \nu}$ satisfy a sourced wave equation [Eq.~\eqref{eq:ST-EOM}], whose leading-order solution for a two-body inspiral is~\cite{Will:1994fb}
\begin{align}
\theta^{ij} &= 2 \left(1 + \gamma\right) \frac{\mu}{R} \left(v_{12}^{ij} - {\cal{G}}_{12} m \frac{x^{ij}}{r^{3}} \right)\,,
\\
\frac{\psi}{\phi_{0}} &= \left(1 - \gamma\right) \frac{\mu}{R} \left[ \Gamma \left(n_{i} v_{12}^{i}\right)^{2} - {\cal{G}}_{12} \Gamma \frac{m}{r^{3}} \left(n_{i} x^{i}\right)^{2} - \frac{m}{r}  \left({\cal{G}}_{12} \Gamma + 2 \Lambda\right) - 2 S n_{i} v_{12}^{i} \right]\,,
\end{align}
where $R$ is the distance to the detector, $n^{i}$ is a unit vector pointing toward the detector, $r$ is the magnitude of relative position vector $x^{i} \equiv x_{1}^{i} - x_{2}^{i}$, with $x_{a}^{i}$ the trajectory of body $a$, $\mu = m_{1} m_{2}/m$ is the reduced mass and $m = m_{1} + m_{2}$ is the total mass, $v_{12}^{i} \equiv  v_{1}^{i} - v_{2}^{i}$ is the relative velocity vector and we have defined the shorthands
\begin{align}
\Gamma &\equiv 1 - 2 \frac{m_{1} s_{2} + m_{2} s_{1}}{m}\,,
\qquad
S \equiv s_{2} - s_{1}\,,
\label{eq:S-def}
\\
\Lambda &\equiv{\cal{G}}_{12} \left(1 - s_{1} - s_{2}\right) - \left(2 + \omega_{\rm BD}\right)^{-1} \left[\left(1- 2 s_{1}\right) s_{2}' + \left(1 - 2 s_{2}\right)s_{1}'\right]\,.
\end{align}
We have here also introduced multi-index notation, such that $A^{ij\ldots} = A^{i} A^{j} \ldots$. Such a solution is derived using the Lorenz gauge condition $\theta^{\mu \nu}{}_{,\nu} = 0$ and in a post-Newtonian expansion, where we have left out subleading terms of relative order $v_{12}^{2}$ or $m/r$.  

Given the new metric perturbation $\theta^{ij}$, one can reconstruct the gravitational wave $h^{ij}$ metric perturbation, and from this, the response function, associated with the quasi-circular inspiral of compact binaries. After using Kepler's third law to simplify expressions [$\omega = ({\cal{G}}_{12} m/r^{3})^{1/2}$, where $\omega$ is the orbital angular frequency and $m$ is the total mass and $r$ is the orbital separation], one finds for a ground-based L-shaped detector~\cite{Chatziioannou:2012rf}:
\begin{align}
\label{eq:BD-h(t)}
h(t) &=- \frac{{\cal{M}}_{c}}{R} \left(2 \pi {\cal{M}}_{c} F\right)^{2/3} e^{-2 i \Phi}
\left\{\left[F_{+} \left(1 + \cos^{2}{\iota} \right) + 2 i F_{\times} \cos{\iota}\right] 
\left[1 - \frac{1 - \gamma}{2} \left(1 + \frac{4}{3} S^{2}\right)\right] 
\right.
\nonumber \\
&- \left.
\frac{1-\gamma}{2} \Gamma F_{\rm b} \sin^{2}{\iota}\right\}
- \eta^{1/5}  \frac{{\cal{M}}_{c}}{R}  \left(2 \pi {\cal{M}}_{c} F\right)^{1/3} e^{-i \Phi} 
S \left(1 - \gamma\right)F_{\rm b} \sin{\iota}
\nonumber \\
&- \frac{{\cal{M}}_{c}}{R} \left(2 \pi {\cal{M}}_{c} F\right)^{2/3} 
\frac{1-\gamma}{2} F_{\rm b} \left(\Gamma + 2 \Lambda\right)\,,
\end{align} 
where $\eta \equiv \mu/m$ is the symmetric mass ratio, ${\cal{M}}_{c} \equiv \eta^{3/5} m$ is the chirp mass, $\iota$ is the inclination angle, and where we have used the beam-pattern functions in Eq.~\eqref{IFOresponse}. In Eq.~\eqref{eq:BD-h(t)} and henceforth, we linearize all expressions in $1 - \gamma \ll 1$. Jordan--Fierz--Brans--Dicke theory predicts the generic excitation of three polarizations: the usual plus and cross polarizations, and a breathing, scalar mode. We see that the latter contributes to the response at two, one and zero times the orbital frequency. One should note that all of these corrections arise during the generation of gravitational waves, and not due to a propagation effect. In fact, gravitational waves travel at the speed of light (and the graviton remains massless) in standard Jordan--Fierz--Brans--Dicke theory.  

The quantities $\Phi$ and $F$ are the orbital phase and frequency respectively, which are to be found by solving the differential equation
\begin{equation}
\frac{dF}{dt} = \left(1 - \gamma\right) S^{2} \frac{\eta^{2/5}}{\pi} {\cal{M}}_{c}^{-2}  (2 \pi {\cal{M}}_{c} F)^{3}+ \frac{48}{5 \pi} {\cal{M}}_{c}^{-2}  \left(2 \pi {\cal{M}}_{c} F\right)^{11/3} \left[1 - \frac{1-\gamma}{2} \left(1  - \frac{\Gamma^{2}}{6} + \frac{4}{3} S^{2}\right)\right] \ldots\,,
\label{eq:Fdot-BD}
\end{equation}
where the ellipses stand for higher-order terms in the post-Newtonian approximation. In this expression, and henceforth, we have kept only the leading-order dipole term and all known post-Newtonian, General Relativity terms. If one wished to include higher post-Newtonian order Brans-Dicke terms, one would have to include monopole contributions as well as post-Newtonian corrections to the dipole term. The first term in Eq.~\eqref{eq:Fdot-BD} corresponds to dipole radiation, which is activated by the scalar mode. That is, the scalar field carries energy away from the system modifying the energy balance law to~\cite{Will:1994fb,Scharre:2001hn,Will:2004xi}
\begin{equation}
\dot{E}_{\rm BD} = -\frac{2}{3} {\cal{G}}_{12}^{2} \eta^{2} \frac{m^{4}}{r^{4}} \left(1 - \gamma\right) S^{2} - \frac{32}{5} {\cal{G}}_{12}^{2} \eta^{2} \left(\frac{m}{r}\right)^{5} \left[1 - \frac{1-\gamma}{2} \left(1 - \frac{\Gamma^{2}}{6} \right) \right] + \ldots\,,
\end{equation}
where the ellipses stand again for higher-order terms in the post-Newtonian approximation. Solving the frequency evolution equation perturbatively in $1/\omega_{\rm BD} \ll 1$, one finds
\begin{align}
\frac{256}{5} \frac{t_{c} - t}{{\cal{M}}_{c}} &= u^{-8} \left[ 1 - \frac{1}{12} \left(1 - \gamma\right) S^{2} \eta^{2/5} u^{-2} + \dots\right]\,,
\\
\Phi &= -\frac{1}{64 \pi} \left(\frac{256}{5} \frac{t_{c} - t}{{\cal{M}}_{c}}\right)^{5/8} \left[1 - \frac{5}{224} \left(1 - \gamma\right) S^{2} \eta^{2/5} \left(\frac{256}{5} \frac{t_{c} - t}{{\cal{M}}_{c}}\right)^{1/4}  + \dots \right]\,.
\end{align}
where we have defined $u \equiv (2 \pi {\cal{M}}_{c} F)^{1/3}$. In deriving these equations, we have neglected the last term in Eq.~\eqref{eq:Fdot-BD}, as this is a constant that can be reabsorbed into the chirp mass. Notice that since the two definitions of chirp mass differ only by a term of ${\cal{O}}(\omega_{\rm BD}^{-1})$, the first term of Eq.~\eqref{eq:Fdot-BD} is not modified.

One of the main ingredients that goes into parameter estimation is the Fourier transform of the response function. This can be estimated in the so-called stationary-phase approximation, for a simple, non-spinning, quasi-circular inspiral. In this approximation, one assumes the phase is changing much more rapidly than the amplitude~\cite{Bender,Cutler:1994ys,Droz:1999qx,Yunes:2009yz}. One finds~\cite{Chatziioannou:2012rf}
\begin{align}
\tilde{h}(f) &=  {\cal{A}}_{\rm BD}  \left(\pi {\cal{M}}_{c} f\right)^{-7/6} 
\left[1 - \frac{5}{96} \frac{S^{2}}{\omega_{\rm BD}} \eta^{2/5} \left(\pi {\cal{M}}_{c} f\right)^{-2/3}\right] e^{-i \Psi_{\rm BD}^{(2)}}
+ \gamma_{\rm BD}  \left(\pi {\cal{M}}_{c} f\right)^{-3/2} e^{-i \Psi_{\rm BD}^{(1)}}
\end{align}
where we have defined the amplitudes
\begin{align}
{\cal{A}}_{\rm BD} &\equiv \left(\frac{5 \pi}{96}\right)^{1/2} \frac{{\cal{M}}_{c}^{2}}{R} \left[F_{+}^{2} \left(1 + \cos^{2}\iota\right)^{2} + 4 F_{\times}^{2} \cos^{2}{\iota}
- F_{+} F_{\rm b} \left(1 - \cos^{4}{\iota}\right) \frac{\Gamma}{\omega_{\rm BD}}\right]^{1/2}
\\ 
\gamma_{\rm BD} &\equiv - \left(\frac{5 \pi}{48}\right)^{1/2} \frac{{\cal{M}}_{c}^{2}}{R} \eta^{1/5} \frac{S}{\omega_{\rm BD}} F_{\rm b} \sin{\iota}\,.
\end{align}
and the Fourier phase
\begin{align}
\Psi_{\rm BD}^{(\ell)} &= - 2 \pi f t_{c} + \ell \Phi_{c}^{(\ell)} + \frac{\pi}{4} - \frac{3 \ell}{256} \left(\frac{2 \pi {\cal{M}}_{c} f}{\ell}\right)^{-5/3} \sum_{n=0}^{7} \left(\frac{2 \pi {\cal{M}}_{c} f}{\ell}\right)^{n/3} \left(c_{n}^{\rm PN} + l_{n}^{\rm PN} \ln f\right) 
\nonumber \\
&+ \frac{5 \ell}{7168} \frac{S^{2}}{\omega_{\rm BD}} \eta^{2/5} \left(\frac{2 \pi {\cal{M}}_{c} f}{\ell}\right)^{-7/3}\,,
\label{eq:BD-phase}
\end{align}
where the Brans--Dicke correction is kept only to leading order in $\omega_{\rm BD}^{-1}$ and $v$, while $(c_{n}^{\rm PN}, l_{n}^{\rm PN})$ are post-Newtonian General Relativity coefficients (see e.g.~\cite{Klein:2013qda}). In writing the Fourier response in this way, we had to redefine the phase of coalescence via
\begin{align}
\Phi_{c}^{(\ell)} = \Phi_{c} - \delta_{\ell,2} \left\{ \arctan\left[\frac{2 \cos{\iota} \; F_{\times}}{\left(1 + \cos^{2}{\iota}\right) F_{+}}\right] + \frac{\Gamma}{\omega_{\rm BD}} \frac{\cos{\iota} \left(1 - \cos^{2}{\iota}\right) F_{\times} F_{\rm b}}{\left(1 + \cos^{2}{\iota}\right)^{2} F_{+}^{2} + 4 \cos^{2}{\iota} F_{\times}^{2}} \right\}\,,
\end{align}
where $\delta_{\ell,m}$ is the Kronecker delta and $\Phi_{c}$ is the GR phase of coalescence (defined as an integration constant when the frequency diverges). Of course, in this calculation we have neglected amplitude corrections that arise purely in General Relativity, if one were to carry out the post-Newtonian approximation to higher order.

Many studies have been carried out to determine the level at which such corrections to the waveform could be measured or constrained once a gravitational wave has been detected. The first such study was carried out by Will~\cite{Will:1994fb}, who determined that given a LIGO detection at signal-to-noise ratio $\rho = 10$ of a $(1.4,3) M_{\odot}$ black-hole/neutron-star non-spinning, quasi-circular inspiral, one could constrain $\omega_{\rm BD} > 10^{3}$. Scharre and Will~\cite{Scharre:2001hn} carried out a similar analysis but for a LISA detection with $\rho = 10$ of a $(1.4,10^{3}) M_{\odot}$ intermediate-mass black-hole/neutron-star, non-spinning, quasi-circular inspiral, and found that one could constrain $\omega_{\rm BD} > 2.1 \times 10^{4}$. Such an analysis was then repeated by Will and Yunes~\cite{Will:2004xi} but as a function of the classic LISA instrument. They found that the bound is independent of the LISA arm length, but inversely proportional to the LISA position noise error, if the position error noise dominates over laser shot noise. All such studies considered an angle-averaged signal that neglected the spin of either body, assumptions that were relaxed by Berti, et al.~\cite{Berti:2004bd,Berti:2005qd}. They carried out Monte-Carlo simulations over all sky position of signals that included spin-orbit precession to find that the projected bound with LISA deteriorates to $\omega_{\rm BD} > 0.7 \times 10^{4}$ for the same system and signal-to-noise ratio. This was confirmed and extended by Yagi, et al.~\cite{Yagi:2009zm}, who in addition to spin-orbit precession allowed for non-circular (eccentric) inspirals. In fact, when eccentricity is included, the bound deteriorates even further to $\omega_{\rm BD} > 0.5 \times 10^{4}$. The same authors also found that similar gravitational wave observations with the next-generation detector DECIGO could constrain $\omega_{\rm BD} > 1.6 \times 10^{6}$. Similarly, for a non-spinning neutron star/black hole binary, the future ground-based detector, the Einstein Telescope (ET)~\cite{Punturo:2010zz}, could place constraints about 5 times stronger than the Cassini bound, as shown in~\cite{Arun:2013bp}. 

All such projected constraints are to be compared with the current Solar System bound of $\omega_{\rm BD} > 4 \times 10^{4}$ placed through the tracking of the Cassini spacecraft~\cite{Bertotti:2003rm}. Table~\ref{table:comparison-BD} presents all such bounds for ease of comparison\epubtkFootnote{All LISA bounds refer to the classic LISA configuration.}, normalized to signal-to-noise ratio of $10$. As it should be clear, it is unlikely that LIGO observations will be able to constrain $\omega_{\rm BD}$ better than current Solar System bounds. In fact, even LISA would probably not be able to do better than the Cassini bound. Table~\ref{table:comparison-BD} also shows that the inclusion of more complexity in the waveform seems to dilute the level at which $\omega_{\rm BD}$ can be constrained. This is because the inclusion of eccentricity and spin forces one to introduce more parameters in the waveform, without these modifications truly adding enough waveform complexity to break the induced degeneracies. One would then expect that the inclusion of amplitude modulation due to precession and higher harmonics should break such degeneracies, at least partially, as was found for massive binary black holes~\cite{Lang:1900bz,Lang:2011je}. Even then, however, it seems reasonable to expect that only third-generation detectors will be able to constrain $\omega_{\rm BD}$ beyond Solar System levels. 

 \begin{table}[htbp]
 \caption{Comparison of Proposed Tests of Scalar-Tensor Theories}
 \label{table:comparison-BD}
  \centering
  \begin{tabular}{cccl}
    \toprule
	\textbf{Reference} & \textbf{Binary Mass}  & $\omega_{\rm BD} [10^{4}]$ & \textbf{Properties}\\
	\midrule
 	\cite{Bertotti:2003rm} & x & $4$ & Solar System\\
	\midrule
	\midrule
	\cite{Will:1994fb} & $(1.4,3) M_{\odot}$ & $0.1$ & LIGO, Fisher, Ang. Ave.\\
	&&& circular, non-spinning \\
	\midrule
	\cite{Scharre:2001hn} & $(1.4,10^{3}) M_{\odot}$ & $24$ & LISA, Fisher, Ang. Ave. \\
	&&& circular, non-spinning \\
	\midrule
	\cite{Will:2004xi} & $(1.4,10^{3}) M_{\odot}$ & $20$ & LISA study, Fisher, \\
	&&& Ang. Ave., circular, non-spinning \\
	\midrule
	\cite{Berti:2004bd} & $(1.4,10^{3}) M_{\odot}$ & $0.7$ & LISA, Fisher, Monte-Carlo.\\
	&&& circular, w/spin-orbit \\
	\midrule
	\cite{Yagi:2009zm} & $(1.4,10^{3}) M_{\odot}$ & $0.5$ & LISA, Fisher, Monte-Carlo\\
	&&& eccentric, spin-orbit \\
	\midrule
	\cite{Yagi:2009zz} & $(1.4,10) M_{\odot}$ & $160$ & DECIGO, Fisher, Monte-Carlo\\
	&&& eccentric, spin-orbit \\
	\midrule
	\cite{Arun:2013bp} & $(1.4,10) M_{\odot}$ & $10$ & ET, Fisher, Ang. Ave.\\
	&&& circular, non-spinning \\
	\bottomrule
  \end{tabular}
\end{table}
%I have rescaled Yagi's DECIGO results to SNR=10 (since it goes inversely with $\rho) 
% 6944*10/sqrt(200)=4860 and 2.3 10^{6} 10/sqrt(200) = 1.6 10^{6}

The main reason that Solar System constraints of Jordan--Fierz--Brans--Dicke theory cannot be beaten with gravitational wave observations is that the former are particularly well-suited to constrain weak-field deviations of General Relativity. One might have thought that scalar-tensor theories constitute strong-field tests of Einstein's theory, but this is not quite true, as argued in Section~\ref{sec:ST}. One can see this clearly by noting that scalar-tensor theory predicts dipolar radiation, which dominates at low velocities over the General Relativity prediction (precisely the opposite behavior that one would expect from a strong-field modification to Einstein's theory).  

One should note, however, that all the above analysis considered only the inspiral phase of coalescence, usually truncating their study at the innermost stable-circular orbit. The merger and ringdown phases, where most of the gravitational wave power resides, have so far been mostly neglected. One might expect that an increase in power will be accompanied by an increase in signal-to-noise ratio, thus allowing us to constrain $\omega_{\rm BD}$ further, as this scales with 1/SNR~\cite{Keppel:2010qu}. Moreover, during merger and ringdown, dynamical strong-field gravity effects in scalar-tensor theories could affect neutron star parameters and their oscillations~\cite{Sotani:2005qx}, as well as possibly induce spontaneous scalarization~\cite{Barausse:2012da}. All of these non-linear effects could easily lead to a strengthening of projected bounds. To date, however, no detailed analysis has attempted to determine how well one could constrain scalar-tensor theories using full information about the entire coalescence of a compact binary. 

The subclass of scalar-tensor models described by Jordan--Fierz--Brans--Dicke theory is not the only type of model that can be constrained with gravitational wave observations. In the extreme mass-ratio limit, for binaries consisting of a stellar-mass compact object spiraling into a supermassive black hole, Yunes et al.~\cite{Yunes:2011aa} have recently showed that generic scalar-tensor theories reduce to either massless or massive Jordan--Fierz--Brans--Dicke theory. Of course, in this case the sensitivities need to be calculated from the equations of structure within the full scalar-tensor theory. The inclusion of a scalar field mass leads to an interesting possibility: floating orbits~\cite{Cardoso:2011xi}. Such orbits arise when the small compact object experiences superradiance, leading to resonances in the scalar flux that can momentarily counteract the gravitational wave flux, leading to a temporarily stalled orbit that greatly modifies the orbital phase evolution. These authors showed that if an extreme mass-ratio inspiral is detected with a template consistent with General Relativity, this alone allows us to rule out a large region of $(m_{s},\omega_{\rm BD})$ phase space, where $m_{s}$ is the mass of the scalar (see Figure~1 in~\cite{Yunes:2011aa}). This is because if such an inspiral had gone through a resonance, a General Relativity template would be grossly different from the signal. Such bounds are dramatically stronger than the current most stringent bound $\omega_{\rm BD} > 4 \times 10^{4}$ and $m_{s} < 2.5 \times 10^{-20} \; {\rm{eV}}$ obtained from Cassini measurements of the Shapiro time-delay in the Solar System~\cite{Alsing:2011er}. Even if resonances are not hit, Berti et al.~\cite{Berti:2012bp} have estimated that second-generation ground-based interferometers could constrain the combination $m_{s}/(\omega_{\rm BD})^{1/2} \lesssim 10^{-15} \; {\rm{eV}}$ with the observation of gravitational waves from neutron-star/binary inspirals at signal-to-noise ratio of $10$. These bounds can also be stronger than current constraints, especially for large scalar mass.

Lastly one should mention possible gravitational wave constraints on other types of scalar tensor theories. Let us first consider Brans--Dicke type scalar-tensor theories, where the coupling constant is allowed to vary. Will~\cite{Will:1994fb} has argued that the constraints described in Table~\ref{table:comparison-BD} go through, with the change
\begin{equation}
\frac{2{\cal{G}}_{1,2}}{2 + \omega_{\rm BD}} \to \frac{2{\cal{G}}_{1,2}}{2 + \omega_{\rm BD}} \left[1  + \frac{2 \omega_{\rm BD}'}{(3 + 2 \omega_{\rm BD})^{2}} \right]^{2}\,,
\end{equation}
where $\omega'_{\rm BD} \equiv d\omega_{\rm BD}/d\phi$. In the $\omega_{\rm BD} \gg 1$ limit, this implies the replacement $\omega_{\rm BD} \to \omega_{\rm BD} [1 + \omega_{\rm BD}'/(2 \omega_{\rm BD}^{2})]^{-2}$. Of course, this assumes that there is neither a potential nor a geometric source driving the evolution of the scalar field, and is not applicable for theories where spontaneous scalarization is present~\cite{Damour:1992we}. 

Another interesting scalar-tensor theory to consider is that studied by Damour and Esposito-Farese~\cite{Damour:1992we,Damour:1993hw}. As explained in Section~\ref{sec:ST}, this theory is defined by the action of Eq.~\eqref{ST-gen-action} with the conformal factor $A(\psi) = e^{\beta \psi^{2}}$. In standard Brans--Dicke theory, only mixed binaries composed of a black hole and a neutron star lead to large deviations from General Relativity due to dipolar emission. This is because dipole emission is proportional to the difference in sensitivities of the binary components. For neutron--star binaries with similar masses, this difference is close to zero, while for black holes it is identically zero (see Eqs.~\eqref{eq:S-def} and~\eqref{eq:BD-phase}). In the theory considered by Damour and Esposito-Farese, however, when the gravitational energy is large enough, as in the very late inspiral, non-linear effects can lead to drastic modifications from the General Relativity expectation, such as spontaneously scalarization~\cite{Barausse:2012da}. Unfortunately, most of this happens at rather high frequency, and thus, it is not clear whether such effects are observable with current ground-based detectors.

%------------------------------------------------------------------------------------------------------------------------------
\subsubsection{Modified quadratic gravity}
\label{sec:direct-test-MQG}

Black holes exist in the classes of modified quadratic gravity that have been so far considered. In non-dynamical theories (when $\beta = 0$ and the scalar-fields are constant, refer to Eq.~\eqref{exactaction}), Stein and Yunes~\cite{Yunes:2011we} have shown that all metrics that are Ricci tensor flat are also solutions to the modified field equations (see also~\cite{Psaltis:2007cw}). This is not so for dynamical theories, since then the $\vartheta$ field is sourced by curvature, leading to corrections to the field equations proportional to the Riemann tensor and its dual. 

In dynamical Chern--Simons gravity, stationary and spherically symmetric spacetimes are still described by General Relativity  solutions, but stationary and axisymmetric spacetimes are not. Instead, they are represented by~\cite{Yunes:2009hc,Konno:2009kg} 
\begin{equation}
ds^{2}_{\rm CS} = ds^{2}_{\rm Kerr} + \frac{5}{4} \frac{\alpha_{\rm CS}^{2}}{\beta \kappa} \frac{a}{r^{4}} \left(1 + \frac{12}{7} \frac{M}{r}  + \frac{27}{10}\frac{M^{2}}{r^{2}} \right) \sin^{2}{\theta} d\theta dt + {\cal{O}}(a^{2}/M^{2})\,, 
\end{equation}
with the scalar field
\begin{equation}
\vartheta_{\rm CS} = \frac{5}{8} \frac{\alpha_{\rm CS}}{\beta} \frac{a}{M}  \frac{\cos{\theta}}{r^{2}} \left(1 + \frac{2 M}{r}  + \frac{18 M^{2}}{5 r^{2}} \right) + {\cal{O}}(a^{3}/M^{3})\,. 
\end{equation}
where $ds^{2}_{\rm Kerr}$ is the line element of the Kerr metric and we recall that $\alpha_{\rm CS} = -4 \alpha_{4}$ in the notation of Section~\ref{subsec:MQG}. These expressions are obtained in Boyer--Lindquist coordinates and in the small-rotation/small-coupling limit to ${\cal{O}}(a/M)$ in~\cite{Yunes:2009hc,Konno:2009kg} and to ${\cal{O}}(a^{2}/M^{2})$ in~\cite{Yagi:2012ya}. The linear-in-spin corrections modify the frame-dragging effect and they are of 3.5 post-Newtonian order. The quadratic-in-spin corrections modify the quadrupole moment, which induces 2 post-Newtonian order corrections to the binding energy. The stability of these black holes, however, has not yet been demonstrated. 

In Einstein-Dilaton-Gauss--Bonnet gravity, stationary and spherically symmetric spacetimes are described, in the small-coupling approximation, by the line element~\cite{Yunes:2011we}
\begin{equation}
ds^{2}_{\rm EDGB} = -f_{\rm Schw} \left(1 + h\right) dt^{2} + f^{-1}_{\rm Schw} \left(1 + k \right)  dr^{2} + r^{2} d\Omega^{2}\,,
\end{equation}
in Schwarzschild coordinates, where $d \Omega^{2}$ is the line element on the two-sphere, $f_{\rm Schw}=1 - 2 M/r$ is the Schwarzschild factor and we have defined
\begin{align}
h &= \frac{\alpha_{3}^{2}}{\beta \kappa M^{4}} \frac{1}{3 f_{\rm Schw}} \frac{M^{3}}{r^{3}} \left(1  + 26 \frac{M}{r} + \frac{66}{5} \frac{M^{2}}{r^{2}} + \frac{96}{5} \frac{M^{3}}{r^{3}} -  80 \frac{M^{4}}{r^{4}}\right)\,,
\\
k &= -\frac{\alpha_{3}^{2}}{\beta \kappa M^{4}} \frac{1}{f_{\rm Schw}} \frac{M^{2}}{r^{2}} \left[ 1 +  \frac{M}{r} + \frac{52}{3} \frac{M^{2}}{r^{2}} + 2 \frac{M^{3}}{r^{3}} + \frac{16}{5} \frac{M^{4}}{r^{4}} - \frac{368}{3} \frac{M^{5}}{r^{5}} \right]\,,
\end{align}
while the corresponding scalar field is 
\begin{equation}
\vartheta_{\rm EDGB} = \frac{\alpha_{3}}{\beta} \frac{2}{M r} \left(1 + \frac{M}{r} + \frac{4}{3} \frac{M^{2}}{r^{2}}\right)\,.
\label{eq:theta-EDGB}
\end{equation}
This solution is not restricted just to Einstein-Dilaton-Gauss--Bonnet gravity, but it is also the most general, stationary and spherically symmetric solution in quadratic gravity. This is because all terms proportional to $\alpha_{1,2}$ are proportional to the Ricci tensor, which vanishes in vacuum General Relativity, while the $\alpha_{4}$ term does not contribute in spherical symmetry (see~\cite{Yunes:2011we} for more details). 
Linear slow-rotation corrections to this solution have been found in~\cite{Pani:2011gy}. Although the stability of these black holes has not yet been demonstrated, other dilatonic black hole solutions obtained numerically  (equivalent to those in Einstein-Dilaton-Gauss--Bonnet theory in the limit of small fields)~\cite{Kanti:1995vq} have been found to be stable under axial perturbations~\cite{Kanti:1997br,Torii:1998gm,Pani:2009wy}.

Neutron stars also exist in quadratic modified gravity. In dynamical Chern--Simons gravity, the mass-radius relation remains unmodified to first-order in the slow-rotation expansion, but the moment of inertia changes to this order~\cite{Yunes:2009ch,AliHaimoud:2011fw}, while the quadrupole moment and the mass measured at spatial infinity change to quadratic order in spin~\cite{Yagi:2013mbt}. This is because the mass-radius relation, to first order in slow-rotation, depends on the spherically-symmetric part of the metric, which is unmodified in dynamical Chern--Simons gravity. In Einstein-Dilaton-Gauss--Bonnet gravity, the mass-radius relation is modified~\cite{Pani:2011xm}. As in General Relativity, these functions must be solved for numerically and they depend on the equation of state. 

Gravitational waves are also modified in quadratic modified gravity. In dynamical Chern--Simons gravity, Garfinkle et al.~\cite{Garfinkle:2010zx} have shown that the propagation of such waves on a Minkowski background remains unaltered, and thus, all modifications arise during the generation stage. In Einstein-Dilaton-Gauss--Bonnet theory, no such analysis of the propagation of gravitational waves has yet been carried out. Yagi et al.~\cite{Yagi:2011xp} studied the generation mechanism in both theories during the quasi-circular inspiral of comparable-mass, spinning black holes in the post-Newtonian and small-coupling approximations. They found that a standard post-Newtonian analysis fails for such theories because the assumption that black holes can be described by a distributional stress-energy tensor without any further structure fails. They also found that since black holes acquire scalar hair in these theories, and this scalar field is anchored to the curvature profiles, as black holes move, the scalar fields must follow the singularities, leading to dipole scalar-field emission. 

During a quasi-circular inspiral of spinning black holes in dynamical Chern--Simons gravity, the total gravitational wave energy flux carried out to spatial infinity (equal to minus the rate of change of a binary's binding energy by the balance law) is modified from the General Relativity expectation to leading order by~\cite{Yagi:2011xp}  
\begin{equation}
\frac{\delta \dot{E}^{\rm CS}_{\rm spin}}{\dot{E}_{\rm GR}} =\zeta_{4} \eta^{-2}  \left\{\frac{25}{1536} \left[\bar{\Delta} +2 \left< \left(\bar{\Delta} \cdot \hat{v}_{12}\right) \right> \right] + \frac{75}{16} \frac{a_{1} a_{2}}{m^{2}} \left< \hat{S}_{1}^{i} \hat{S}_{2}^{j} \left(2 \hat{v}_{12}^{i} \hat{v}_{12}^{j} - 2 \hat{n}_{12}^{i} \hat{n}_{12}^{j}\right)\right>\right\} v_{12}^{4}\,,
\label{eq:Edot-CS-spin}
\end{equation}
due to scalar field radiation and corrections to the metric perturbation that are of magnetic-type, quadrupole form. In this equation, $\dot{E}_{\rm GR} = (32/5) \eta^{2} v_{12}^{10}$ is the leading order General Relativity prediction for the total energy flux, $\zeta_{4} = \alpha_{4}^{2}/(\beta \kappa m^{4})$ is the dimensionless Chern--Simons coupling parameter, $v_{12}$ is the magnitude of the relative velocity with unit vector $\hat{v}_{12}^{i}$, $\bar{\Delta}^{i} = (m_{2}/m_{1}) (a_{1}/m) \hat{S}^{i}_{1} - (m_{1}/m_{2}) (a_{2}/m) \hat{S}^{i}_{2}$, where $a_{A}$ is the Kerr spin parameter of the $A$th black hole and $\hat{S}^{i}_{A}$ is the unit vector in the direction of the spin angular momentum, the unit vector $\hat{n}_{12}^{i}$ points from body one to two, and the angle brackets stand for an average over several gravitational wave wavelengths. If the black holes are not spinning, then the correction to the scalar energy flux is greatly suppressed~\cite{Yagi:2011xp} 
\begin{equation}
\frac{\delta \dot{E}^{\rm CS}_{\rm no-spin}}{\dot{E}_{\rm GR}} = \frac{2}{3} \delta_{m}^{2} \zeta_{4} v_{12}^{14}\,,
\end{equation}
where we have defined the reduced mass difference $\delta_{m} \equiv (m_{1} - m_{2})/m$. Notice that this is a 7 post-Newtonian order correction, instead of a 2 post-Newtonian correction as in Eq.~\eqref{eq:Edot-CS-spin}. In the non-spinning limit, the dynamical Chern--Simons correction to the metric tensor induces a 6 post-Newtonian order correction to the gravitational energy flux~\cite{Yagi:2011xp}, which is consistent with the numerical results of~\cite{Pani:2011xj}. 

On the other hand, in Einstein-Dilaton-Gauss--Bonnet gravity, the corrections to the energy flux are~\cite{Yagi:2011xp} 
\begin{equation}
\frac{\delta \dot{E}^{\rm EDGB}_{\rm no-spin}}{\dot{E}_{\rm GR}} = \frac{5}{96} \eta^{-4} \delta_{m}^{2} \zeta_{3} v_{12}^{-2}\,,
\end{equation}
which is a $-1$ post-Newtonian correction. This is because the scalar field $\vartheta_{\rm EDGB}$ behaves like a monopole (see Eq.~\eqref{eq:theta-EDGB}), and when such a scalar monopole is dragged by the black hole, it emits electric-type, dipole scalar radiation. Any hairy black hole with monopole hair will thus emit dipolar radiation, leading to $-1$ post-Newtonian corrections in the energy flux carried to spatial infinity.  

Such modifications to the energy flux modify the rate of change of the binary's binding energy through the balance law, $\dot{E} = -\dot{E}_{\rm b}$, which in turn modify the rate of change of the gravitational wave frequency and phase, $\dot{F} = - \dot{E} \; (dE_{\rm b}/dF)^{-1}$. For dynamical Chern--Simons gravity (when the spins are aligned with the orbital angular momentum) and for Einstein-Dilaton-Gauss--Bonnet theory (in the non-spinning case), the Fourier transform of the gravitational wave response function in the stationary phase approximation becomes~\cite{Yagi:2011xp,Yagi:2012vf} 
\begin{equation}
\tilde{h}_{\rm dCS,EDGB} = \tilde{h}_{\rm GR} e^{i \beta_{\rm dCS,EDGB} u^{b_{\rm dCS,EDGB}}}\,,
\end{equation}
where $\tilde{h}_{\rm GR}$ is the Fourier transform of the response in General Relativity, $u \equiv (\pi {\cal{M}}_{c} f)^{1/3}$ with $f$ the gravitational wave frequency and~\cite{Yagi:2011xp,Yagi:2012vf}   
\begin{align}
\label{eq:beta-dCS}
\beta_{\rm dCS} &= \frac{1549225}{11812864} \frac{\zeta_{4}}{\eta^{14/5}} \left[ \left( 1-\frac{47953}{61969}\eta \right) \chi_s^2 +\left( 1-\frac{199923}{61969}\eta \right) \chi_a^2  - 2 \delta_m \chi_s \chi_a \right]\,,
\qquad
b_{\rm dCS} = -1\,,
\\
\beta_{\rm EDGB} &= -\frac{5}{7168} \zeta_{3} \eta^{-18/5} \delta_{m}^{2} \,, 
\qquad
b_{\rm EDGB} = -7\,,
\end{align}
where we have defined the symmetric and antisymmetric spin combinations $\chi_{s,a} \equiv \left(a_{1}/m_{1} \pm a_{2}/m_{2}\right)/2$. We have here neglected any possible amplitude correction, but we have included both deformations to the binding energy and Kepler's third law, in addition to changes in the energy flux, when computing the phase correction. In Einstein-Dilaton-Gauss--Bonnet theory, however, the binding energy is modified at higher post-Newtonian order, and thus, corrections to the energy flux control the modifications to the gravitational wave response function. 

From the above analysis, it should be clear that the corrections to the gravitational wave observable in quadratic modified gravity are always proportional to the quantity $\zeta_{3,4} \equiv \xi_{3,4}/m^{4} = \alpha_{3,4}^{2}/(\beta \kappa m^{4})$. Thus, any measurement that is consistent with General Relativity will allow a constraint of the form $\zeta_{3,4} < N \delta$, where $N$ is a number of order unity, and $\delta$ is the accuracy of the measurement. Solving for the coupling constants of the theory, such a measurement would lead to $\xi_{3,4}^{1/4} < (N \delta)^{1/4} m$~\cite{Sopuerta:2009iy}. Therefore, constraints on quadratic modified gravity will weaken for systems with larger characteristic mass. This can be understood by noticing that the corrections to the action scale with positive powers of the Riemann tensor, while this scales inversely with the mass of the object, ie.~the smaller a compact object is, the larger its curvature. Such an analysis then automatically predicts that LIGO will be able to place stronger constraints than LISA-like missions on such theories, because LIGO operates in the 100~Hz frequency band, allowing for the detection of stellar-mass inspirals, while LISA-like missions operate in the mHz band, and are limited to supermassive black holes inspirals.      

How well can these modifications be measured with gravitational wave observations? Yagi, et al.~\cite{Yagi:2011xp} predicted, based on the results of Cornish et al.~\cite{Cornish:2011ys}, that a sky-averaged LIGO gravitational wave observation with signal-to-noise ratio of $10$ of the quasi-circular inspiral of non spinning black holes with masses $(6,12) \; M_{\odot}$ would allow a constraint of $\xi_{3}^{1/4} \lesssim 20\mathrm{\ km}$, where we recall that $\xi_{3} = \alpha_{3}^{2}/(\beta \kappa)$. A similar sky-averaged, eLISA observation of a quasi-circular, spin-aligned black hole inspiral with masses $(10^{6},3\times10^{6}) \; M_{\odot}$ would constrain $\xi_{3}^{1/4} < 10^{7}\mathrm{\ km}$~\cite{Yagi:2011xp}. The loss in constraining power comes from the fact that the constraint on $\xi_{3}$ will scale with the total mass of the binary, which is six orders of magnitude larger for space-borne sources. These constraints are not stronger than current bounds from the existence of compact objects~\cite{Pani:2011xm} ($\xi_{3} < 26\mathrm{\ km}$) and from the change in the orbital period of the low-mass x-ray binary $A0620-00$ ($\xi_{3} < 1.9\mathrm{\ km}$)~\cite{Yagi:2012gb}, but they are independent of the nature of the object and sample the theory in a different energy scale. In dynamical Chern--Simons gravity, one expects similar projected gravitational wave constraints on $\xi_{4}$, namely $\xi_{4}^{1/4} < {\cal{O}}(M)$, where $M$ is the total mass of the binary system in kilometers. Therefore, for binaries detectable with ground-based interferometers, one expects constraints of order $\xi_{4}^{1/4} < 10\mathrm{\ km}$. In this case, such a constraint would be roughly six orders of magnitude stronger than current LAGEOS bounds~\cite{AliHaimoud:2011fw}. Dynamical Chern--Simons gravity cannot be constrained with binary pulsar observations, since the theory's corrections to the post-Keplerian observables are too high post-Newtonian order, given the current observational uncertainties~\cite{Yagi:2013mbt}. The gravitational wave constraint, however, is more difficult to achieve in the dynamical Chern--Simons case, because the correction to the gravitational wave phase is degenerate with spin. Yagi, et al.~\cite{Yagi:2012vf}, however, argued that precession should break this degeneracy, and if a signal with sufficiently high signal-to-noise ratio is observed, such bounds would be possible. One must be careful, of course, to check that the small-coupling approximation is still satisfied when saturating such a constraint~\cite{Yagi:2012vf}. 

%------------------------------------------------------------------------------------------------------------------------------
\subsubsection{Non-commutative geometry}

Black holes exist in non-commutative geometry theories, as discussed in Section~\ref{sec:NCG}. What is more, the usual Schwarzschild and Kerr solutions of General Relativity persist in these theories. This is not because such solutions have vanishing Weyl tensor, but because the quantity $\nabla^{\alpha \beta} C_{\mu \alpha \nu \beta}$ happens to vanish for such metrics. Similarly, one would expect that the two-body, post-Newtonian metric that describes a binary black hole system should also satisfy the non-commutative geometry field equations, although this has not been proven explicitly. Similarly, although neutron star spacetimes have not yet been considered in non-commutative geometries, it is likely that if such spacetimes are stationary and satisfy the Einstein equations, they will also satisfy the modified field equations. Much more work on this is still needed to establish all of these concepts on a firmer basis.  

Gravitational waves exist in non-commutative gravity. Their generation for a compact binary system in a circular orbit was analyzed by Nelson et al, in~\cite{Nelson:2010rt,Nelson:2010ru}. They began by showing that a transverse-traceless gauge exists in this theory, although the transverse-traceless operator is slightly different from that in General Relativity. They then proceeded to solve the modified field equations for the metric perturbation [Eq.~\eqref{eq:NCG-h-eq}] via a Green's function approach:
\begin{equation}
h^{ik} = 2 \beta \int \frac{dt'}{\sqrt{(t - t')^{2} - |r|^{2}}} \ddot{I}^{ik}(t') {\cal{J}}_{1}(\beta \sqrt{(t - t')^{2} - |r|^{2}})\,,
\label{eq:h-int-eq-NCG}
\end{equation}
where recall that $\beta^{2} = (-32 \pi \alpha_{0})^{-1}$ acts like a mass term, the integral is taken over the entire past light cone, ${\cal{J}}_{1}(\cdot)$ is the Bessel function of the first kind, $|r|$ is the distance from the source to the observer and the quadrupole moment is defined as usual:
\begin{equation}
I^{ik} = \int d^{3}x \; T^{00}_{\rm mat} x^{ik} \,,
\end{equation}
where $T^{00}$ is the time-time component of the matter stress-energy tensor. Of course, this is only the first term in an infinite multipole expansion. 

Although the integral in Eq.~\eqref{eq:h-int-eq-NCG} has not yet been solved in the post-Newtonian approximation, Nelson et al.~\cite{Nelson:2010rt,Nelson:2010ru} did solve for its time derivative to find
\begin{subequations}
\begin{align}
\dot{h}^{xx} &= -\dot{h}^{yy} = 32 \beta \mu r_{12}^{2} \Omega^{4} \left[\sin{(2 \phi)} f_{c}\left(\beta |r|,\frac{\Omega}{\beta}\right) + \cos{(2 \phi)} f_{s}\left(\beta |r|,\frac{2 \Omega}{\beta}\right)\right]\,,
\\
\dot{h}^{xy} &= - 32 \beta \mu r_{12}^{2} \Omega^{4} \left[\sin{\left(2 \phi - \frac{\pi}{2}\right)} f_{c}\left(\beta |r|,\frac{\Omega}{\beta}\right) + \cos{\left(2 \phi - \frac{\pi}{2}\right)} f_{s}\left(\beta |r|,\frac{2 \Omega}{\beta}\right)\right]\,, 
\end{align}
\label{h-NCG}
\end{subequations}
% 
% \omega^{ij} = 2 \omega except for \omega^{zz} = 0.
% A^{xx} = - A^{yy} = -A^{xy} = 12 \mu r_{12}^{2} \omega^{3}, 	A^{zz} = - \mu r_{12}^{2}
where $\Omega = 2 \pi F$ is the orbital angular frequency and we have defined
\begin{align}
f_{s}(x,z) &= \int_{0}^{\infty} \frac{ds}{\sqrt{s^{2} + x^{2}}} {\cal{J}}_{1}(s) \sin\left(z \sqrt{s^{2} + x^{2}}\right)\,,
\\
f_{c}(x,z) &= \int_{0}^{\infty} \frac{ds}{\sqrt{s^{2} + x^{2}}} {\cal{J}}_{1}(s) \cos\left(z \sqrt{s^{2} + x^{2}}\right)\,.
\end{align}
and one has assumed that the binary is in the $x$-$y$ plane and the observer is on the $z$-axis. If one expands these expressions about $\beta = \infty$ however, one recovers the General Relativity solution to leading order, plus corrections that decay faster than $1/r$. This then automatically implies that such modifications to the generation mechanism will be difficult to observe for sources at astronomical distances.  

Given such a solution, one can compute the flux of energy carried by gravitational waves to spatial infinity. Stein and Yunes~\cite{Stein:2010pn} have shown that in quadratic gravity theories, this flux is still given by
\begin{equation}
\dot{E} = \frac{\kappa}{2} \int d\Omega r^{2} \left< \dot{\bar{h}}_{\mu \nu} \dot{\bar{h}}^{\mu \nu}\right>\,,
\label{eq:Edot-NCG}
\end{equation}
where $\bar{h}_{\mu \nu}$ is the trace-reversed metric perturbation, the integral is taken over a 2-sphere at spatial infinity, and we recall that the angle brackets stand for an average over several wavelengths. Given the solution in Eq.~\eqref{h-NCG}, one finds that the energy flux is
\begin{equation}
\dot{E} = \frac{9}{20} \mu^{2} r_{12}^{2} \Omega^{4} \beta^{2} 
\left[|r|^{2} f_{c}^{2}\left(\beta |r|, \frac{2 \Omega}{\beta}\right) + |r|^{2} f_{s}^{2}\left(\beta |r|, \frac{2 \Omega}{\beta}\right) \right]\,.
\label{eq:NCG-Edot}
\end{equation}
The asymptotic expansion of the term in between square brackets about $\beta = \infty$ is
\begin{equation}
|r|^{2} \left[ f_{c}^{2}\left(\beta |r|, \frac{2 \Omega}{\beta}\right) + f_{s}^{2}\left(\beta |r|, \frac{2 \Omega}{\beta}\right) \right]\sim
|r|^{2} \left\{ \frac{1}{\beta^{2} |r|^{2}} \left[1 + {\cal{O}}\left(\frac{1}{|r|}\right) \right] \right\}\,,
\end{equation}
which then leads to an energy flux identical to that in General Relativity, as any subdominant term goes to zero when the 2-sphere of integration is taken to spatial infinity. In that case, there are no modifications to the rate of change of the orbital frequency. Of course, if one were not to expand about $\beta = \infty$, then the energy flux would lead to certain resonances at $\beta = 2 \Omega$, but the energy flux is only well-defined at future null infinity. 

The above analysis was used by Nelson, et al.~\cite{Nelson:2010rt,Nelson:2010ru} to compute the rate of change of the orbital period of binary pulsars, in the hopes of using this to constrain $\beta$. Using data from the double binary pulsar, they stipulated an order-of-magnitude constraint of $\beta \geq 10^{-13} \; {\rm{m}}^{-1}$. Such an analysis, however, could be revisited to relax a few assumptions used in~\cite{Nelson:2010rt,Nelson:2010ru}. First, binary pulsar constraints on modified gravity theories require the use of at least three observables. These observables can be, for example, the rate of change of the period $\dot{P}$, the line of nodes $\dot{\Omega}$ and the perihelion shift $\dot{w}$. Any one observable depends on the parameters $(m_{1},m_{2})$ in General Relativity or $(m_{1},m_{2},\beta)$ in non-commutative geometries, where $m_{1,2}$ are the component masses. Therefore, each observable corresponds to a surface of co-dimension one, ie.~a two-dimensional surface or sheet in the three-dimensional space $(m_{1},m_{2},\beta)$. If the binary pulsar observations are consistent with Einstein's theory, then all sheets will intersect at some point, within a certain uncertainty volume given by the observational error. The simultaneous fitting of all these observables is what allows one to place a bound on $\beta$. The analysis of~\cite{Nelson:2010rt,Nelson:2010ru} assumed that all binary pulsar observables were known, except for $\beta$, but degeneracies between $(m_{1},m_{2},\beta)$ could potentially dilute constraints on these quantities. Moreover, this analysis should be generalized to eccentric and inclined binaries, since binary pulsars are known to not be on exactly circular orbits. 

But perhaps the most important modification that ought to be made has to do with the calculation of the energy flux itself. The expression for $\dot{E}$ in Eq.~\eqref{eq:Edot-NCG} in terms of derivatives of the metric perturbation derives from the effective gravitational wave stress-energy tensor, obtained by perturbatively expanding the action or the field equations and averaging over several wavelengths (the so-called Isaacson procedure~\cite{Isaacson:1968ra,Isaacson:1968gw}). In modified gravity theories, the definition of the effective stress-energy tensor in terms of the metric perturbation is usually modified, as found for example in~\cite{Stein:2010pn}. In the case of non-commutative geometries, Stein and Yunes~\cite{Stein:2010pn} showed that Eq.~\eqref{eq:Edot-NCG} still holds, provided one considers fluxes at spatial infinity. The analysis of~\cite{Nelson:2010rt,Nelson:2010ru}, however, evaluated this energy flux at a fixed distance, instead of taking the $r \to \infty$ limit. 

The balance law relates the rate of change of a binary's binding energy with the gravitational wave flux emitted by the binary, but for it to hold, one must require the following: (i) that the binary be isolated and it possess a well-defined binding energy; (ii) the total stress-energy of the spacetime satisfies a local covariant conservation law. If (ii) holds, one can use this conservation law to relate the rate of change of the volume integral of the energy density, ie.~the energy flux, to the volume integral of the current density, which can be rewritten as an integral over the boundary of the volume through Stokes' theorem. Since in principle one can choose any integration volume, any physically meaningful result should be independent of the surface of that volume. This is indeed the case in General Relativity, provided one takes the integration $2$-sphere to spatial infinity. Presumably, if one included all the relevant terms in $\dot{E}$, without taking the limit to $i^{0}$, one would still find a result that is independent of the surface of this two-sphere. This, however, has not yet been verified. Therefore, the analysis of~\cite{Nelson:2010rt,Nelson:2010ru} should be taken as an interesting first step toward understanding possible changes in the gravitational wave metric perturbation in non-commutative geometries.    

Not much beyond this has been done regarding non-commutative geometries and gravitational waves. In particular, one lacks a study of whether gravitational wave propagation is modified, what the final response function is, which of course depends on the time-evolution of all propagating gravitational wave degrees of freedom, and whether there are only the two usual dynamical degrees of freedom in the metric perturbation. 

%------------------------------------------------------------------------------------------------------------------------------
\subsection{Generic tests}

%------------------------------------------------------------------------------------------------------------------------------
\subsubsection{Massive graviton theories and Lorentz violation}
\label{generic-tests:MG-LV}

Several massive graviton theories have been proposed to later be discarded due to ghosts, non-linear or radiative instabilities. Thus, little work has gone into studying whether black holes and neutron stars in these theories persist and are stable, and how the generation of gravitational waves is modified. Such questions will depend on the specific massive gravity model considered, and of course, if a Vainshtein mechanism is employed, then there will not be any modifications. 

A few generic properties of such theories, however, can still be stated. One of them is that the non-dynamical (near-zone) gravitational field will be corrected, leading to Yukawa-like modifications to the gravitational potential
~\cite{Will:1997bb}
\begin{equation}
V_{\rm MG}(r) = \frac{M}{r} e^{-r/\lambda_{g}}\,,
\qquad {\rm{or}} \qquad
V_{\rm MG}(r) = \frac{M}{r} \left(1 + \gamma_{\rm MG} e^{-r/\lambda_{g}} \right)\,,
\label{eq:Yukawa}
\end{equation}
where $r$ is the distance from the source to a field point. The latter parameterization, for example, arises in gravitational theories with compactified extra dimensions~\cite{Kehagias:1999my}. Such corrections lead to a so-called {\emph{fifth force}}, which then in turn allows us to place constraints on $m_{g}$ through Solar System observations~\cite{Talmadge:1988qz}. Nobody has yet considered how such modifications to the near-zone metric could affect the binding energy of compact binaries and their associated gravitational waves. 

Another generic consequence of a graviton mass is the appearance of additional propagating degrees of freedom in the gravitational wave metric perturbation. In particular, one expects scalar, longitudinal modes to be excited (see eg.~\cite{Dilkes:2001av}). This is, for example, the case if the action is of Pauli--Fierz type~\cite{Fierz:1939ix,Dilkes:2001av}. Such longitudinal modes arise due to the non-vanishing of the $\Psi_{2}$ and $\Psi_{3}$ Newman--Penrose scalars, and can be associated with the presence of spin-$0$ particles, if the theory is of Type N in the $E(2)$ classification~\cite{lrr-2006-3}. The specific form of the scalar mode will depend on the structure of the modified field equations, and thus, it is not possible to generically predict its associated contribution to the response function. 

A robust prediction of massive graviton theories relates to how the propagation of gravitational waves is affected. If the graviton has a mass, its velocity of propagation will differ from the speed of light, as given for example in Eq.~\eqref{eq:vg-standard}. Will~\cite{Will:1997bb} showed that such a modification in the dispersion relation leads to a correction in the relation between the difference in time of emission $\Delta t_{e}$ and arrival $\Delta t_{a}$ of two gravitons:
\begin{equation}
\Delta t_{a} = \left(1 + z\right) \left[ \Delta t_{e} + \frac{D}{2 \lambda_{g}^{2}} \left(\frac{1}{f_{e}^{2}} + \frac{1}{f_{e}^{'2}} \right) \right],
\label{eq:t-MG}
\end{equation}
where $z$ is here the redshift, $\lambda_{g}$ is the graviton's Compton wavelength, $f_{e}$ and $f_{e}'$ are the emission frequencies of the two gravitons and $D$ is the distance measure 
\begin{equation}
D = \frac{1 + z}{H_{0}} \int_{0}^{z} \frac{dz'}{(1 + z')^{2} [\Omega_{M} (1 + z')^{3} + \Omega_{\Lambda}]^{1/2}}\,,
\end{equation}
where $H_{0}$ is the present value of the Hubble parameter, $\Omega_{M}$ is the matter energy density and $\Omega_{\Lambda}$ is the vacuum energy density (for a zero spatial-curvature Universe).  

Even if the gravitational wave at the source is unmodified, the graviton time delay will leave an imprint on the Fourier transform of the response function by the time it reaches the detector~\cite{Will:1997bb}. This is because the Fourier phase is proportional to 
\begin{equation}
\Psi \propto 2 \pi \int^{f}_{f_{c}} [t(f) - t_{c}] df'\,,
\end{equation}
where now $t$ is not a constant but a function of frequency as given by Eq.~\eqref{eq:t-MG}. Carrying out the integration, one finds that the Fourier transform of the response function becomes
\begin{equation}
\tilde{h}_{\rm MG} = \tilde{h}_{\rm GR} e^{i \beta_{\rm MG} u^{b_{\rm MG}}}\,,
\label{eq:response-MG}
\end{equation}
where $\tilde{h}_{\rm GR}$ is the Fourier transform of the response function in General Relativity, we recall that $u = (\pi {\cal{M}}_{c} f)^{1/3}$ and we have defined 
\begin{equation}
\beta_{\rm MG} = - \frac{\pi^{2} D {\cal{M}}_{c}}{\lambda_{g}^{2} (1 + z)}\,,
\qquad
b_{\rm MG} = -3\,.
\end{equation}
Such a correction is of 1 post-Newtonian order relative to the leading-order, Newtonian term in the Fourier phase. Notice also that there are no modifications to the amplitude at all.

Numerous studies have considered possible bounds on $\lambda_{g}$. The most stringent Solar System constraint is $\lambda_{g} > 2.8 \times 10^{12}\mathrm{\ km}$ and it comes from observations of Kepler's third law (mainly Mars' orbit), which if the graviton had a mass would be modified by the Yukawa factor in Eq.~\eqref{eq:Yukawa}. Observations of the rate of decay of the period in binary pulsars~\cite{Finn:2001qi,Baskaran:2008za} can also be used to place the more stringent constraint $\lambda > 1.5 \times 10^{14}\mathrm{\ km}$. Similarly, studies of the stability of Kerr black holes in Pauli--Fierz theory~\cite{Fierz:1939ix} have yielded constraints of $\lambda_{g} > 2.4 \times 10^{13} \; {\textrm{km}}$~\cite{Brito:2013wya}. Gravitational wave observations of binary systems could also be used to constrain the mass of the graviton once gravitational waves are detected. One possible test is to compare the times of arrival of coincident gravitational wave and electromagnetic signals, for example in white-dwarf binary systems. Larson and Hiscock~\cite{Larson:1999kg} and Cutler et al.~\cite{Cutler:2002ef} estimated that one could constrain $\lambda_{g} > 3 \times 10^{13}\mathrm{\ km}$ with classic LISA. Will~\cite{Will:1997bb} was the first to consider constraints on $\lambda_{g}$ from gravitational wave observations only. He considered sky-averaged, quasi-circular inspirals and found that LIGO observations of $10 M_{\odot}$ equal-mass black holes would lead to a constraint of $\lambda_{g} > 6 \times 10^{12}\mathrm{\ km}$ with a Fisher analysis. Such constraints are improved to $\lambda_{g} > 6.9 \times 10^{16}\mathrm{\ km}$ with classic LISA observations of $10^{7} M_{\odot}$, equal-mass black holes. This increase comes about because the massive graviton correction accumulates with distance traveled (see Eq.~\eqref{eq:response-MG}). Since classic LISA would have been able to observe sources at Gpc scales with high signal-to-noise ratio, its constraints on $\lambda_{g}$ would have been similarly stronger than what one would achieve with LIGO observations. Will's study was later generalized by Will and Yunes~\cite{Will:2004xi}, who considered how the detector characteristics affected the possible bounds on $\lambda_{g}$. They found that this bound scales with the square-root of the LISA arm length and inversely with the square root of the LISA acceleration noise. The initial study of Will was then expanded by Berti et al.~\cite{Berti:2004bd}, Yagi and Tanaka~\cite{Yagi:2009zm}, Arun and Will~\cite{Arun:2009pq}, Stavridis and Will~\cite{Stavridis:2010zz} and  Berti, et al.~\cite{Berti:2011jz} to allow for non sky-averaged responses, spin-orbit and spin-spin coupling, higher-harmonics in the gravitational wave amplitude, eccentricity and multiple detections. Although the bound deteriorates on average for sources that are not optimally oriented relative to the detector, the bound improves when one includes spin couplings, higher harmonics, eccentricity, and multiple detections as the additional information and power encoded in the waveform increases, helping to break parameter degeneracies. All of these studies, however, neglected the merger and ringdown phases of the coalescence, an assumption that was relaxed by Keppel and Ajith~\cite{Keppel:2010qu}, leading to the strongest projected bounds $\lambda_{g} > 4 \times 10^{17}\mathrm{\ km}$. Moreover, all studies until then had computed bounds using a Fisher analysis prescription, an assumption relaxed by del Pozzo et al.~\cite{DelPozzo:2011pg}, who found that a Bayesian analysis with priors consistent with Solar System experiments leads to bounds stronger than Fisher ones by roughly a factor of two. All of these results are summarized in Table~\ref{table:comparison-MG}, normalizing everything to signal-to-noise ratio of 10. In summary, projected constraints on $\lambda_{g}$ are generically stronger than current Solar System or binary pulsar constraints by several orders of magnitude, given a LISA observation of massive black hole mergers. Even an Adv.~LIGO observation would do better than current Solar System constraints by a factor between a few~\cite{DelPozzo:2011pg} to an order of magnitude~\cite{Keppel:2010qu}, depending on the source.  

 \begin{table}[htbp]
 \caption{Comparison of Proposed Tests of Massive Graviton Theories. Ang.~Ave.~stands for an angular average over all sky locations.}
 \label{table:comparison-MG}
  \centering
  \begin{tabular}{cccl}
    \toprule
	\textbf{Reference} & \textbf{Binary Mass}  & $\lambda_{g} [10^{15}\mathrm{\ km}]$ & \textbf{Properties}\\
	\midrule
 	\cite{Talmadge:1988qz} & x & 0.0028 & Solar System Dynamics\\
	\midrule
	\cite{Finn:2001qi} & x & $1.6 \times 10^{-5}$ & Binary pulsar orbital period \\
	&&& in Visser's theory~\cite{Visser:1997hd} \\	
	\midrule 
	\cite{Brito:2013wya} & x & 0.024 & Stability of Black holes \\
	&&& in Pauli-Fierz theory~\cite{Fierz:1939ix} \\
	\midrule
	\midrule
	\cite{Will:1997bb} & $(10,10)\,M_{\odot}$ & 0.006 & LIGO, Fisher, Ang. Ave.\\
	&&& circular, non-spinning \\
	\midrule
	\cite{Will:1997bb} & $(10^{7},10^{7})\,M_{\odot}$ & 69 & LISA, Fisher, Ang. Ave.\\
	&&& circular, non-spinning \\
	\midrule
	\cite{Larson:1999kg,Cutler:2002ef} & $(0.5,0.5)\,M_{\odot}$ & 0.03 & LISA, WD-WD, coincident\\
	&&& with electromagnetic signal \\
	\midrule
	\cite{Will:2004xi} & $(10^{7},10^{7})\,M_{\odot}$ & 50 & LISA, Fisher, Ang. Ave. \\
	&&& circular, non-spinning \\
	\midrule
	\cite{Berti:2004bd} & $(10^{6},10^{6})\,M_{\odot}$ & 10 & LISA, Fisher, Monte-Carlo.\\
	&&& circular, w/spin-orbit \\
	\midrule	
	\cite{Arun:2009pq} & $(10^{5},10^{5})\,M_{\odot}$ & 10 & LISA, Fisher, Ang.~Ave\\
	&&& higher-harmonics, circular, non-spinning \\
	\midrule	
	\cite{Yagi:2009zm} & $(10^{6},10^{7})\,M_{\odot}$ & 22 & LISA, Fisher, Monte-Carlo\\
	&&& eccentric, spin-orbit \\
	\midrule
	\cite{Yagi:2009zz} & $(10^{6},10^{7})\,M_{\odot}$ & 2.4 & DECIGO, Fisher, Monte-Carlo\\
	&&& eccentric, spin-orbit \\
	\midrule
	\cite{Stavridis:2010zz} & $(10^{6},10^{6})\,M_{\odot}$ & 50 & LISA, Fisher, Monte-Carlo\\
	&&& circular, w/spin modulations \\
	\midrule
	\cite{Keppel:2010qu} & $(10^{7},10^{7})\,M_{\odot}$ & 400 & LISA, Fisher, Ang.~Ave.~\\
	&&& circular, non-spinning, w/merger \\	
	\midrule
	\cite{DelPozzo:2011pg} & $(13,3)\,M_{\odot}$ & 0.006\,--\,0.014 & LIGO, Bayesian, Ang.~Ave.~\\
	&&& circular, non-spinning \\	
	\midrule
	\cite{Berti:2011jz} & $(13,3)\,M_{\odot}$ & 30 & eLISA, Fisher, Monte-Carlo\\
	&&& multiple detections, circular, non-spinning \\	
	\bottomrule
  \end{tabular}
\end{table}
%
% For the Cutler paper, I've used a (1/2,1/2) Msun WD-WD and alpha = 3, as the state in the paper.
% Solar System is 2.8 \times 10^{12} km, based on Solar System dynamics.
% For the Yagi paper, they get 3.1 10^{21} cm = 3.1 10^{16} km, but one has to rescale Yagi's results to SNR=10, so
% since the bound goes inversely with $\rho, the multiplication factor is 10/sqrt[200], given a result of 2.2 10^{16} km. 
% Similarly, for DECIGO 3.35 10^{20}cm*10/sqrt(200)= 2.5 10^{15} km.

Before proceeding, we should note that the correction to the propagation of gravitational waves due to a non-zero graviton mass are not exclusive to binary systems. In fact, any gravitational wave that propagates a significant distance from the source will suffer from the time delays described in this section. Binary inspirals are particularly useful as probes of this effect because one knows the functional form of the waveform, and thus, one can employ matched filtering to obtain a strong constraint. But in principle, one could use gravitational wave bursts from supernovae or other sources.

We have so far concentrated on massive graviton theories, but, as discussed in Section~\ref{sec:MG-LV}, there is a strong connection between such theories and Lorentz-violation. Modifications to the dispersion relation are usually a result of a modification of the Lorentz group or its action in real or momentum space. For this reason, it is interesting to consider generic Lorentz-violating-inspired, modified dispersion relations of the form of Eq.~\eqref{eq:vg-LV}, or more precisely~\cite{Mirshekari:2011yq}
\begin{equation}
\frac{v_{g}^{2}}{c^{2}} = 1 - A E^{\alpha_{\rm LV} - 2}\,,
\label{eq:gen-disp-rel}
\end{equation}
where $\alpha_{\rm LV}$ controls the structure of the modification and $A$ its amplitude. When $\alpha_{\rm LV}=0$ and $A = m_{g}^{2} c^{2}$ one recovers the standard modified dispersion relation of Eq.~\eqref{eq:vg-standard}. Equation~\eqref{eq:gen-disp-rel} introduces a generalized time delay between subsequent gravitons of the form~\cite{Mirshekari:2011yq}
\begin{equation}
\Delta t_{a} = (1 + z) \left[\Delta t_{e} + \frac{D_{\alpha_{\rm LV}}}{2 \lambda_{a}^{2-\alpha_{\rm LV}}} \left(\frac{1}{f_{e}^{2-\alpha_{\rm LV}}} - \frac{1}{f_{e}'{}^{2 - \alpha_{\rm LV}}} \right) \right]\,,
\end{equation}
where we have defined $\lambda_{A} \equiv h_{p} A^{1/(\alpha_{\rm LV}-2)}$, with $h_{p}$ Planck's constant, and the generalized distance measure~\cite{Mirshekari:2011yq}
\begin{equation}
D_{\alpha_{\rm LV}} = \frac{(1 + z)^{1-\alpha_{\rm LV}}}{H_{0}} \int_{0}^{z} \frac{(1 + z')^{\alpha_{\rm LV}-2}}{\left[\Omega_{M} (1 + z')^{3} + \Omega_{\Lambda}\right]^{1/2}} dz'\,.  
\end{equation}
Such a modification then leads to the following correction to the Fourier transform of the response function~\cite{Mirshekari:2011yq}
\begin{equation}
\tilde{h}_{\rm LV} = \tilde{h}_{\rm GR} e^{i \beta_{\rm LV} u^{b_{\rm LV}}}\,,
\label{eq:response-LV}
\end{equation}
where $\tilde{h}_{\rm GR}$ is the Fourier transform of the response function in General Relativity and we have defined~\cite{Mirshekari:2011yq}
\begin{equation}
\beta_{\rm LV}^{\alpha_{\rm LV}\neq1} = - \frac{\pi^{2-\alpha_{\rm LV}}}{1-\alpha_{\rm LV}} \frac{D_{\alpha_{\rm LV}}}{\lambda_{A}^{2-\alpha_{\rm LV}}} \frac{{\cal{M}}_{c}^{1-\alpha_{\rm LV}}}{(1 + z)^{1-\alpha_{\rm LV}}}\,,
\qquad
b_{\rm LV}^{\alpha_{\rm LV}\neq1} = 3(\alpha_{\rm LV}-1)\,.
\end{equation}
The case $\alpha_{\rm LV}=1$ is special leading to the Fourier phase correction~\cite{Mirshekari:2011yq}
\begin{equation}
\delta \Psi_{\alpha_{\rm LV}=1} = \frac{3\pi D_{1}}{\lambda_{A}} \ln{u}\,.
\end{equation}
The reason for this is that when $\alpha_{\rm LV}=1$ the Fourier phase is proportional to the integral of $1/f$, which then leads to a natural logarithm. 

Different $\alpha_{\rm LV}$ limits deserve further discussion here. When $\alpha_{\rm LV} = 0$, one of course recovers the standard massive graviton result with the mapping $\lambda_{g}^{-2} \to \lambda_{g}^{-2} + \lambda_{A}^{-2}$. When $\alpha_{\rm LV} = 2$, the dispersion relation is identical to that in Eq.~\eqref{eq:vg-standard}, but with a redefinition of the speed of light, and should thus be unobservable. Indeed, in this limit the correction to the Fourier phase in Eq.~\eqref{eq:response-LV} becomes linear in frequency, and this is $100\%$ degenerate with the time of coalescence parameter in the standard General Relativity Fourier phase. Finally, relative to the standard General Relativity terms that arise in the post-Newtonian expansion of the Fourier phase, the new corrections are of $(1 + 3\alpha_{\rm LV}/2)$ post-Newtonian order. Then, if LIGO gravitational wave observations were incapable of discerning between a 4 post-Newtonian and a 5 post-Newtonian waveform, then such observations would not be able to see the modified dispersion effect if $\alpha_{\rm LV} > 2$. Mirshekari et al.~\cite{Mirshekari:2011yq} confirmed this expectation with a Fisher analysis of non-spinning, comparable mass quasi-circular inspirals. They found that for $\alpha_{\rm LV} = 3$, one can place very weak bounds on $\lambda_{A}$, namely $A <10^{-7} \; {\rm{eV}}^{-1}$ with a LIGO observation of a $(1.4,1.4)M_{\odot}$ neutron star inspiral, $A <  0.2 \; {\rm{eV}}^{-1}$ with an enhanced-LISA or NGO observation of a $(10^{5},10^{5}) M_{\odot}$ black hole inspiral, assuming a signal-to-noise ratio of $10$ and $100$ respectively. A word of caution is due here, though, as these analyses neglect any Lorentz-violating correction to the generation of gravitational waves, including the excitation of additional polarization modes. One would expect that the inclusion of such effects would only strengthen the bounds one could place on Lorentz-violating theories, but this must be done on a theory by theory basis.

%------------------------------------------------------------------------------------------------------------------------------
\subsubsection{Variable \textit{G} theories and large extra-dimensions}
\label{sec:generic-tests-G-ED}

The lack of a particular Lagrangian associated with variable $G$ theories, excluding scalar-tensor theories, or extra dimensions makes it difficult to ascertain whether black hole or neutron star binaries exist in such theories. Whether this is so will depend on the particular variable $G$ model considered. In spite of this, if such binaries do exist, the gravitational waves emitted by such systems will carry some generic modifications relative to the General Relativity expectation. 

Most current tests of the variability of Newton's gravitational constant rely on electromagnetic observations of massive bodies, such as neutron stars. As discussed in Section~\ref{sec:Variable-G}, scalar-tensor theories can be interpreted as variable-$G$ theories, where the variability of $G$ is really a variation in the coupling between gravity and matter. Newton's constant, however, serves the more fundamental role of defining the relationship between geometry or length and energy, and such a relationship is not altered in most scalar-tensor theories, unless the scalar fields are allowed to vary on a cosmological scale (background, homogeneous scalar solution). 

For this reason, one might wish to consider a possible temporal variation of Newton's constant in pure vacuum spacetimes, such as in binary black hole inspirals. Such temporal variation would encode $\dot{G}/G$ at the time and location of the merger event. Thus, once a sufficiently large number of gravitational wave events has been observed and found consistent with General Relativity, one could reconstruct a constraint map that bounds $\dot{G}/G$ along our past light-cone (as a function of redshift and sky position). Since our past-light cone with gravitational waves would have extended to roughly redshift $10$ with classic LISA (limited by the existence of merger events at such high redshifts), such a constraint map would have been much more complete than what one can achieve with current tests at redshift almost zero. Big Bang nucleosynthesis constraints also allow us to bound a linear drift in $\dot{G}/G$  from $z \gg 10^{3}$ to zero, but these become degenerate with limits on the number of relativistic species. Moreover, these bounds exploit the huge lever-arm provided by integrating over cosmic time, but they are insensitive to local, oscillatory variations of $G$ with periods much less than the cosmic observation time. Thus, gravitational-wave constraint maps would test one of the pillars of General Relativity: local position invariance. This principle (encoded in the equivalence principle) states that the laws of physics (and thus the fundamental constants of nature) are the same everywhere in the Universe. 

Let us then promote $G$ to a function of time of the form~\cite{Yunes:2009bv}
\begin{equation}
\label{eq:Gdot-parametrization}
G(t,x,y,z) \approx G_{\rm c} + \dot{G}_{\rm c} \left(t_{c} - t\right)\,, 
\end{equation}
where $G_{\rm c} = G(t_{c},x_{c},y_{c},z_{c})$ and $\dot{G}_{\rm c} = (\partial G/\partial t)(t_{c},x_{c},y_{c},z_{c})$ are constants, and the sub-index $c$ means that these quantities are evaluated at coalescence. Clearly, this is a Taylor expansion to first order in time and position about the coalescence event $(t_{c},x^{i}_{c})$, which is valid provided the spatial variation of $G$ is much smaller than its temporal variation, ie. $|\nabla^{i} G| \ll \dot{G}$, and the characteristic period of the temporal variation is longer than the observation window (at most, $T_{\rm obs} \leq 3$ years for classic LISA), so that $\dot{G}_{\rm c} T_{\rm obs} \ll G_{\rm c}$. Similar parameterization of $G(t)$ have been used to study deviations from Newton's second law in the Solar System~\cite{Dirac:1937ti,1988NuPhB.302..645W,1989RvMP...61....1W,2003RvMP...75..403U}. One can thus think of this modification as the consequence of some effective theory that could represent the predictions of several different alternative theories.  

The promotion of Newton's constant to a function of time changes the rate of change of the orbital frequency, which then directly impacts the gravitational wave phase evolution. To leading order, Yunes, et al.~\cite{Yunes:2009bv} find
\begin{equation}
\dot{F} = \dot{F}_{\rm GR} + \frac{195}{256\pi} {\cal{M}}_{c}^{-2} x^{3} \eta^{3/5} (\dot{G}_{c} {\cal{M}}_{c})\,,
\end{equation}
where $\dot{F}_{\rm GR}$ is the rate of change of the orbital frequency in General Relativity, due to the emission of gravitational waves and $x = (2 \pi M F)^{1/3}$. Such a modification to the orbital frequency evolution leads to the following modification~\cite{Yunes:2009bv} to the Fourier transform of the response function in the stationary-phase approximation~\cite{Bender,Cutler:1994ys,Droz:1999qx,Yunes:2009yz}
\begin{equation}
\tilde{h} = \tilde{h}_{\rm GR} \left(1 + \alpha_{\dot{G}} u^{a_{\dot{G}}} \right) e^{i \beta_{\dot{G}} u^{b_{\dot{G}}}}\,,
\label{eq:h-Gdot}
\end{equation}
where we recall again that $u = (\pi {\cal{M}}_{c} f)^{1/3}$ and have defined the constant parameters~\cite{Yunes:2009bv}
\begin{equation}
\alpha_{\dot{G}} = - \frac{5}{512} \frac{\dot{G}_{c}}{G_{c}} \left(G_{c} {\cal{M}}_{z}\right)\,,
\qquad 
\beta_{\dot{G}} = - \frac{25}{65536} \frac{\dot{G}_{c}}{G_{c}} \left(G_{c} {\cal{M}}_{z}\right)\,,
\qquad 
a = -8\,,
\qquad
b = - 13\,,
\end{equation}
to leading-order in the post-Newtonian approximation. We note that this corresponds to a correction of $-4$ post-Newtonian order in the phase, relative to the leading-order term, and that the corrections are independent of the symmetric mass ratio, scaling only with the redshifted chirp mass ${\cal{M}}_{z}$. Due to this, one expects the strongest effects to be seen in low-frequency gravitational waves, such as those one could detect with LISA or DECIGO/BBO. 

Given such corrections to the gravitational wave response function, one can investigate the level to which a gravitational wave observation consistent with General Relativity would allow us to constrain $\dot{G}_{c}$. Yunes, et al.~\cite{Yunes:2009bv} carried out such a study and found that for comparable-mass black hole inspirals of total redshifted mass $m_{z} = 10^{6} M_{\odot}$ with LISA, one could constrain $\dot{G}_{c}/G_{c} \lesssim 10^{-9} \; {\rm{yr}}^{-1}$ or better to redshift $10$ (assuming a signal-to-noise ratio of $10^{3})$. Similar constraints are possible with observations of extreme mass-ratio inspirals.  The constraint is strengthened when one considers intermediate mass black hole inspirals, where one would be able to achieve a bound of $\dot{G}_{c}/G_{c} \lesssim 10^{-11} \; {\rm{yr}}^{-1}$. Although this is not as stringent as the strongest constraints from other observations (see Section~\ref{sec:Variable-G}), we recall that gravitational wave constraints would measure local variations at the source, as opposed to local variations at zero redshift or integrated variations from the very early Universe.

The effect of promoting Newton's constant to a function of time is degenerate with several different effects. One such effect is a temporal variability of the black hole masses, ie.~if $\dot{m} \neq 0$. Such time-variation could be induced by gravitational leakage into the bulk in certain brane-world scenarios~\cite{Johannsen:2008tm}, as we explained in Section~\ref{sec:Variable-G}. For a black hole of mass $M$, the rate of black hole evaporation is given by
\begin{equation}
\frac{dM}{dt} = -2.8 \times 10^{-7} \left(\frac{1 M_{\odot}}{M}\right)^{2} \left(\frac{\ell}{10 \; \mu{\rm{m}}}\right)^{2} M_{\odot} {\rm{yr}}^{-1}\,,
\end{equation}
where $\ell$ is the size of the large extra dimension. As expected, such a modification to a binary black hole inspiral will lead to a correction to the Fourier transform of the response function that is identical in structure to that of Eq.~\eqref{eq:h-Gdot}, but the parameters $(\beta_{\dot{G}},b_{\dot{G}}) \to (\beta_{\rm ED},b_{\rm ED})$ change to~\cite{Yagi:2011yu}
\begin{equation}
\beta_{\rm ED} = -8.378 \times 10^{-8} \left(\frac{\ell}{{\cal{M}}_{c}}\right)^{4} \left(1 - \frac{26}{3} \eta + 34 \eta^{2}\right)\,,
\qquad
b_{\rm ED} = -13 \,.
\end{equation}
A similar expression is found for a neutron star/black-hole inspiral, except that the $\eta$-dependent factor in between parenthesis is corrected. 

Given a gravitational wave detection consistent with General Relativity, one could then in principle place an upper bound on $\ell$. Yagi et al.~\cite{Yagi:2011yu} carried out a Fisher analysis and found that a $1$-year LISA detection would constrain $\ell \leq 10^{3} \; \mu{\rm{m}}$ with a $(10,10^{5})M_{\odot}$ binary inspiral at signal-to-noise ratio of $100$. This constraint is roughly two orders of magnitude weaker than current table-top experiment constraints~\cite{Adelberger:2006dh}. Moreover, the constraint weakens somewhat for more generic inspirals, due to degeneracies between $\ell$ and eccentricity and spin. A similar observation with the third generation detector DECIGO/BBO, however, should be able to beat current constraints by roughly one order of magnitude. Such a constraint could be strengthened by roughly one order of magnitude further, if one included the statistical enhancement in parameter estimation due to detection of order $10^{5}$ sources by DECIGO/BBO.

Another way to place a constraint on $\ell$ is to consider the effect of mass loss in the orbital dynamics~\cite{McWilliams:2009ym}. When a system losses mass, the evolution of its semi-major axis $a$ will acquire a correction of the form $\dot{a} = -(\dot{M}/M) a$, due to conservation of specific orbital angular momentum. There is then a critical semi-major axis $a_{c}$ at which this correction balances the semi-major decay rate due to gravitational wave emission. McWilliams~\cite{McWilliams:2009ym} argues that systems with $a < a_{c}$ are then gravitational-wave dominated and will thus inspiral, while systems with $a > a_{c}$ will be mass-loss dominated and will thus outspiral. If a gravitational wave arising from an inspiraling binary is detected at a given semi-major axis, then $\ell$ is automatically constrained to about ${\cal{O}}(20 \; \mu{\rm{m}})$. Yagi, et al.~\cite{Yagi:2011yu} extended this analysis to find that such a constraint is weaker than what one could achieve via matched filtering with a waveform of the form in Eq.~\eqref{eq:h-Gdot}, using the DECIGO detector.   

The $\dot{G}$ correction to the gravitational wave phase evolution is also degenerate with cosmological acceleration. That is, if a gravitational wave is generated at high-redshift, its phase will be affected by the acceleration of the Universe. To zeroth-order, the correction is a simple redshift of all physical scales. However, if one allows the redshift to be a function of time 
\begin{equation}
z \sim z_{c} + \dot{z}_{c} (t - t_{c}) \sim z_{c} + H_{0} \left[ \left(1 + z_{c}\right)^{2} - \left(1 + z_{c}\right)^{5/2} \Omega_{M}^{1/2} \right] (t - t_{c})\,,
\end{equation}
then the observed waveform at the detector becomes structurally identical to Eq.~\eqref{eq:h-Gdot} but with the parameters
\begin{equation}
\beta_{\dot{z}} = \frac{25}{32768} \dot{z}_{c} {\cal{M}}_{z}\,,
\qquad
b_{\dot{z}} = -13\,.
\end{equation}
Using the measured values of the cosmological parameters from the WMAP analysis~\cite{Komatsu:2008hk,Dunkley:2008ie}, however, one finds that this effect is roughly $10^{-3}$ times smaller than that of a possible $\dot{G}$ correction at the level of the possible bounds quoted above~\cite{Yunes:2009bv}. Of course, if one could in the future constrain $\dot{G}$ better by 3 orders of magnitude, possible degeneracies with $\dot{z}$ would become an issue.

A final possible degeneracy arises with the effect of a third body~\cite{Yunes:2010sm}, accretion disk migration~\cite{Kocsis:2011dr,Yunes:2011ws} and the interaction of a binary with a circumbinary accretion disk~\cite{Hayasaki:2012qn}. All of these effects introduce corrections to the gravitational wave phase of negative PN order, just like the effect of a variable gravitational constant. Degeneracies of this type, however, are only expected to affect a small subset of binary black hole observations, namely those with a third body sufficiently close to the binary, or a sufficiently massive accretion disk.  

%------------------------------------------------------------------------------------------------------------------------------
\subsubsection{Parity violation}
\label{sec:generic-tests-PV}

As discussed in Section~\ref{sec:GPV} the simplest action to model parity violation in the gravitational interaction is given in Eq.~\eqref{CS-action}. Black holes and neutron stars exist in this theory, albeit non-rotating. A generic feature of this theory is that parity violation imprints onto the propagation of gravitational waves, an effect that has been dubbed {\emph{amplitude birefringence}}. Such birefringence is not to be confused with optical or electromagnetic birefringence, in which the gauge boson interacts with a medium and is doubly-refracted into two separate rays. In amplitude birefringence, right- (left)-circularly polarized gravitational waves are enhanced or suppressed (suppressed or enhanced) relative to the General Relativity expectation as they propagate~\cite{Jackiw:2003pm,Lue:1998mq,Alexander:2007kv,Yunes:2008bu,Alexander:2009tp,Yunes:2010yf}.

One can understand amplitude birefringence in gravitational wave propagation due to a possible non-commutativity of the parity operator and the Hamiltonian. The Hamiltonian is the generator of time evolution, and thus, one can write~\cite{Yunes:2010yf}
\begin{equation}
\label{eq:matrix}
\left( \begin{array}{c} {h_{+,k}(t)} \\ h_{\times,k}(t) \end{array} \right)
= e^{-i f t}
\left( \begin{array}{rr} u_{c} & iv \\ -iv & u_{c} \end{array} \right)
\left( \begin{array}{c} h_{+,k}(0) \\ h_{\times,k}(0) \end{array} \right)\,,
\end{equation}
where $f$ is the gravitational wave angular frequency, $t$ is time, and $h_{+,\times,k}$ are the gravitational wave Fourier components with wavenumber $k$.  The quantity $u_{c}$ models possible background curvature effects, with $u_{c} = 1$ for propagation on a Minkowski metric, and $v$ proportional to redshift for propagation on a Friedman-Robertson-Walker metric~\cite{Laguna:2009re}. The quantity $v$ models possible parity-violating effects, with $v = 0$ in General Relativity. One can rewrite the above equation in terms of right- and left-circular polarizations, $h_{\rm{R,L}} = (h_+ \pm ih_\times) /\sqrt{2}$ to find
\begin{equation}
\label{eq:matrix2}
\left( \begin{array}{c} h_{\rm{R},k}(t) \\ h_{\rm{L},k}(t) \end{array} \right)
= e^{-i f t}
\left( \begin{array}{rr} u_{c} + v & 0 \\ 0 & u_{c}-v \end{array} \right)
\left( \begin{array}{c} h_{\rm{R},k}(0) \\ h_{\rm{L},k}(0) \end{array} \right).
\end{equation}
Amplitude birefringence has the effect of modifying the eigenvalues of the diagonal propagator matrix for right and left-polarized waves, with right modes amplified or suppressed and left modes suppressed or enhanced relative to General Relativity, depending on the sign of $v$. In addition to these parity-violating propagation effects, parity violation should also leave an imprint in the generation of gravitational waves. Such effects, however, need to be analyzed on a theory by theory basis. Moreover, the propagation distance-independent nature of generation effects should make them easily distinguishable from the propagation effects we consider here. 

The degree of parity violation, $v$, can be expressed entirely in terms of the waveform observables via~\cite{Yunes:2010yf}
\begin{equation}
v = \frac{1}{2} \left(\frac{h_{\rm R}}{h_{\rm R}^{\rm GR}} - \frac{h_{\rm L}}{h_{\rm L}^{\rm GR}} \right) = \frac{i}{2} \left( \delta \phi_{\rm L} - \delta \phi_{\rm R}\right)\,,
\end{equation}
where $h_{\rm R,L}^{\rm GR}$ is the General Relativity expectation for a right- or left-polarized gravitational wave. In the last equality we have also introduced the notation $\delta \phi \equiv \phi - \phi^{\rm GR}$, where $\phi^{\rm GR}$ is the General Relativity gravitational wave phase and 
\begin{equation}
h_{\rm R,L} = h_{0,\rm R,L} e^{-i \left[\phi(\eta) - \kappa_{i} \chi^{i}\right]}\,,
\end{equation}
where $h_{0, \rm R,L}$ is a constant factor, $\kappa$ is the conformal wave number and $(\eta,\chi^{i})$ are conformal coordinates for propagation in a Friedmann-Robertson-Walker Universe. The precise form of $v$ will depend on the particular theory under consideration. For example, in non-dynamical Chern-Simons gravity with a field $\vartheta = \vartheta(t)$, and in an expansion about $z \ll 1$, one finds~\cite{Yunes:2010yf}
\begin{equation}
v =  \frac{\alpha}{\kappa} \pi f z \left(\dot{\vartheta}_{0} - \frac{\ddot{\vartheta}_{0}}{H_{0}}\right) 
=  \frac{\alpha}{\kappa}  \pi f D \left(H_{0}\dot{\vartheta}_{0} - \ddot{\vartheta}_{0} \right)\,,
\label{eq:v-def}
\end{equation}
where $\vartheta_{0}$ is the Chern-Simons scalar field at the detector, with $\alpha$ the Chern-Simons coupling constant [see e.g.~Eq.~\eqref{CS-action}], $z$ is redshift, $D$ is the comoving distance and $H_{0}$ is the value of the Hubble parameter today and $f$ is the observed gravitational wave frequency. When considering propagation on a Minkowski background, one obtains the above equation in the limit as $\dot{a} \to 0$, so the second term dominates, where $a$ is the scale-factor. To leading-order in a curvature expansion, the parity-violating coefficient $v$ will always be linear in frequency, as shown in Eq.~\eqref{eq:v-def}. For more general parity violation and flat-spacetime propagation, $v$ will be proportional to $(f D) f^{a} \alpha$, where $\alpha$ is a coupling constant of the theory (or a certain derivative of a coupling field) with units of $[{\rm{Length}}]^{a}$ (in the previous case, $a =0$, so the correction was simply proportional to $f D \alpha$, where $\alpha \propto \ddot{\vartheta}$).

How does such parity violation affect the waveform? By using Eq.~\eqref{eq:matrix2} one can easily show that the Fourier transform of the response function becomes~\cite{Alexander:2007kv,Yunes:2008bu,Yunes:2010yf}
\begin{equation}
\tilde{h}_{\rm PV} = \left(F_{+} - i \; v \;F_{\times} \right) \tilde{h}_{+} + \left(F_{\times} + i \; v \; F_{+} \right) \tilde{h}_{\times}\,.
\end{equation}
One can of course rewrite this in terms of a real amplitude correction and a real phase correction. Expanding in $v \ll 1$ to leading order, we find~\cite{Yunes:2010yf}
\begin{equation}
\tilde{h}_{\rm PV} = \tilde{h}^{\rm GR} \left(1 + v \; \delta Q_{\rm PV} \right) e^{i v^{2} \delta \psi_{\rm PV}}\,,
\label{eq:hpar-PV}
\end{equation}
where $\tilde{h}_{\rm GR}$ is the Fourier transform of the response function in General Relativity and we have defined
\begin{align}
Q_{\rm GR} &= \sqrt{F_{+}^{2} \left(1 + \cos^{2}{\iota}\right)^{2} + 4 \cos^{2}{\iota} F_{\times}^{2}}\,,
\\
\delta Q_{\rm PV} &= \frac{2 \left(1 + \cos^{2}{\iota}\right) \cos{\iota} \left(F_{+}^{2} + F_{\times}^{2}\right)}{Q_{\rm GR}^{2}}\,,
\\
\delta \psi_{\rm PV} &= \frac{2 \cos{\iota} \left(1 + \cos^{2}{\iota}\right) \left(1 - \cos^{2}{\iota}\right)^{2} \left(F_{+}^{2} + F_{\times}^{2}\right) F_{+} F_{\times}}{Q_{\rm GR}^{4}}\,.
\end{align}
We see then that amplitude birefringence modifies both the amplitude and the phase of the response function. Using the non-dynamical Chern-Simons expression for $v$ in Eq.~\eqref{eq:v-def}, we can rewrite Eq.~\eqref{eq:hpar-PV} as~\cite{Yunes:2010yf}
\begin{equation}
\tilde{h}_{\rm PV} =  \tilde{h}^{\rm GR} \left(1 + \alpha_{\rm PV} u^{a_{\rm PV}}\right) e^{i \beta_{\rm PV} u^{b_{\rm PV}}}\,,
\end{equation}
where we have defined the coefficients
\begin{align}
\label{eq:alpha-PV}
\alpha_{\rm PV} &= 
\left(\frac{D}{\cal{M}}_{c}\right) 
\left[\frac{2 \left(1 + \cos^{2}{\iota}\right) \cos{\iota} \left(F_{+}^{2} + F_{\times}^{2}\right)}{Q_{\rm GR}^{2}}\right]
 \frac{\alpha}{\kappa}  \left(H_{0}\dot{\vartheta}_{0} - \ddot{\vartheta}_{0} \right) 
\,,
\qquad
a_{\rm PV} = 3\,,
\\
\label{eq:beta-PV}
\beta_{\rm PV} &=  
\left(\frac{D}{\cal{M}}_{c}\right)^{2} 
\left[ \frac{2 \cos{\iota} \left(1 + \cos^{2}{\iota}\right) \left(1 - \cos^{2}{\iota}\right)^{2} \left(F_{+}^{2} + F_{\times}^{2}\right) F_{+} F_{\times}}{Q_{\rm GR}^{4}}\right]
\frac{\alpha}{\kappa} \left(H_{0}\dot{\vartheta}_{0} - \ddot{\vartheta}_{0} \right)^{2}
\,,
\qquad
b_{\rm PV} = 6\,,
\end{align}
where we recall that $u = (\pi {\cal{M}}_{c} f)^{1/3}$. The phase correction corresponds to a term of 5.5 post-Newtonian order relative to the Newtonian contribution, and, it scales quadratically with the Chern--Simons coupling field $\vartheta$, which is why it was left out in~\cite{Yunes:2010yf}.  The amplitude correction, on the other hand, is of $1.5$ post-Newtonian order relative to the Newtonian contribution. Since both of these appear as positive-order, post-Newtonian corrections, there is a possibility of degeneracy between them and standard waveform template parameters.  

Given such a modification to the response function, one can ask whether such parity violation is observable with current detectors. Alexander et al.~\cite{Alexander:2007kv,Yunes:2008bu} argued that a gravitational wave observation with LISA would be able to constrain an integrated measure of $v$, because LISA can observe massive black hole mergers to cosmological distances, while amplitude birefringence accumulates with distance traveled. For such an analysis, one cannot Taylor expand $\vartheta$ about its present value, and instead, one finds that
\begin{equation}
\frac{1 + v}{1 - v} = e^{2 \pi f \zeta(z)}\,,
\end{equation}
where we have defined
\begin{align}
\zeta(z) &= \frac{\alpha H_{0}}{\kappa} \int_{0}^{z} dz \left(1 + z\right)^{5/2} \left[\frac{7}{2} \frac{d\vartheta}{dz} + \left(1 + z\right) \frac{d^{2}\vartheta}{dz^{2}} \right]\,.
\end{align}
We can solve the above equation to find
\begin{equation}
v = \frac{e^{2 \pi f \zeta(z)}  -1}{1 + e^{2 \pi f \zeta(z)}} \sim  \pi f \zeta(z)\,,
\end{equation}
where in the second equality we have linearized about $v\ll1$ and $f \zeta \ll1$. Alexander et al.~\cite{Alexander:2007kv,Yunes:2008bu} realized that this induces a time-dependent change in the inclination angle (ie.~the apparent orientation of the binary's orbital angular momentum with respect to the observer's line-of-sight), since the latter can be defined by the ratio $h_{\rm R}/h_{\rm L}$. They then carried out a simplified Fisher analysis and found that a LISA observation of the inspiral of two massive black holes with component masses $10^{6} M_{\odot} (1 + z)^{-1}$ at redshift $z = 15$  would allow us to constrain the integrated dimensionless measure $\zeta < 10^{-19}$ to $1\sigma$. One might worry that such an effect would be degenerate with other standard General Relativity processes that induce similar time-dependencies, such as spin-orbit coupling. However, this time-dependence is very different from that of the parity-violating effect, and thus, Alexander et al.~\cite{Alexander:2007kv,Yunes:2008bu} argued that these effects would be weakly correlated. 

Another test of parity violation was proposed by Yunes et al.~\cite{Yunes:2010yf}, who considered the coincident detection of a gravitational wave and a gamma-ray burst with the SWIFT~\cite{Gehrels:2004am} and GLAST/Fermi~\cite{Carson:2006af} gamma-ray satellites, and the ground-based LIGO~\cite{Abbott:2007kv} and Virgo~\cite{Acernese:2007zze} gravitational wave detectors. If the progenitor of the gamma-ray burst is a neutron-star/neutron-star merger, the gamma-ray jet is expected to be highly collimated. Therefore, an electromagnetic observation of such an event implies that the binary's orbital angular momentum at merger must be pointing along the line of sight to Earth, leading to a strongly circularly-polarized gravitational wave signal and to maximal parity violation. If the gamma-ray burst observation were to provide an accurate sky location, one would be able to obtain an accurate distance measurement from the gravitational wave signal alone. Moreover, since GLAST/Fermi observations of gamma-ray bursts occur at low redshift, one would also possess a purely electromagnetic measurement of the distance to the source. Amplitude birefringence would manifest itself as a discrepancy between these two distance measurements. Therefore, if no discrepancy is found, the error ellipse on the distance measurement would allow us to place an upper limit on any possible gravitational parity violation. Because of the nature of such a test, one is here constraining generic parity violation over distances of hundreds of Mpc, along the light-cone on which the gravitational waves propagate. 

The coincident gamma-ray burst/gravitational wave test compares favorably to the pure LISA test, with the sensitivity to parity violation being about 2\,--\,3 orders of magnitude better in the former case. This is because, although the fractional error in the gravitational wave distance measurement is much smaller for LISA than for LIGO, since it is inversely proportional to the signal-to-noise ratio, the parity violating effect also depends on the gravitational wave frequency, which is much larger for neutron-star inspirals than massive black hole coalescences. Mathematically, the simplest models of gravitational parity violation will lead to a signature in the response function that is proportional to the gravitational wave wavelength\epubtkFootnote{Even if it is not linear, the effect should scale with positive powers of $\lambda_{\rm GW}$. It is difficult to think of any parity violating theory that would lead to an inversely proportional relation.} $\lambda_{\rm GW} \propto D f$. Although the coincident test requires small distances and low signal-to-noise ratios (by roughly 1\,--\,2 orders of magnitude), the frequency is also larger by a factor of 5\,--\,6 orders of magnitude for the LIGO-Virgo network. 

The coincident gamma-ray burst/gravitational wave test also compares favorably to current Solar System constraints. Using the motion of the LAGEOS satellites, Smith et al.~\cite{Smith:2007jm} have placed the $1\sigma$ bound $\dot{\vartheta}_{0} < 2000 \; {\rm{km}}$ assuming $\ddot{\vartheta}_{0} = 0$. A similar assumption leads to a $2\sigma$ bound of $\dot{\vartheta}_{0} < 200 \; {\rm{km}}$ with a coincident gamma-ray burst/gravitational wave observation. Moreover, the latter test also allows us to constrain the second time-derivative of the scalar field. Finally, a LISA observation would constrain the integrated history of $\vartheta$ along the past light-cone on which the gravitational wave propagated.  These tests, however, are not as stringent as the recently proposed test by Dyda et al.~\cite{Dyda:2012rj}, $\dot{\vartheta}_{0} < 10^{-7} \; {\rm{km}}$, assuming the effective theory cut-off scale is less than $10 \; {\rm{eV}}$ and obtained by demanding that the energy density in photons created by vacuum decay over the lifetime of the Universe not violate observational bounds. 

The coincident test is somewhat idealistic in that there are certain astrophysical uncertainties that could hamper the degree to which we could constrain parity violation. One of the most important uncertainties relates to our knowledge of the inclination angle, as gamma-ray burst jets are not necessarily perfectly aligned with the line of sight. If the inclination angle is not known {\emph{a priori}}, it will become degenerate with the distance in the waveform template, decreasing the accuracy to which the luminosity could be extracted from a pure gravitational wave observation by at least a factor of two. Even after taking such uncertainties into account, Yunes et al.~\cite{Yunes:2010yf} found that $\dot{\vartheta}_{0}$ could be constrained much better with gravitational waves than with current Solar System observations.

%------------------------------------------------------------------------------------------------------------------------------
\subsubsection{Parameterized post-Einsteinian framework}
\label{subsubsection:ppE}

One of the biggest disadvantages of a top-down or direct approach toward testing General Relativity is that one must pick a particular theory from the beginning of the analysis. However, given the large number of possible modifications to Einstein's theory and the lack of a particularly compelling alternative, it is entirely possible that none of these will represent the correct gravitational theory in the strong field. Thus, if one carries out a top-down approach, one will be forced to make the assumption that we, as theorists, know which modifications of gravity are possible and which are not~\cite{Yunes:2009ke}. The parameterized post-Einsteinian (ppE) approach is a framework developed specifically to alleviate such a bias by allowing the data to select the correct theory of Nature through the systematic study of statistically significant anomalies. 

For detection purposes, one usually expects to use match filters that are consistent with General Relativity. But if General Relativity happened to be wrong in the strong field, it is possible that a General Relativity template will still extract the signal, but with the wrong parameters. That is, the best fit parameters obtained from a matched filtering analysis with General Relativity templates will be biased by the assumption that General Relativity is sufficiently accurate to model the entire coalescence. This {\emph{fundamental bias}} could lead to a highly distorted image of the gravitational wave Universe. In fact, recent work by Vallisneri and Yunes~\cite{Vallisneri:2013rc} indicates that such fundamental bias could indeed be present in observations of neutron star inspirals, if General Relativity is not quite the right theory in the strong-field. 

One of the primary motivations for the development of the ppE scheme was to alleviate fundamental bias, and one of its most dangerous incarnations: {\emph{stealth-bias}}~\cite{Cornish:2011ys}. If General Relativity is not the right theory of nature, yet all our future detections are of low signal-to-noise ratio, we may estimate the wrong parameters from a matched-filtering analysis, yet without being able to identify that there is a non-General Relativity anomaly in the data. Thus, stealth bias is nothing but fundamental bias hidden by our limited signal-to-noise ratio observations. Vallisneri and Yunes~\cite{Vallisneri:2013rc} have found that such stealth-bias is indeed possible in a certain sector of parameter space, inducing errors in parameter estimation that could be larger than statistical ones, without us being able to identify the presence of a non-General Relativity anomaly. 

%------------------------------------
%% \vspace{0.3cm}
%% \hspace{-0.7cm} {\bf{{5.3.4.1 Historical Development}}}
%% \vspace{0.2cm}

\subsubsection*{Historical development}

The ppE scheme was designed in close analogy with the parameterized post-Newtonian (ppN) framework, developed in the 1970s to test General Relativity with Solar System observations (see e.g.~\cite{lrr-2006-3} for a review). In the Solar System, all direct observables depend on a single quantity, the metric, which can be obtained by a small-velocity/weak-field post-Newtonian expansion of the field equations of whatever theory one is considering. Thus, Will and Nordtvedt~\cite{Nordtvedt:1968qs,1971ApJ...163..611W,1972ApJ...177..757W,1972ApJ...177..775N,1973ApJ...185...31W} proposed the generalization of the Solar System metric into a {\emph{meta-metric}} that could effectively interpolate between the predictions of many different alternative theories. This meta-metric depends on the product of certain Green function potentials and ppN parameters. For example, the spatial-spatial components of the meta-metric take the form
\begin{equation}
g_{ij} = \delta_{ij} \left(1 + 2 \gamma U + \dots\right)
\end{equation}
where $\delta_{ij}$ is the Kronecker delta, $U$ is the Newtonian potential and $\gamma$ is one of the ppN parameters, which acquires different values in different theories: $\gamma = 1$ in General Relativity, $\gamma = (1 + \omega_{\rm BD}) (2 + \omega_{\rm BD})^{-1} \sim 1 - \omega_{\rm BD}^{-1}$ in Jordan--Fierz--Brans--Dicke theory, etc. Therefore, any Solar System observable could then be written in terms of system parameters, such as the masses of the planets, and the ppN parameters. An observation consistent with General Relativity allows for a bound on these parameters, thus simultaneously constraining a large class of modified gravity theories. 

The idea behind the ppE framework was to develop a formalism that allowed for similar generic tests but with gravitational waves instead of Solar System observations. The first such attempt was by Arun, et al.~\cite{Arun:2006yw,Mishra:2010tp} who considered the quasi-circular inspiral of compact objects. They suggested the waveform template family 
\begin{equation}
\tilde{h}_{\rm{PNT}} = \tilde{h}^{\rm GR} e^{i \beta_{\rm PNT} u^{b_{\rm PN}}}\,.
\label{eq:rppE}
\end{equation}
This waveform depends on the standard system parameters that are always present in General Relativity waveforms, plus one theory parameter $\beta_{\rm PNT}$ that is to be constrained. The quantity $b_{\rm PN}$ is a number chosen by the data analyst and is restricted to be equal to one of the post-Newtonian predictions for the phase frequency exponents, ie.~$b_{\rm PN}=(-5,-3,-2,-1,\ldots)$.

The template family in Eq.~\eqref{eq:rppE} allows for the so-called {\emph{post-Newtonian tests of General Relativity}}, ie.~consistency checks of the signal with the post-Newtonian expansion. For example, let us imagine that a gravitational wave has been detected with sufficient signal-to-noise ratio that the chirp mass and mass ratio have been measured from the Newtonian and 1 post-Newtonian terms in the waveform phase. One can then ask whether the 1.5 post-Newtonian term in the phase is consistent with these values of chirp mass and mass ratio. Put another way, each term in the phase can be thought of as a curve in $({\cal{M}}_{c},\eta)$ space. If General Relativity is correct, all these curves should intersect inside some uncertainty box, just like when one tests General Relativity with binary pulsar data. From that standpoint, these tests can be thought of as null-tests of General Relativity and one can ask: given an event, is the data consistent with the hypothesis $\beta_{\rm rppE} = 0$ for the restricted set of frequency exponents $b_{\rm PN}$?

A Fisher and a Bayesian data analysis study of how well $\beta_{\rm
PNT}$ could be constrained given a certain $b_{\rm PN}$ was carried
out in~\cite{Mishra:2010tp,Huwyler:2011iq,Li:2011cg}. Mishra et
al~\cite{Mishra:2010tp} considered the quasi-circular inspiral of
non-spinning compact objects and showed that Ad.~LIGO observations
would allow one to constrain $\beta_{\rm PNT}$ to 6\% up to the 1.5
post-Newtonian order correction ($b_{\rm PN}=-2$). Third-generation
detectors, such as ET, should allow for better constraints on all
post-Newtonian coefficients to roughly 2\%. Clearly, the higher the
value of $b_{\rm PN}$, the worse the bound on $\beta_{\rm PNT}$
because the power contained in higher frequency exponent terms decreases, i.e., the number of useful additional cycles induced by the $\beta_{\rm PNT} u^{b_{\rm PN}}$ term decreases as $b_{\rm PN}$ increases. Huwyler et al.~\cite{Huwyler:2011iq} repeated this analysis but for LISA observations of the quasi-circular inspiral of black hole binaries with spin precession. They found that the inclusion of precessing spins forces one to introduce more parameters in the waveform, which dilutes information and weakens constraints on $\beta_{\rm PNT}$ by as much as a factor of 5. Li et al.~\cite{Li:2011cg} carried out a Bayesian analysis of the odds-ratio between General Relativity and restricted ppE templates given a non-spinning, quasi-circular compact binary inspiral observation with Ad.~LIGO and Virgo. They calculated the odds ratio for each value of $b_{\rm PN}$ listed above and then combined all of this into a single probability measure that allows one to quantify how likely the data is to be consistent with General Relativity.  

%------------------------------------
%% \vspace{0.3cm}
%% \hspace{-0.7cm} {\bf{{5.3.4.2 The Simplest ppE Model}}}
%% \vspace{0.2cm}

\subsubsection*{The simplest ppE model}

One of the main disadvantages of the post-Newtonian template family in Eq.~\eqref{eq:rppE} is that it is not rooted on a theoretical understanding of modified gravity theories. To alleviate this problem, Yunes and Pretorius~\cite{Yunes:2009ke} re-considered the quasi-circular inspiral of compact objects. They proposed a more general {\emph{ppE template}} family through generic deformations of the $\ell=2$ harmonic of the response function in Fourier space :
\begin{equation}
\tilde{h}_{\rm ppE, insp, 1}^{(\ell=2)} = \tilde{h}^{\rm GR}  \left(1 + \alpha_{\rm ppE} u^{a_{\rm ppE}}\right) e^{i \beta_{\rm ppE} u^{b_{\rm ppE}}}\,, 
\label{eq:fullppE}
\end{equation}
where now $(\alpha_{\rm ppE},a_{\rm ppE},\beta_{\rm ppE},b_{\rm ppE})$ are all free parameters to be fitted by the data, in addition to the usual system parameters. This waveform family reproduces all predictions from known modified gravity theories: when $(\alpha_{\rm ppE},\beta_{\rm ppE}) = (0,0)$, the waveform reduces exactly to General Relativity, while for other parameters one reproduces the modified gravity predictions of Table~\ref{table:ppEpars}.

\begin{table}[htbp]
\caption[Parameters that define the deformation of the response
function in a variety of modified gravity theories.]{Parameters that
define the deformation of the response function in a variety of
modified gravity theories. The notation $\cdot$ means that a value for
this parameter is irrelevant, as its amplitude is zero.}
\label{table:ppEpars} 
\centering
\begin{tabular}{p{2.75cm}p{2cm}p{0.5cm}p{5cm}p{1.5cm}}
\toprule
Theory  & $\alpha_{\rm ppE}$ & $a_{\rm ppE}$ & $\beta_{\rm ppE}$ & $b_{\rm ppE}$ \\ 
\midrule
Jordan--Fierz--Brans--Dicke & $-\frac{5}{96} \frac{S^{2}}{\omega_{\rm BD}} \eta^{2/5}$ & $-2$ & $-\frac{5}{3584} \frac{S^{2}}{\omega_{\rm BD}} \eta^{2/5}$ & $-7$ \\ 
\midrule
Dissipative Einstein-Dilaton-Gauss--Bonnet gravity & 0 & $\cdot$ & $-\frac{5}{7168}  \zeta_{3} \eta^{-18/5} \delta_{m}^{2} $ & $-7$ \\
\midrule
Massive Graviton & $0$ & $\cdot$ & $- \frac{\pi^{2} D {\cal{M}}_{c}}{\lambda_{g}^{2} (1 + z)}$ & $-3$ \\ 
\midrule
Lorentz Violation & $0$ & $\cdot$ & $-\frac{\pi^{2 - \gamma_{\rm LV}}}{(1 - \gamma_{\rm LV})} \frac{D_{\gamma_{\rm LV}}}{\lambda_{\rm LV}^{2 - \gamma_{\rm LV}}} \frac{{\cal{M}}_{c}^{1- \gamma_{\rm LV}}}{(1  + z)^{1- \gamma_{\rm LV}}}$ & $-3 \gamma_{\rm LV} - 3$ \\  
\midrule
$G(t)$ Theory & $-\frac{5}{512} \dot{G} {\cal{M}}_{c}$ & $-8 $ & $-\frac{25}{65536} \dot{G}_{c} {\cal{M}}_{c}$ & $-13$ \\ 
\midrule
Extra Dimensions & $\cdot$ & $\cdot$ & $-\frac{75}{2554344} \frac{dM}{dt} \eta^{-4} (3- 26 \eta + 24 \eta^{2} )$ & $-13$ \\
\midrule
Non-Dynamical Chern--Simons Gravity & $\alpha_{\rm PV}$ & $3$ & $\beta_{\rm PV}$ & $6$ \\ 
\midrule
Dynamical Chern--Simons Gravity & $0$ & $\cdot$ & $\beta_{\rm dCS}$ & $-1$ \\ 
\bottomrule
\end{tabular}
\end{table}

In Table~\ref{table:ppEpars}, recall that $S$ is the difference in the square of the sensitivities and $\omega_{\rm BD}$ is the Brans--Dicke coupling parameter (see Section~\ref{sec:direct-test-BD} and we have here neglected the scalar mode), $\zeta_{3}$ is the coupling parameter in Einstein-Dilaton-Gauss--Bonnet theory (see Section~\ref{sec:direct-test-MQG}), where we have here included both the dissipative and the conservative corrections, $D$ is a certain distance measure and $\lambda_{g}$ is the Compton wavelength of the graviton (see Section~\ref{generic-tests:MG-LV}), $\lambda_{\rm LV}$ is a distance scale at which Lorentz-violation becomes important and $\gamma_{\rm LV}$ is the graviton momentum exponent in the deformation of the dispersion relation (see Section~\ref{generic-tests:MG-LV}), $\dot{G}_{c}$ is the value of the time derivative of Newton's constant at coalescence and  $dM/dt$ is the mass loss due to enhanced Hawking radiation in extra-dimensional scenarios (see Section~\ref{sec:generic-tests-G-ED}), $\beta_{\rm dCS}$ is given in Eq.~\eqref{eq:beta-dCS} and $(\alpha_{\rm PV},\beta_{\rm PV})$ are given in Eqs.~\eqref{eq:alpha-PV} and~\eqref{eq:beta-PV} of Section~\ref{sec:generic-tests-PV}. 

Although there are only a few modified gravity theories where the leading-order post-Newtonian correction to the Fourier transform of the response function can be parameterized by post-Newtonian waveforms of Eq.~\eqref{eq:rppE}, all such predictions can be modeled with the ppE templates of Eq.~\eqref{eq:fullppE}. In fact, only massive graviton theories, certain classes of Lorentz-violating theories and dynamical Chern--Simons gravity lead to waveform corrections that can be parameterized via Eq.~\eqref{eq:rppE}. For example, the lack of amplitude corrections in Eq.~\eqref{eq:rppE} does not allow for tests of gravitational parity violation or non-dynamical Chern--Simons gravity. 

This, however, does not imply that Eq.~\eqref{eq:fullppE} can parameterize all possible deformations of General Relativity. First, Eq.~\eqref{eq:fullppE} can be understood as a single-parameter deformation away from Einstein's theory. If the correct theory of Nature happens to be a deformation of General Relativity with several parameters (eg.~several coupling constants, mass terms, potentials, etc.~), then Eq.~\eqref{eq:fullppE} will only be able to parameterize the one that leads to the most useful-cycles. This was recently verified by Sampson et al.~\cite{Sampson:2013lpa}. Second, Eq.~\eqref{eq:fullppE} assumes that the modification can be represented as a power series in velocity, with possibly non-integer values. Such an assumption does not allow for possible logarithmic terms, which are known to arise due to non-linear memory interactions at sufficiently high-post-Newtonian order. It also does not allow for interactions that are screened, eg.~in theories with massive degrees of freedom. Nonetheless, the parameterization in Eq.~\eqref{eq:fullppE} will still be able to signal that the detection is not a pure Einstein event, at the cost of biasing their true value.  

The inspiral ppE model of Eq.~\eqref{eq:fullppE} is motivated not only from examples of modified gravity predictions, but from generic modifications to the physical quantities that drive the inspiral: the binding energy or Hamiltonian and the radiation-reaction force or the fluxes of the constants of the motion. Yunes and Pretorius~\cite{Yunes:2009ke} and Chatziioannou, et al.~\cite{Chatziioannou:2012rf} considered generic modifications of the form
\begin{align}
E &= \frac{\mu}{2} \frac{m}{r} \left[1 + A_{\rm ppE} \left(\frac{m}{r}\right)^{p_{\rm ppE}}\right]\,,
\qquad
\dot{E} = \dot{E}_{\rm GR} \left[1 + B_{\rm ppE} \left(\frac{m}{r}\right)^{q_{\rm ppE}}\right]\,,
\end{align}
where $(p,q) \in \mathbb{Z}$, since otherwise one would lose analyticity in the limit of zero-velocities for circular inspirals, and where $(A,B)$ are parameters that depend on the modified gravity theory and, in principle, could depend on dimensionless quantities like the symmetric mass ratio. Such modifications lead to the following corrections to the SPA Fourier transform of the $\ell=2$ time-domain response function for a quasi-circular binary inspiral template (to leading order in the deformations and in post-Newtonian theory)
\begin{align}
\tilde{h} &= A \left(\pi {\cal{M}}_{c} f\right)^{-7/6} e^{-i \Psi_{\rm GR}}  \left[1 - \frac{B_{\rm ppE}}{2} \eta^{-2q_{\rm ppE}/5} \left(\pi {\cal{M}}_{c} f\right)^{2q_{\rm ppE}} 
\right. 
\nonumber \\ 
&+ \left.
 \frac{A_{\rm ppE}}{6} \left(6 + 4 p_{\rm ppE} - 5 p_{\rm ppE}^{2}\right) \eta^{-2p_{\rm ppE}/5} \left(\pi {\cal{M}}_{c} f\right)^{2p_{\rm ppE}}\right] e^{-i \delta\Psi_{\rm ppE}}\,,
\label{eq:h-new}
\\
\delta \Psi_{\rm ppE} &= \frac{5}{32} A \frac{5 p_{\rm ppE}^{2} - 2 p_{\rm ppE} - 6}{(4-p_{\rm ppE})(5-2p_{\rm ppE})} \eta^{-2p_{\rm ppE}/5} \left(\pi {\cal{M}}_{c} f\right)^{2p_{\rm ppE}-5} 
\nonumber \\ 
&+ \frac{15}{32} \frac{B_{\rm ppE}}{(4-q_{\rm ppE})(5-2q_{\rm ppE})} \eta^{-2q_{\rm ppE}/5} \left(\pi {\cal{M}}_{c} f\right)^{2q_{\rm ppE}-5}\,.
\end{align}
Of course, usually one of these two modifications dominates over the other, depending on whether $q>p$ or $p<q$. In Jordan--Fierz--Brans--Dicke theory, for example, the radiation-reaction correction dominates as $q<p$. If, in addition to these modifications in the generation of gravitational waves, one also allows for modifications in the propagation, one is then led to the following template family~\cite{Chatziioannou:2012rf}
\begin{align}
\tilde{h}_{\rm ppE, insp, 2}^{(\ell=2)} = {\cal{A}} \left(\pi {\cal{M}}_{c} f\right)^{-7/6} e^{-i \Psi_{\rm GR}} \left[1 + c \beta_{\rm ppE} \left(\pi {\cal{M}}_{c} f\right)^{b_{\rm ppE}/3+5/3} \right] e^{2 i \beta_{\rm ppE} u^{b_{\rm ppE}}} e^{i \kappa_{\rm ppE} u^{k_{\rm ppE}}}\,,
\label{eq:fullppE-2}
\end{align}
Here, $(b_{\rm ppE},\beta_{\rm ppE})$ and $(k_{\rm ppE},\kappa_{\rm ppE})$ are ppE parameters induced by modifications to the generation and propagation of gravitational waves respectively, where still $(b_{\rm ppE},k_{\rm ppE}) \in \mathbb{Z}$, while $c$ is fully determined by the former set via
\be
\label{c-cons}
c_{\rm{cons}} = -\frac{16}{15} \frac{(3 - b) (42 b + 61 + 5 b^{2})}{5 b^{2} + 46 b + 81}\,,
\ee
if the modifications to the binding energy dominate, 
\be
\label{c-diss}
c_{\rm{diss}} = -\frac{16}{15} (3 - b) b\,,
\ee
if the modifications to the energy flux dominate, or
\be
c_{\rm{both}} = -\frac{32}{15} \frac{b(3 -b) (44b + 71 + 5 b^{2})}{5b^{2} + 46 b + 81}\,.
\label{c-both}
\ee
if both corrections enter at the same post-Newtonian order. Noticing again that if only a single term in the phase correction dominates in the post-Newtonian approximation (or both will enter at the same post-Newtonian order), one can map Eq.~\eqref{eq:h-new} to Eq.~\eqref{eq:fullppE} by a suitable redefinition of constants. 

%------------------------------------
%% \vspace{0.3cm}
%% \hspace{-0.7cm} {\bf{{5.3.4.3 More Complex ppE Models}}}
%% \vspace{0.2cm}

\subsubsection*{More complex ppE models}

Of course, one can introduce more ppE parameters to increase the complexity of the waveform family, and thus, Eq.~\eqref{eq:fullppE} should be thought of as a minimal choice. In fact, one expects any modified theory of gravity to introduce not just a single parametric modification to the amplitude and the phase of the signal, but two new functional degrees of freedom:
\begin{equation}
\alpha_{\rm ppE}u^{a_{\rm ppE}} \to \delta A_{\rm ppE}(\lambda^{a},\theta^{a};u)\,,
\qquad
\beta_{\rm ppE} u^{b_{\rm ppE}} \to \delta \Psi_{\rm ppE}(\lambda^{a},\theta^{a};u)\,,
\label{eq:gen-ppE}
\end{equation}
where these functions will depend on the frequency $u$, as well as on system parameters $\lambda^{a}$ and theory parameters $\theta^{a}$. In a post-Newtonian expansion, one expects these functions to reduce to leading-order to the left-hand sides of Eq.~\eqref{eq:gen-ppE}, but also to acquire post-Newtonian corrections of the form 
\begin{align}
\delta A_{\rm ppE}(\lambda^{a},\theta^{a};u) &= \alpha_{\rm ppE}(\lambda^{a},\theta^{a})u^{a_{\rm ppE}} \sum_{n} \alpha_{n,\rm ppE}(\lambda^{a},\theta^{a}) u^{n}\,,
\\
\delta \Psi_{\rm ppE}(\lambda^{a},\theta^{a};u) &= \beta_{\rm ppE}(\lambda^{a},\theta^{a}) u^{b_{\rm ppE}} \sum_{n} \beta_{n,\rm ppE}(\lambda^{a},\theta^{a}) u^{n}\,,
\end{align}
where here the structure of the series is assumed to be of the form $u^{n}$ with $u>0$.  Such a model, also suggested by Yunes and Pretorius~\cite{Yunes:2009ke}, would introduce too many new parameters that would dilute the information content of the waveform model. Recently, Sampson et al.~\cite{Sampson:2013lpa} demonstrated that the simplest ppE model of Eq.~\eqref{eq:fullppE} suffices to signal a deviation from General Relativity, even if the injection contains three terms in the phase. 

In fact, this is precisely one of the most important differences between the ppE and ppN frameworks. In ppN, it does not matter how many ppN parameters are introduced, because the observations are of very high signal--to--noise ratio, and thus, templates are not needed to extract the signal from the noise. On the other hand, in gravitational wave astrophysics, templates are essential to make detections and do parameter estimation. Spurious parameters in these templates that are not needed to match the signal will deteriorate the accuracy to which {\emph{all}} parameters can be measured because of an Occam penalty. Thus, in gravitational wave astrophysics and data analysis one wishes to minimize the number of theory parameters when testing General Relativity~\cite{Cornish:2011ys,Sampson:2013lpa}. One must then find a balance between the number of additional theory parameters to introduce and the amount of bias contained in the templates. 

At this junction, one must emphasize that frequency exponents in the
amplitude and phase correction were above assumed to be integers,
i.e., $(a_{\rm ppE},b_{\rm ppE},n) \in \mathbb{Z}$. This must be the
case if these corrections arise due to modifications that can be
represented as integer powers of the momenta or velocity. We are not
aware of any theory that predicts corrections proportional to
fractional powers of the velocity for circular inspirals. Moreover,
one can show that theories that introduce non-integer powers of the
velocity into the equations of motion will lead to issues with
analyticity at zero velocity and a breakdown of uniqueness of
solutions~\cite{Chatziioannou:2012rf}. In spite of this, modified
theories can introduce logarithmic terms, that for example enter at
high post-Newtonian order in General Relativity due to non-linear
propagation effects (see, e.g., \cite{Blanchet:2006zz} and references therein). Moreover, certain modified gravity theories introduce {\emph{screened}} modifications that become ``active'' only above a certain frequency. Such effects would be modeled through a Heaviside function, for example needed when dealing with massive Brans--Dicke gravity~\cite{Detweiler:1980uk,Cardoso:2011xi,Alsing:2011er,Yunes:2011aa}. Even these non-polynomial injections, however, would be detectable with the simplest ppE model. In essence, one finds similar results as if one were trying to fit a 3-parameter injection with the simplest 1-parameter ppE model~\cite{Sampson:2013lpa}.

One can of course also generalize the inspiral ppE waveform families to more general orbits, for example through the inclusion of spins aligned or counter-aligned with the orbital angular momentum. More general inspirals would still lead to waveform families of the form of Eq.~\eqref{eq:fullppE} or~\eqref{eq:fullppE-2}, but where the parameters $(\alpha_{\rm ppE},\beta_{\rm ppE})$ would now depend on the mass ratio, mass difference, and the spin parameters of the black holes. With a single detection, one cannot break the degeneracy in the ppE parameters and separately fit for its system parameter dependencies. Given multiple detections, however, one should be able to break such a degeneracy, at least to a certain degree~\cite{Cornish:2011ys}. Such breaking of degeneracies begins to become possible when the number of detections exceeds the number of additional parameters required to capture the physical parameter dependencies of $(\alpha_{\rm ppE},\beta_{\rm ppE})$. 

PpE waveforms can be extended to account for the merger and ringdown phases of coalescence. Yunes and Pretorius have suggested the following template family to account also for this~\cite{Yunes:2009ke}
\ba
\tilde{h}_{\rm ppE,full}^{(\ell=2)} =
 \begin{cases}
\tilde{h}_{\rm ppE} & \text{$f < f_{\rm IM}$}, \\
  \gamma u^{c} e^{i (\delta + \epsilon u)}& \text{$f_{\rm IM} < f < f_{\rm MRD}$}, \\
   \zeta \frac{\tau}{1 + 4 \pi^{2} \tau^{2} \kappa \left(f - f_{\rm RD}\right)^{d}} 
& \text{$f > f_{\rm MRD}$},
\end{cases}
\label{main-eq}
\ea
where the subscript $IM$ and $MRD$ stand for inspiral-merger and merger-ringdown. The merger phase ($f_{\rm IM} < f < f_{\rm MRD}$) is here modeled as an interpolating region between the inspiral and ringdown, where the merger parameters $(\gamma,\delta)$ are set by continuity and differentiability and the ppE merger parameters $(c, \epsilon)$ should be fit for. In the ringdown phase ($f > f_{\rm MRD}$), the response function is modeled as a single-mode generalized Lorentzian, with real and imaginary dominant frequencies $f_{\rm RD}$ and $\tau$, ringdown parameter $\zeta$ also set by continuity and differentiability, and the ppE ringdown parameters $(\kappa,d)$ are to be fit for. The transition frequencies $(f_{\rm IM},f_{\rm MRD})$ can either be treated as ppE parameters or set via some physical criteria, such as at light-ring frequency and the fundamental ringdown frequency, respectively.

Recently, there has been effort to generalize the ppE templates to allow for the excitation of non-GR gravitational wave polarizations. Modifications to only the two General Relativity polarizations map to corrections to terms in the time-domain Fourier transform that are proportional to the $\ell=2$ harmonic of the orbital phase. Arun suggested, however, that if additional polarizations are present, other terms proportional to the $\ell=0$ and $\ell=1$ harmonic will also arise~\cite{Arun:2012hf}. Chatziioannou, Yunes and Cornish~\cite{Chatziioannou:2012rf} have found that the presence of such harmonics can be captured through the more complete single-detector template family
\begin{align}
\label{eq:fullppE-with-amp}
\tilde{h}_{\rm ppE, insp}^{{\rm{all}} \, \ell}(f) &= {\cal{A}} \; \left(\pi {\cal{M}}_{c} f\right)^{-7/6} e^{-i \Psi^{(2)_{\rm GR}}} \left[1 + c \; \beta_{\rm ppE} \left(\pi {\cal{M}}_{c} f\right)^{b_{\rm ppE}/3+5/3} \right] 
e^{2 i \beta_{\rm ppE} u_{2}^{b_{\rm ppE}}} e^{2 i k_{\rm ppE} u_{2}^{\kappa_{\rm ppE}}}
\nonumber \\
&+ \gamma_{\rm ppE} \; u_{1}^{-9/2} e^{-i \Psi^{(1)_{\rm GR}}}
e^{i \beta_{\rm ppE} u_{1}^{b_{\rm ppE}}} e^{2 i k_{\rm ppE} u_{1}^{\kappa_{\rm ppE}}}\,,
\\
\Psi^{(\ell)}_{\rm GR} &=-2 \pi f t_{c} + \ell \Phi_{c}^{(\ell)} + \frac{\pi}{4} - \frac{3 \ell}{256 u_{\ell}^{5}} \sum_{n=0}^{7} u_{\ell}^{n/3} \left(c_{n}^{\rm PN} + l_{n}^{\rm PN} \ln u_{\ell}\right)\,,
\end{align}
where we have defined $u_{\ell} = (2 \pi {\cal{M}}_{c} f/\ell)^{1/3}$. The ppE theory parameters are now $\vec{\theta} = (b_{\rm ppE}, \beta_{\rm ppE}, k_{\rm ppE}, \kappa_{\rm ppE}, \gamma_{\rm ppE}, \Phi_{c}^{(1)})$. Of course, one may ignore $(k_{\rm ppE},\kappa_{\rm ppE})$ altogether, if one wishes to ignore propagation effects. Such a parameterization recovers the predictions of Jordan--Fierz--Brans--Dicke theory for a single-detector response function~\cite{Chatziioannou:2012rf}, as well as Arun's analysis for generic dipole radiation~\cite{Arun:2012hf}. 

One might worry that the corrections introduced by the $\ell=1$ harmonic, ie.~terms proportional to $\gamma_{\rm ppE}$ in Eq.~\eqref{eq:fullppE-with-amp}, will be degenerate with post-Newtonian corrections to the amplitude of the $\ell=2$ mode (not displayed in Eq.~\eqref{eq:fullppE-with-amp}). This, however, is clearly not the case, as the latter scale as $(\pi {\cal{M}}_{c} f)^{-7/6 + n/3}$ with $n$ an integer greater than $0$, while the $\ell=1$ mode is proportional to $(\pi {\cal{M}}_{c} f)^{-3/2}$, which would correspond to a $(-0.5)$ post-Newtonian order correction, ie.~$n=-1$. On the other hand, the ppE amplitude corrections to the $\ell=2$ mode, ie.~terms proportional to $\beta_{\rm ppE}$ in the amplitude of Eq.~\eqref{eq:fullppE-with-amp}, can be degenerate with such post-Newtonian corrections when $b_{\rm ppE}$ is an integer greater than $-4$. 

%------------------------------------
%% \vspace{0.3cm}
%% \hspace{-0.7cm} {\bf{{5.3.4.4 Applications of the ppE Formalism}}}
%% \vspace{0.2cm}

\subsubsection*{Applications of the ppE formalism}

The two models in Eq.~\eqref{eq:fullppE} and~\eqref{eq:fullppE-2} answer different questions. The latter contains a stronger prior (that ppE frequency exponents be integers), and thus, it is ideal for fitting a particular set of theoretical models. On the other hand, Eq.~\eqref{eq:fullppE} with continuous ppE frequency exponents allows one to search for {\emph{generic}} deviations that are statistically significant, without imposing such theoretical priors. That is, if a deviation from General Relativity is present, then Eq.~\eqref{eq:fullppE} is more likely to be able to fit it, than Eq.~\eqref{eq:fullppE-2}. If one prioritizes the introduction of the least number of new parameters, Eq.~\eqref{eq:fullppE} with $(a_{\rm ppE},b_{\rm ppE}) \in \mathbb{R}$ can still recover deviations from General Relativity, even if the latter cannot be represented as a correction proportional to an integer power of velocity. 

Given these ppE waveforms, how should they be used in a data analysis pipeline? The main idea behind the ppE framework is to match filter or perform Bayesian statistics with ppE enhanced template banks to allow the data to select the best-fit values of $\theta^{a}$. As discussed in~\cite{Yunes:2009ke,Cornish:2011ys} and then later in~\cite{Li:2011cg}, one might wish to first run detection searches with General Relativity template banks, and then, once a signal has been found, do a Bayesian model selection analysis with ppE templates.  The first such Bayesian analysis was carried out by Cornish et al.~\cite{Cornish:2011ys}, who concluded that an Ad.~LIGO detection at signal-to-noise ratio of $20$ for a quasi-circular, non-spinning black hole inspiral would allow us to constrain $\alpha_{\rm ppE}$ and $\beta_{\rm ppE}$ much better than existent constraints for sufficiently strong-field corrections, eg.~$b_{\rm ppE} > -5$. This is because for lower values of the frequency exponents, the corrections to the waveform are weak-field and better constrained with binary pulsar observations~\cite{Yunes:2010qb}. The large statistical study of Li, et al.~\cite{Li:2011cg} uses a reduced set of ppE waveforms and investigates our ability to detect deviations of GR when considering a catalogue of Advanced LIGO/Virgo detections. Of course, the disadvantage of such a pipeline is that it requires a first detection, and if the gravitational interaction is too different from General Relativity's prediction, it is possible that a search with General Relativity templates might miss the signal all together; we dim this possibility to be less likely. 

A built-in problem with the ppE and the ppN formalisms is that if a non-zero ppE or ppN parameter is detected, then one cannot necessarily map it back to a particular modified gravity action. On the contrary, as suggested in Table~\ref{table:ppEpars}, there can be more than one theory that predicts structurally similar corrections to the Fourier transform of the response function. For example, both Jordan--Fierz--Brans--Dicke theory and the dissipative sector of Einstein-Dilaton-Gauss--Bonnet theory predict the same type of leading-order correction to the waveform phase. However, if a given ppE parameter is measured to be non-zero, this could provide very useful information as to the type of correction that should be investigated further at the level of the action. The information that could be extracted is presented in Table~\ref{table:ppEpars-interpretation}, which is derived from knowledge of the type of corrections that lead to Table~\ref{table:ppEpars}. 

\begin{table}[htbp]
\caption{Interpretation of non-zero ppE parameters.}
\label{table:ppEpars-interpretation}
\centering
\begin{tabular}{p{1cm}p{1cm}p{7cm}}
\toprule
$a_{\rm ppE}$ & $b_{\rm ppE}$ & Interpretation\\ 
\midrule
$1$ & $\cdot$ & Parity Violation \\
\midrule
$-8$ & $-13$ & Anomalous Acceleration, Extra Dimensions, Violation of Position Invariance \\
\midrule
$\cdot$ & $-7$ & Dipole Gravitational Radiation, Electric Dipole Scalar Radiation \\
\midrule
$\cdot$ & $-3$ & Massive Graviton Propagation \\
\midrule
$\propto$ spin & $-1$ & Magnetic Dipole Scalar Radiation, Quadrupole Moment Correction, Scalar Dipole Force \\
\bottomrule
\end{tabular}
\end{table}

Moreover, if a follow-up search is done with the ppE model in Eq.~\eqref{eq:fullppE-2}, one could infer whether the correction is one due to modifications to the generation or the propagation of gravitational waves. In this way, a non-zero ppE detection could inform theorist of what type of General Relativity modification is preferred by Nature. 

%------------------------------------
%% \vspace{0.3cm}
%% \hspace{-0.7cm} {\bf{{5.3.4.5 Degeneracies}}}
%% \vspace{0.2cm}

\subsubsection*{Degeneracies}

Much care must be taken, however, to avoid confusing a ppE theory modification with some other systematic, such as an astrophysical, a mismodeling or an instrumental effect. Instrumental effects can be easily remedied by requiring that several instruments, with presumably unrelated instrumental systematics, independently derive a posterior probability for $(\alpha_{\rm ppE},\beta_{\rm ppE})$ that peaks away from zero. Astrophysical uncertainties can also be alleviated by requiring that different events lead to the same posteriors for ppE parameters (after breaking degeneracies with system parameters). Astrophysically, however, there are a limited number of scenarios that could lead to corrections in the waveforms that are large enough to interfere with these tests. For comparable mass ratio inspirals, this is usually not a problem as the inertia of each binary component is too large for any astrophysical environment to affect the orbital trajectory~\cite{Hayasaki:2012qn}. Magnetohydrodynamic effects could affect the merger of neutron-star binaries, but this usually occurs outside of the sensitivity band of ground-based interferometers. In extreme mass-ratio inspirals, however, the small compact object can be easily nudged away by astrophysical effects, such as the presence of an accretion disk~\cite{Yunes:2011ws,Kocsis:2011dr} or a third supermassive black hole~\cite{Yunes:2010sm}. These astrophysical effects, however, present the interesting feature that they correct the waveform in a form similar to Eq.~\eqref{eq:fullppE} but with $b_{\rm ppE} < -5$. This is because the larger the orbital separation, the stronger the perturbations of the astrophysical environment, either because the compact object gets closer to the third body or because it leaves the inner edge of the accretion disk and the disk density increases with separation. Such effects, however, are not likely to be present in all sources observed, as few extreme mass-ratio inspirals are expected to be embedded in an accretion disk or sufficiently close to a third body ($\lesssim 0.1 \; {\rm{pc}}$) for the latter to have an effect on the waveform. 

Perhaps the most dangerous systematic is mismodeling, which is due to the use of approximation schemes when constructing waveform templates. For example, in the inspiral one uses the post-Newtonian approximation, series expanding and truncating the waveform at a given power of orbital velocity. Moreover, neutron stars are usually modeled as test-particles (with a Dirac distributional density profile), when in reality they have a finite radius, which will depend on its equation of state. Such finite-size effects enter at 5 post-Newtonian order (the so-called effacement principle~\cite{Hawking:1987en,Damour:1987pa}), but with a post-Newtonian coefficient that can be rather large~\cite{mora-will,berti-iyer-will,Flanagan:2007ix}. Ignorance of the post-Newtonian series beyond 3 post-Newtonian order can lead to systematics in the determination of physical parameters and possibly also to confusion when carrying out ppE-like tests. Much more work is needed to determine the systems and signal-to-noise ratios for which such systematics are truly a problem.  

%%%%%%%%%%%%%%%%%%%%%%%%
\subsubsection{Searching for Non-Tensorial Gravitational Wave Polarizations}

Another way to search for generic deviations from General Relativity is to ask whether any gravitational wave signal detected contains more than the two traditional polarizations expected in General Relativity. A general approach to answer this question is through null streams, as discussed in Section~\ref{sec:Stoch-Anal}. This concept was first studied by G\"ursel and Tinto~\cite{Guersel:1989th} and later by Chatterji et al.~\cite{Chatterji:2006nh} with the aim to separate false-alarm events from real detections. Chatziioannou, et al.~\cite{Chatziioannou:2012rf} proposed the extension of the idea of null streams to develop null tests of General Relativity, which was proposed using stochastic gravitational wave backgrounds in~\cite{Nishizawa:2009bf,Nishizawa:2009jh} and recently implemented in~\cite{Hayama:2012au} to reconstruct the independent polarization modes in time-series data of a ground-based detector network.

Given a gravitational wave detection, one can ask whether the data is consistent with two polarizations by constructing a null stream through the combination of data streams from 3 or more detectors. As explained in Section~\ref{sec:Stoch-Anal}, such a null stream should be consistent with noise in General Relativity, while it would present a systematic deviation from noise if the gravitational wave metric perturbation possessed more than 2 polarizations. Notice that such a test would not require a template; if one were parametrically constructed, such as in~\cite{Chatziioannou:2012rf}, more powerful null tests could be applied to such a null steam. In the future, we expect several gravitational wave detectors to be online: the two Ad.~LIGO ones in the United States, Ad.~VIRGO in Italy, LIGO-India in India, and KAGRA in Japan. Given a gravitational wave observation that is detected by all $5$ detectors, one can then construct $3$ enhanced General Relativity-null streams, each with power in a signal null direction. 

%------------------------------------------------------------------------------------------------------------------------------
\subsubsection{I-Love-Q Tests}

Neutron stars in the slow-rotation limit can be characterized by their mass and radius (to zeroth-order in spin), by their moment of inertia (to first-order in spin), and by their quadrupole moment and Love numbers (to second-order in spin). One may expect these quantities to be quite sensitive to the neutron star's internal structure, which can be parameterized by its equation of state, ie.~the relation between its internal pressure and its internal energy density. Since the equation of state cannot be well-constrained at super-nuclear densities in the laboratory, one is left with a variety of possibilities that predict different neutron star mass-radius relations. 

Recently, however, Yagi and Yunes~\cite{Yagi:2013bca,Yagi:2013awa} have demonstrated that there are relations between the moment of inertia ($I$), the Love numbers ($\lambda)$, and the quadrupole moment ($Q$), the {\emph{I-Love-Q relations}} that are essentially insensitive to the equation of state. Figure~\ref{fig:I-Love-Q} shows two of these relations (the normalized I-Love and Q-Love relations -- see caption) for a variety of equations of state, including APR~\cite{APR}, SLy~\cite{SLy,shibata-fitting}, Lattimer--Swesty with nuclear incompressibility of 220MeV (LS220)~\cite{LS, ott-EOS}, Shen~\cite{Shen1,Shen2,ott-EOS}, the latter two with temperature of 0.01MeV and an electron fraction of 30$\%$, and polytropic equations of state with indices of $n=0.6$, $0.8$ and $1$\footnote{Notice that these relations are independent of the polytropic constant $K$, where $p=K \rho^{(1+1/n)}$, as shown in~\cite{Yagi:2013awa}.}. The bottom panels show the difference between the numerical results and the analytical, fitting curve. Observe that all equations of state lead to the same I-Love and Q-Love relations, with discrepancies smaller than $1\%$ for realistic neutron-star masses. These results have been recently verified in~\cite{Maselli:2013mva} through the post-Newtonian-Affine approach~\cite{Ferrari:2011as,Maselli:2012zq}, which proves the I-Love-Q relations hold not only during the inspiral, but also close to plunge and merger.  

%%%%%%%%%
\epubtkImage{}{%
\begin{figure}[htbp]
\centerline{
 \includegraphics[width=7.25cm,clip=true]{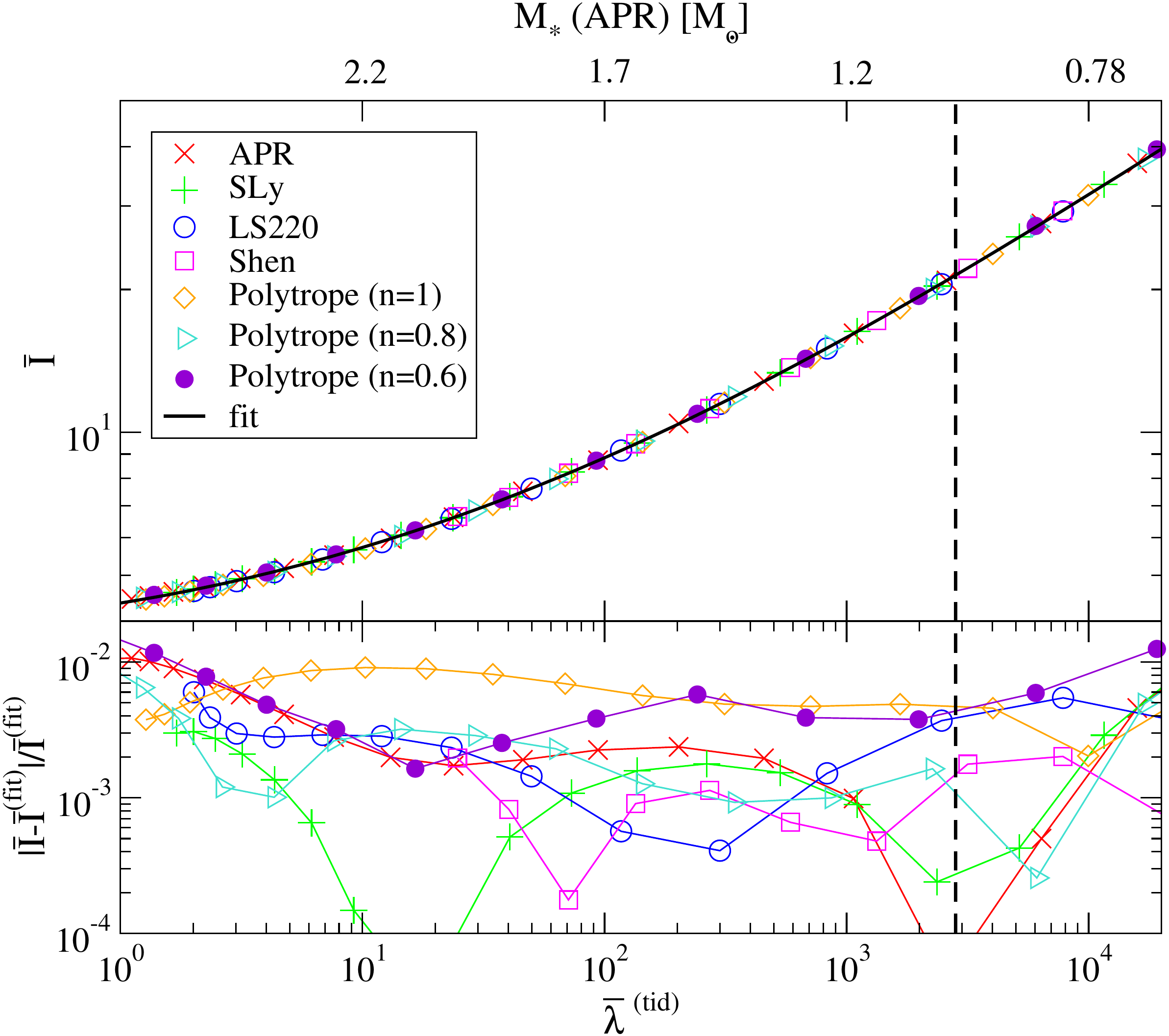}
  \includegraphics[width=7.25cm,clip=true]{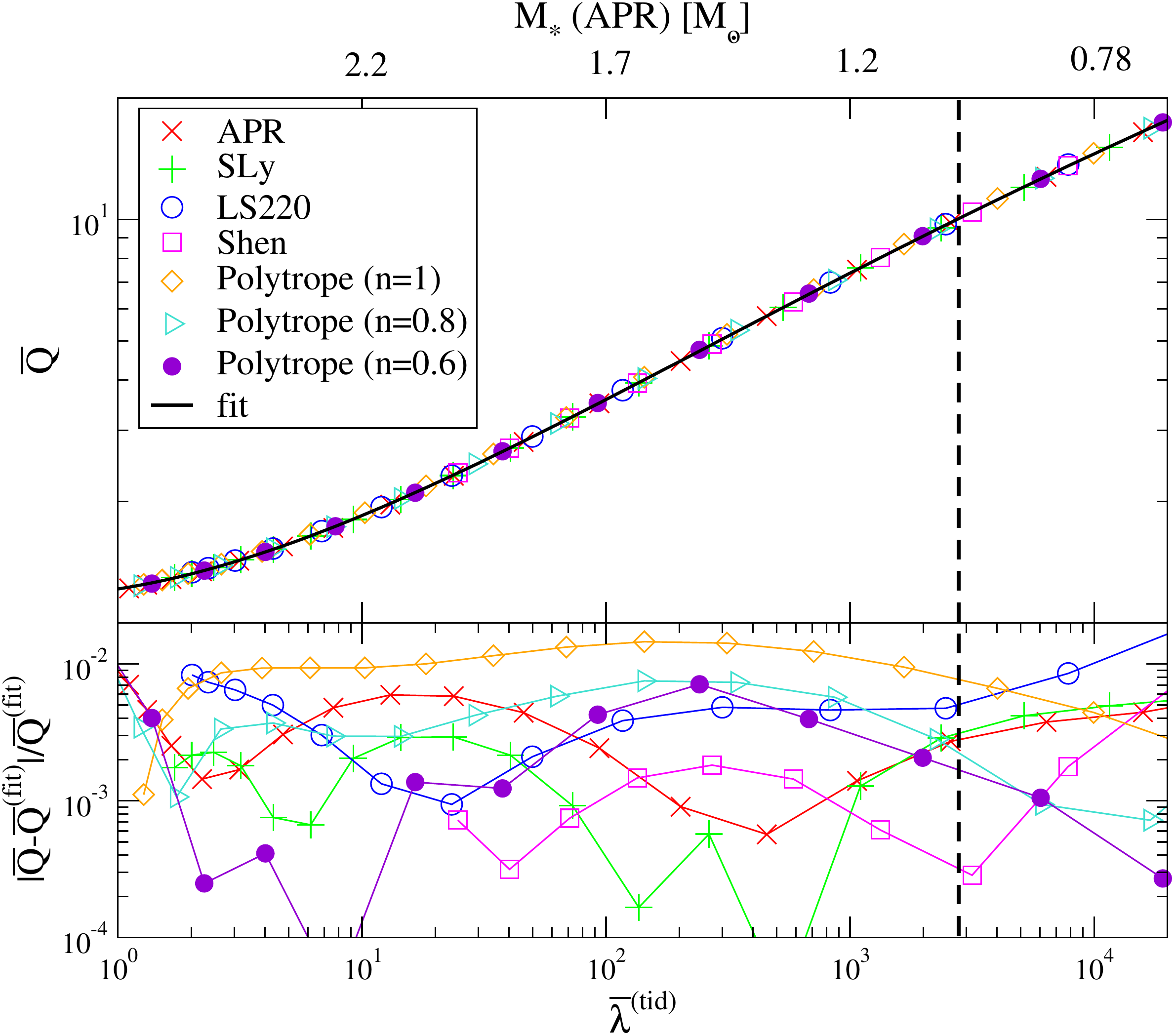}
}
\caption{(Top) Fitting curves (solid curve) and numerical results (points) of the universal I-Love (left) and Q-Love (right) relations for various equations of state, normalized as $\bar{I} = I/M_{\rm NS}^{3}$, $\bar{\lambda}^{(\rm tid)} = \lambda^{(\rm tid)}/M_{\rm NS}^{5}$ and $\bar{Q} = -Q^{(\rm rot)}/[M_{\rm NS}^{3} (S/M_{\rm NS}^{2})^{2}]$, $M_{\rm NS}$ is the neutron-star mass,  $\lambda^\mathrm{(tid)}$ is the tidal Love number, $Q^\mathrm{(rot)}$ is the rotation-induced quadrupole moment, and $S$ is magnitude of the neutron star spin angular momentum. The neutron star central density is the parameter varied along each curve, or equivalently the NS compactness. The top axis shows the neutron star mass for the APR equation of state, with the vertical dashed line showing $M_{\rm NS} = 1 M_{\odot}$. (Bottom) Relative fractional errors between the fitting curve and the numerical results. Observe that these relations are essentially independent of the equation of state, with loss of universality at the 1\% level.}
\label{fig:I-Love-Q} 
\end{figure}}
%%%%%%%%% 

Given the independent measurement of any two members of the I-Love-Q trio, one could carry out a (null) model-independent and equation-of-state-independent test of General Relativity~\cite{Yagi:2013bca,Yagi:2013awa}. For example, assume that electromagnetic observations of the double binary pulsar J0737-3039 have measured the moment of inertia to 10\% accuracy~\cite{lattimer-schutz,kramer-wex,kramer-double-pulsar}. The slow-rotation approximation is perfectly valid for this double binary pulsar, due to its relatively long spin period. Assume further that a gravitational wave observation of a binary neutron star inspiral, with individual masses similar to that of the primary in J0737-3039, manages to measure the neutron star tidal Love number to 60\% accuracy~\cite{Yagi:2013bca,Yagi:2013awa}. These observations then lead to an error box in the I-Love plane, which must contain the curve in the left-panel of Fig.~\ref{fig:I-Love-Q}. 

A similar test could be carried out by using data from only binary pulsar observations or only gravitational wave detections. In the case of the latter, one would have to simultaneously measure or constrain the value of the quadrupole moment and the Love number, since the moment of inertia is not measurable with gravitational wave observations. In the case of the former, one would have to extract the moment of inertia and the quadrupole moment, the latter of which will be difficult to measure. Therefore, the combination of electromagnetic and gravitational wave observations would be the ideal way to carry out such tests. 

Such a test of General Relativity, of course, is powerful only as long as modified gravity theories predict I-Love-Q relations that are not degenerated with the General Relativistic ones. Yagi and Yunes~\cite{Yagi:2013bca,Yagi:2013awa} investigated such a relation in dynamical Chern--Simons gravity to find that such degeneracy is only present in the limit $\zeta_{\rm CS} \to 0$. That is, for any finite value of $\zeta_{\rm CS}$, the dynamical Chern--Simons I-Love-Q relation differs from that of General Relativity, with the distance to the General Relativity expectation increasing for larger $\zeta_{\rm CS}$. Yagi and Yunes~\cite{Yagi:2013bca,Yagi:2013awa} predicted that a test similar to the one described above could constrain dynamical Chern--Simons gravity to roughly $\xi_{\rm CS}^{1/4} < 10 M_{\rm NS} \sim 15 \; {\rm{km}}$, where recall that $\xi_{\rm CS} = \alpha_{\rm CS}^{2}/(\beta \kappa)$.  

The test described above, of course, only holds provided the I-Love-Q relations are valid, which in turn depends on the assumptions made in deriving them. In particular, Yagi and Yunes~\cite{Yagi:2013bca,Yagi:2013awa} assumed that the neutron stars are uniformly and slowly-rotating, as well as only slightly tidally deformed by their rotational velocity or companion. These assumptions would not be valid for newly-born neutron stars, which are probably differentially rotating and doing so fast. The gravitational waves emitted by neutron star inspirals, however, are expected to have binary components that are old and not rapidly spinning, by the time they enter the detector sensitivity band~\cite{bildsten-cutler}. Some short-period, millisecond pulsars may spin at a non-negligible rate, for which the normalized moment of inertia, quadrupole moment and Love number would not be independent of the rotational angular velocity. If then, however, the above tests should still be possible, since binary pulsar observations would also automatically determine the rotational angular velocity, for which a unique I-Love-Q relation should exist in General Relativity.

%%%%%%%%%%%
\subsection{Tests of the No-Hair Theorems}

Another important class of generic tests of General Relativity are those that concern the so-called no-hair theorems. Since much work has been done on this area, we have decided to separate this topic from the main generic tests subsection. In what follows, we describe what these theorems are and the possible tests one could carry out with gravitational wave observations emitted by black hole binary systems. 

%------------------------------------------------------------------------------------------------------------------------------
\subsubsection{The No-Hair Theorems}

The {\emph{no-hair}} theorems state that the only stationary, vacuum solution to the Einstein equations that is non-singular outside the event horizon is completely characterized by three quantities: its mass $M$, its spin $S$ and its charge $Q$. This conclusion is arrived at by combining several different theorems. First, Hawking~\cite{Hawking:1971vc,Hawking:1971tu} proved that a stationary black hole must have an event horizon with a spherical topology and that it must be either static or axially symmetric. Israel~\cite{Israel:1967wq,Israel:1967za} then proved that the exterior gravitational field of such static black holes is uniquely determined by $M$ and $Q$ and it must be given by the Schwarzschild or the Reissner--Nordstr\"om metrics. Carter~\cite{Carter:1971zc} constructed a similar proof for uncharged stationary, axially-symmetric black holes, where this time black holes fall into disjoint families, not deformable into each other and with an exterior gravitational field uniquely determined by $M$ and $S$. Robinson~\cite{Robinson:1975bv} and Mazur~\cite{Mazur:1982db} later proved that such black holes must be described by either the Kerr or the Kerr--Newman metric. See also~\cite{Misner:1973cw,Poisson} for more details. 

The no-hair theorems apply under a restrictive set of conditions. First, the theorems only apply in stationary situations. Black hole horizons can be tidally deformed in dynamical situations, and if so, Hawking's theorems~\cite{Hawking:1971vc,Hawking:1971tu} about spherical horizon topologies do not apply. This then implies that all other theorems described above also do not apply, and thus, dynamical black holes will generically have hair. Second, the theorems only apply in vacuum. Consider, for example, an axially symmetric black hole in the presence of a non-symmetrical matter distribution outside the event horizon. One might naively think that this would tidally distort the event horizon, leading to a rotating, stationary black hole that is not axisymmetric. Hawking and Hartle~\cite{Hawking:1972hy}, however, showed that in such a case the matter distribution torques the black hole forcing it to spin down, thus leading to a non-stationary scenario. If the black hole is non-stationary, then again the no-hair theorems do not apply by the arguments described at the beginning of this paragraph, and thus non-isolated black holes can have hair. Third, the theorems only apply within General Relativity, ie.~through the use of the Einstein equations. Therefore, it is plausible that black holes in modified gravity theories or in General Relativity with singularities outside any event horizons (naked singularities) will have hair.  

The no-hair theorems imply that the exterior gravitational field of isolated, stationary, uncharged and vacuum black holes (in General Relativity and provided the spacetime is regular outside all event horizons) can be written as an infinite sum of mass and current multipole moments, where only two of them are independent: the mass monopole moment $M$ and the current dipole moment $S$. One can extend these relations to include charge, but astrophysical black holes are expected to be essentially neutral due to charge accretion. If the no-hair theorems hold, all other multipole moments can be determined from~\cite{Geroch:1970cd,Geroch:1970cc,hansen}
\be
M_{\ell}+{\rm i}S_{\ell}=M({\rm i}a)^{\ell}\,,
\label{kerrmult}
\ee
where $M_{\ell}$ and $S_{\ell}$ are the $\ell$th mass and current multipole moments. Even if the black hole progenitor was not stationary or axisymmetric, the no-hair theorems guarantee that any excess multipole moments will be shed-off during gravitational collapse~\cite{Price:1971fb,Price:1972pw}. Eventually, after the black hole has settled down and reached an equilibrium configuration, it will be described purely in terms of $M_{0} = M$ and $S_{1} = S = M a^{2}$, where $a$ is the Kerr spin parameter.

An astrophysical observation of a hairy black hole would not imply that the no-hair theorems are wrong, but rather than one of the assumptions made in deriving these theorems is not appropriate to describe Nature. As described above, the three main assumptions are stationarity, vacuum and that General Relativity and the regularity condition hold. Astrophysical black holes will generically be hairy due to a violation of the first two assumptions, since they will neither be perfectly stationary, nor exist in a perfect vacuum. Astrophysical black holes will always suffer small perturbations by other stars, electromagnetic fields, other forms of matter, like dust, plasma or dark matter, etc, which will induce non-zero deviations from Eq.~\eqref{kerrmult} and thus evade the no-hair theorems. In all cases of interest, however, such perturbations are expected to be too small to be observable, which is why one argues that even astrophysical black holes should obey the no-hair theorems if General Relativity holds. Put another way, an observation of the violation of the no-hair theorems would be more likely to indicate a failure of General Relativity in the strong-field, than an unreasonably large amount of astrophysical hair. 

Tests of the no-hair theorems come in two flavors: through electromagnetic observations~\cite{Johannsen:2010xs,Johannsen:2010ru,Johannsen:2010bi,Johannsen:2012ng} and through gravitational wave observations~\cite{Ryan:1995wh,Ryan:1997hg,Collins:2004ex,Glampedakis:2005cf,Babak:2006uv,Barack:2006pq,Li:2007qu,Sopuerta:2009iy,Yunes:2009ry,Vigeland:2009pr,Vigeland:2010xe,Gair:2011ym,Vigeland:2011ji,Rodriguez:2011aa}. The former rely on radiation emitted by accelerating particles in an accretion disk around black holes. Such tests, however, are not clean as they require the modeling of complicated astrophysics, with matter and electromagnetic fields. Gravitational wave tests are clean in that respect, but unlike electromagnetic tests, they cannot be carried out yet due to lack of data. Other electromagnetic tests of the no-hair theorems exist, for example through the observation of close stellar orbits around Sgr~A*~\cite{Merritt:2009ex,Merritt:2011ve,Sadeghian:2011ub} and pulsar-black hole binaries~\cite{Wex:1998wt}, but these cannot yet probe the near-horizon, strong-field regime, since electromagnetic observations cannot yet resolve horizon scales. See~\cite{lrr-2008-9} for reviews on this topic.

%------------------------------------------------------------------------------------------------------------------------------
\subsubsection{Extreme Mass-Ratio Tests of The No-Hair Theorem}

Gravitational wave tests of the no-hair theorems require the detection of either extreme mass-ratio inspirals or the ringdown of comparable-mass black hole mergers with future space-borne gravitational wave detectors~\cite{AmaroSeoane:2012km,AmaroSeoane:2012je}. Extreme mass-ratio inspirals consist of a stellar-mass compact object spiraling into a supermassive black hole in a generic orbit within astronomical units from the event horizon of the supermassive object~\cite{AmaroSeoane:2007aw}. These events outlive the observation time of future detectors, emitting millions of gravitational wave cycles, with the stellar-mass compact object essentially acting as a tracer of the supermassive black hole spacetime~\cite{Sotiriou:2004bm}. Ringdown gravitational waves are always emitted after black holes merge and the remnant settles down into its final configuration. During the ringdown, the highly-distorted remnant radiates all excess degrees of freedom and this radiation carries a signature of whether the no-hair theorems hold in its quasi-normal mode spectrum (see e.g.~\cite{Berti:2009kk} for a recent review). 

Both electromagnetic and gravitational wave tests need a metric with which to model accretion disks, quasi-periodic oscillations, or extreme mass-ratio inspirals. One can classify these metrics as {\emph{direct}} or {\emph{generic}}, paralleling the discussion in Section~\ref{sec:Direct-tests}. Direct metrics are exact solutions to a specific set of field equations, with which one can derive observables. Examples of such metrics are the Manko--Novikov metric~\cite{1992CQGra...9.2477M} and the slowly-spinning black hole metric in dynamical Chern--Simons gravity~\cite{Yunes:2009hc}. When computing observables with these metrics, one usually assumes that all radiative and dynamical process (e.g.~the radiation-reaction force) are as predicted in General Relativity. Generic metrics are those that parametrically modify the Kerr spacetime, such that for certain parameter choices one recovers identically the Kerr metric, while for others, one has a deformation of Kerr. Generic metrics can be further classified into two subclasses, Ricci-flat versus non-Ricci-flat, depending on whether they satisfy $R_{\mu \nu} = 0$.

Let us first consider direct metric tests of the no-hair theorem. The most studied direct metric is the Manko--Novikov one, which although an exact, stationary and axisymmetric solution to the vacuum Einstein equations, does not represent a black hole, as the event horizon is broken along the equator by a ring singularity~\cite{1992CQGra...9.2477M}. Just like the Kerr metric, the Manko--Novikov metric possesses an ergoregion, but unlike the former, it also possesses regions of closed time-like curves that overlap the ergoregion. Nonetheless, an appealing property of this metric is that it deviates continuously from the Kerr metric through certain parameters that characterize the higher multiple moments of the solution.

The first geodesic study of Manko--Novikov spacetimes was carried out by Gair, et al.~\cite{Gair:2007kr}. They found that there are two ring-like regions of bound orbits: an outer one where orbits look regular and integrable, as there exists four isolating integrals of the motion; and an inner one where orbits are chaotic and thus ergodic. Gair, et al.~\cite{Gair:2007kr} suggested that orbits that transition from the integrable to the chaotic region would leave a clear observable signature in the frequency spectrum of the emitted gravitational waves. However, they also noted that chaotic regions exist only very close to the central body and are probably not astrophysically accessible. The study of Gair, et al.~\cite{Gair:2007kr} was recently confirmed and followed up by Contopoulos, et al.~\cite{Contopoulos:2011dz}. They studied a wide range of geodesics and found that, in addition to an inner chaotic region and an outer regular region, there are also certain Birkhoff islands of stability. When an extreme mass-ratio inspiral traverses such a region, the ratio of resonant fundamental frequencies would remain constant in time, instead of increasing monotonically. Such a feature would impact the gravitational waves emitted by such a system, and it would signal that the orbit equations are non-integrable and the central object is not a Kerr black hole. 

The study of chaotic motion in geodesics of non-Kerr spacetimes is by no means new. Chaos has also been found in geodesics of Zipoy--Voorhees--Weyl and Curzon spacetimes with multiple singularities~\cite{Sota:1995ms,Sota:1996cv} and in general for Zipoy-Voorhees spacetimes in~\cite{LukesGerakopoulos:2012pq}, of perturbed Schwarzschild spacetimes~\cite{Letelier:1996he}, of Schwarzschild spacetimes with a dipolar halo~\cite{Letelier:1997uv,Letelier:1997fi,Gueron:2001mm} of Erez--Rosen spacetimes~\cite{Gueron:2002jt}, and of deformed generalizations of the Tomimatsy--Sato spacetime ~\cite{Dubeibe:2007hq}. One might worry that such chaotic orbits will depend on the particular spacetime considered, but recently Apostolatos, et al.~\cite{Apostolatos:2009vu} and Lukes--Gerakopoulos, et al.~\cite{LukesGerakopoulos:2010rc} have argued that the Birkhoff islands of stability are a general feature. Although the Kolmogorov, Arnold, and Moser theorem~\cite{Kolmogorov,Arnold,Moser} states that phase orbit tori of an integrable system are only deformed if the Hamiltonian is perturbed, the Poincare--Birkhoff theorem~\cite{Chaos} states that resonant tori of integrable systems actually disintegrate, leaving behind a chain of Birkhoff islands. These islands are only characterized by the ratio of winding frequencies that equals a rational number, and thus, they constitute a distinct and generic feature of non-integrable systems~\cite{Apostolatos:2009vu,LukesGerakopoulos:2010rc}. Given an extreme mass-ratio gravitational wave detection, one can monitor the ratio of fundamental frequencies and search for plateaus in their evolution, which would signal non-integrability. Of course, whether detectors can resolve such plateaus depends on the initial conditions of the orbits and the physical system under consideration (these determine the thickness of the islands), as well as the mass ratio (this determines the radiation-reaction timescale) and the distance and mass of the central black hole (this determines the signal-to-noise ratio).   

Another example of a direct metric test of the no-hair theorem is through the use of the slowly-rotating dynamical Chern--Simons black hole metric~\cite{Yunes:2009hc}. Unlike the Manko--Novikov metric, the dynamical Chern--Simons one does represent a black hole, ie.~it possesses an event horizon, but it evades the no-hair theorems because it is not a solution to the Einstein equations. Sopuerta and Yunes~\cite{Sopuerta:2009iy} carried out the first extreme mass-ratio inspiral analysis when the background supermassive black hole object is taken to be such a Chern--Simons black hole. They used a semi-relativistic model~\cite{Ruffini:1981af} to evolve extreme mass-ratio inspirals and found that the leading-order modification comes from a modification to the geodesic trajectories, induced by the non-Kerr modifications of the background. Because the latter correspond to a strong-field modification to General Relativity, modifications in the trajectories are most prominent for zoom-whirl orbits, as the small compact object zooms around the supermassive black hole in a region of unstable orbits, close to the event horizon.  These modifications were then found to propagate into the gravitational waves emitted, leading to a dephasing that could be observed or ruled out with future gravitational wave observations to roughly the horizon scale of the supermassive black hole, as has been recently confirmed by Canizares et al.~\cite{Canizares:2012ji}. These studies, however, may be underestimates, given that they treat the black hole background in dynamical Chern--Simons gravity only to first-order in spin. 

A final example of a direct metric test of the no-hair theorems is by considering black holes that are not in vacuum. Barausse, et al.~\cite{Barausse:2006vt} studied extreme mass-ratio inspirals in a Kerr black hole background that is perturbed by a self-gravitating, homogeneous torus that is compact, massive and close to the Kerr black hole. They found that the presence of this torus impacts the gravitational waves emitted during such inspirals, but only weakly, making it difficult to distinguish the presence of matter. Yunes, et al.~\cite{Yunes:2011ws} and Kocsis, et al.~\cite{Kocsis:2011dr} carried out a similar study, where this time they considered a small compact object inspiraling completely within a geometrically thin, radiation-pressure dominated accretion disk. They found that disk-induced migration can modify the radiation-reaction force sufficiently so as to leave observable signatures in the waveform, provided the accretion disk is sufficiently dense in the radiation-dominated regime and a gap opens up. These tests of the no-hair theorem, however, will be rather difficult as most extreme mass-ratio inspirals are not expected to be in an accretion disk. 

%Generic, Ricci-Flat
Let us now consider generic metric tests of the no-hair theorem. Generic Ricci-flat deformed metrics will lead to Laplace-type equations for the deformation functions in the far-field since they must satisfy $R_{\mu \nu} = 0$ to linear order in the perturbations. The solution to such an equation can be expanded in a sum of mass and current multipole moments, when expressed in asymptotically Cartesian and mass-centered coordinates~\cite{Thorne:1980ru}. These multipoles can be expressed via~\cite{Collins:2004ex,Vigeland:2009pr,Vigeland:2010xe}
\begin{equation}
M_{\ell} + {\rm i}S_{\ell} = M({\rm i}a)^{{\ell}} + \delta M_{\ell} + {\rm i}\delta S_{\ell}\,,
\label{mult}
\end{equation}
where $\delta M_{\ell}$ and $\delta S_{\ell}$ are mass and current multipole deformations. Ryan~\cite{Ryan:1995wh,Ryan:1997hg} showed that the measurement of three or more multipole moments would allow for a test of the no-hair theorem. Generic non-Ricci flat metrics, on the other hand, will not necessarily lead to Laplace-type equations for the deformation functions in the far field, and thus, the far-field solution and Eq.~\eqref{mult} will depend on a sum of $\ell$ and $m$ multipole moments. 

The first attempt to construct a generic, Ricci-flat metric was by Collins and Hughes~\cite{Collins:2004ex}: the so-called {\emph{bumpy black hole metric}}. In this approach, the metric is assumed to be of the form
\begin{equation}
g_{\mu \nu} = g_{\mu \nu}^{(\rm{Kerr})} + \epsilon \delta g_{\mu \nu}\,,
\end{equation}
where $\epsilon \ll 1$ is a book keeping parameter that enforces that $\delta g_{\mu \nu}$ is a perturbation of the Kerr background. This metric is then required to satisfy the Einstein equations linearized in $\epsilon$, which then leads to differential equations for the metric deformation. Collins and Hughes~\cite{Collins:2004ex} assumed a non-spinning, stationary spacetime, and thus $\delta g_{\mu \nu}$ only possessed two degrees of freedom, both of which were functions of radius only: $\psi_{1}(r)$, which must be a harmonic function and which changes the Newtonian part of the gravitational field at spatial infinity; and $\gamma_{1}(r)$ which is completely determined through the linearized Einstein equations once $\psi_{1}$ is specified. One then has the freedom to choose how to prescribe $\psi_{1}$ and Collins and Hughes investigated~\cite{Collins:2004ex} two choices that corresponded physically to point-like and ring-like naked singularities, thus violating cosmic censorship~\cite{Penrose:1969pc}. Vigeland and Hughes~\cite{Vigeland:2009pr} and Vigeland~\cite{Vigeland:2010xe} then extended this analysis to stationary, axisymmetric spacetimes via the Newman--Janis method~\cite{Newman:1965tw,Drake:1998gf}, showed how such metric deformations modify Eq.~\eqref{mult}, and computed how these bumps imprint themselves onto the orbital frequencies and thus the gravitational waves emitted during an extreme mass-ratio inspiral.  

That the bumps represent unphysical matter should not be a surprise, since by the no-hair theorems, if the bumps are to satisfy the vacuum Einstein equations they must either break stationarity or violate the regularity condition. Naked singularities are an example of the latter. A Lorentz-violating massive field coupled to the Einstein tensor is another example~\cite{Dubovsky:2007zi}. Gravitational wave tests with bumpy black holes must then be understood as {\emph{null tests}}: one assumes the default hypothesis that General Relativity is correct and then sets out to test whether the data rejects or fails to reject this hypothesis (a null hypothesis can never be proven). Unfortunately, however, bumpy black hole metrics cannot parameterize spacetimes in modified gravity theories that lead to corrections in the field equations that are not proportional to the Ricci tensor, such as for example in dynamical Chern--Simons or in Einstein-Dilaton-Gauss--Bonnet modified gravity. 

Other bumpy black hole metrics have also been recently proposed. Glampedakis and Babak~\cite{Glampedakis:2005cf} proposed a different type of stationary and axisymmetric bumpy black hole through the Hartle--Thorne metric~\cite{Hartle:1968si}, with modifications to the quadrupole moment. They then constructed a ``kludge'' extreme mass-ratio inspiral waveform and estimated how well the quadrupole deformation could be measured~\cite{Babak:2006uv}. This metric, however, is valid only when the supermassive black hole is slowly-rotating, as it derives from the Hartle--Thorne ansatz. Recently, Johansen and Psaltis~\cite{Johannsen:2011dh} proposed yet another metric to represent bumpy stationary and spherically symmetric spacetimes. This metric introduces one new degree of freedom, which is a function of radius only and assumed to be a series in $M/r$. Johansen and Psaltis then rotated this metic via the Newman--Janis method~\cite{Newman:1965tw,Drake:1998gf} to obtain a new bumpy metric for axially symmetric spacetimes. Such a metric, however, possesses a naked ring singularity on the equator, and naked singularities on the poles. As before, none of these bumpy metrics can be mapped to known modified gravity black hole solutions, in the Glampedakis and Babak case~\cite{Glampedakis:2005cf} because the Einstein equations are assumed to hold to leading order in the spin, while in the Johansen and Psaltis case~\cite{Johannsen:2011dh} because a single degree of freedom is not sufficient to model the three degrees of freedom contained in stationary and axisymmetric spacetimes~\cite{Stephani:2003tm,Vigeland:2011ji}. 

%Generic, Non-Ricci-Flat
The only generic non-Ricci-flat bumpy black hole metrics so far is that of Vigeland, Yunes and Stein~\cite{Vigeland:2011ji}. They allowed generic deformations in the metric tensor, only requiring that the new metric perturbatively retained the Killing symmetries of the Kerr spacetime: the existence of two Killing vectors associated with stationarity and axisymmetry, as well as the perturbative existence of a Killing tensor (and thus a Carter-like constant), at least to leading order in the metric deformation. Such requirements imply that the geodesic equations in this new background are fully integrable, at least perturbatively in the metric deformation, which then allows one to solve for the orbital motion of extreme mass-ratio inspirals by adapting previously existing tools. Brink~\cite{Brink:2008xx,Brink:2008xy,Brink:2009mq,Brink:2009mt,Brink:2009rf} studied the existence of such a second-order Killing tensor in generic, vacuum, stationary and axisymmetric spacetimes in Einstein's theory and found that these are difficult to construct exactly. By relaxing this exact requirement, Vigeland, Yunes and Stein~\cite{Vigeland:2011ji} found that the existence of a perturbative Killing tensor poses simple differential conditions on the metric perturbation that can be analytically solved. Moreover, they also showed how this new bumpy metric can reproduce all known modified gravity black hole solutions in the appropriate limits, provided these have an at least approximate Killing tensor; thus, these metrics are still vacuum solutions even though $R \neq 0$, since they satisfy a set of modified field equations. Although unclear at this junction, it seems that the imposition that the spacetime retains the Kerr Killing symmetries leads to a bumpy metric that is well-behaved everywhere outside the event horizon (no singularities, no closed-time-like curves, no loss of Lorentz signature)~\cite{Sara}. Recently, Gair and Yunes~\cite{Gair:2011ym} studied how the geodesic equations are modified for a test-particle in a generic orbit in such a spacetime and showed that the bumps are indeed encoded in the orbital motion, and thus, in the gravitational waves emitted during an extreme mass-ratio inspiral.

One might be concerned that such no-hair tests of General Relativity cannot constrain modified gravity theories, because Kerr black holes can also be solutions in the latter~\cite{Psaltis:2007cw}. This is indeed true provided the modified field equations depend only on the Ricci tensor or scalar. In Einstein-Dilaton-Gauss--Bonnet or dynamical Chern--Simons gravity, the modified field equations depend on the Riemann tensor, and thus, Ricci-flat metric need not solve these modified set~\cite{Yunes:2011we}. Moreover, just because the metric background is identically Kerr does not imply that inspiral gravitational waves will be identical to those predicted in General Relativity. All studies carried out to date, be it direct metric tests or generic metric tests, assume that the only quantity that is modified is the metric tensor, or equivalently, the Hamiltonian or binding energy. Inspiral motion, of course, does not depend just on this quantity, but also on the radiation-reaction force that pushes the small object from geodesic to geodesic. Moreover, the gravitational waves generated during such an inspiral depend on the field equations of the theory considered. Therefore, all metric tests discussed above should be considered as partial tests. In general, strong-field modified gravity theories will modify the Hamiltonian, the radiation-reaction force and the wave generation.

%------------------------------------------------------------------------------------------------------------------------------
\subsubsection{Ringdown Tests of The No-Hair Theorem}
\label{sec:rd}

Let us now consider tests of the no-hair theorems with gravitational waves emitted by comparable-mass binaries during the ringdown phase. Gravitational waves emitted during ringdown can be described by a superposition of exponentially damped sinusoids~\cite{Berti:2005ys}:
\begin{equation}
\label{eq:ringdown-waves}
h_{+}(t) + i \; h_{\times}(t) = \frac{M}{r} \sum_{\ell m n} \left\{
{\cal{A}}_{\ell m n} e^{i (\omega_{\ell m n} t + \phi_{\ell m n})} e^{-t/\tau_{\ell m n}} S_{\ell m n}
+
{\cal{A}}_{\ell m n}' e^{i (-\omega_{\ell m n} t + \phi_{\ell m n}')} e^{-t/\tau_{\ell m n}} S_{\ell m n}^{*} \right\}
\,,
\end{equation}
where $r$ is the distance from the source to the detector, the asterisk stands for complex conjugation, the real mode amplitudes ${\cal{A}}_{\ell,m,n}$ and ${\cal{A}}_{\ell,m,n}'$ and the real phases $\phi_{n \ell m}$ and $\phi_{n \ell m}'$ depend on the initial conditions, $S_{\ell m n}$ are spheroidal functions evaluated at the complex quasinormal ringdown frequencies $\omega_{n \ell m} = 2 \pi f_{n \ell m} + i/\tau_{n \ell m}$, and the real physical frequency $f_{n \ell m}$ and the real damping times $\tau_{n \ell m}$ are both functions of the mass $M$ and the Kerr spin parameter $a$ only, provided the no-hair theorems hold. These frequencies and damping times can be computed numerically or semi-analytically, given a particular black hole metric (see~\cite{Berti:2009kk} for a recent review). The Fourier transform of a given $(\ell,m,n)$ mode is~\cite{Berti:2005ys}
\begin{align}
\tilde{h}_{+}^{(\ell,m,n)}(\omega) &= \frac{M}{r} {\cal{A}}_{\ell m n}^{+} \left[e^{i \phi_{\ell m n}^{+}} S_{\ell m n} b_{+}(\omega) + e^{-i \phi_{\ell m n}^{+}} S_{\ell m n}^{*} b_{-}(\omega) \right]\,, 
\\
\tilde{h}_{\times}^{(\ell,m,n)}(\omega) &= \frac{M}{r} {\cal{A}}_{\ell m n}^{\times} \left[e^{i \phi_{\ell m n}^{\times}} S_{\ell m n} b_{+}(\omega) + e^{-i \phi_{\ell m n}^{\times}} S_{\ell m n}^{*} b_{-}(\omega)\right]\,, 
\end{align}
where we have defined ${\cal{A}}_{\ell m n}^{+,\times} e^{i \phi_{\ell m n}^{+,\times}} \equiv {\cal{A}}_{\ell m n} e^{i \phi_{\ell m n}} \pm {\cal{A}}' e^{-i \phi_{\ell m n}'}$ as well as the Lorentzian functions
\begin{equation}
b_{\pm}(\omega) = \frac{\tau_{\ell m n}}{1 + \tau_{\ell m n}^{2} (\omega \pm \omega_{\ell m n})^{2}} 
\end{equation}
Ringdown gravitational waves will all be of the form of Eq.~\eqref{eq:ringdown-waves} provided that the characteristic nature of the differential equation that controls the evolution of ringdown modes is not modified, ie.~provided that one only modifies the potential in the Teukolsky equation or other subdominant terms, which in turn depend on the modified field equations.

Tests of the no-hair theorems through the observation of black hole ringdown date back to Detweiler~\cite{Detweiler:1980gk}, and it was recently worked out in detail by Dreyer, et al.~\cite{Dreyer:2003bv}. Let us first imagine that a single complex mode is detected ${\omega}_{\ell_{1} m_{1} n_{1}}$  and one measures separately its real and imaginary parts. Of course, from such a measurement, one cannot extract the measured harmonic triplet $(\ell_{1},m_{1},n_{1})$, but instead one only measures the complex frequency ${\omega}_{\ell_{1} m_{1} n_{1}}$. This information is not sufficient to extract the mass and spin angular momentum of the black hole because different quintuplets $(M,a,\ell,m,n)$ can lead to the same complex frequency ${\omega}_{\ell_{1} m_{1} n_{1}}$. The best way to think of this is graphically: a given observation of ${\omega}_{\ell_{1} m_{1} n_{1}}^{(1)}$ traces a line in the complex $\Omega_{\ell_{1} m_{1} n_{1}} = M \omega_{\ell_{1} m_{1} n_{1}}^{(1)}$ plane; a given $(\ell,m,n)$ triplet defines a complex frequency $\omega_{\ell m n}$ that also traces a curve in the complex $\Omega_{\ell m n}$ plane; each intersection of the measured line $\Omega_{\ell_{1} m_{1} n_{1}}$ with $\Omega_{\ell m n}$ defines a possible doublet $(M,a)$; since different $(\ell,m,n)$ triplets lead to different $\omega_{\ell m n}$ curves and thus different intersections, one ends up with a set of doublets $S_{1}$, out of which only one represents the correct black hole parameters. We thus conclude that a single mode observation of ringdown gravitational waves is not sufficient to test the no-hair theorem~\cite{Dreyer:2003bv,Berti:2005ys}.

Let us then imagine that one has detected two complex modes, ${\omega}_{\ell_{1} m_{1} n_{1}}$ and ${\omega}_{\ell_{2} m_{2} n_{2}}$. Each detection leads to a separate line ${\Omega}_{\ell_{1} m_{1} n_{1}}$ and ${\Omega}_{\ell_{2} m_{2} n_{2}}$ in the complex plane. As before, each $(n,\ell,m)$ triplet leads to separate curves $\Omega_{\ell m n}$ which will intersect with both ${\Omega}_{\ell_{1} m_{1} n_{1}}$ and ${\Omega}_{\ell_{2} m_{2} n_{2}}$ in the complex plane. Each intersection between $\Omega_{\ell m n}$ and ${\Omega}_{\ell_{1} m_{1} n_{1}}$ leads to a set of doublets $S_{1}$, while each intersection between $\Omega_{\ell m n}$ and ${\Omega}_{\ell_{2} m_{2} n_{2}}$ leads to another set of doublets $S_{2}$. If the no-hair theorems hold, however, sets $S_{1}$ and $S_{2}$ must have at least one element in common. Therefore, a two-mode detection allows for tests of the no-hair theorem~\cite{Dreyer:2003bv,Berti:2005ys}. When dealing with a quasi-circular binary black hole inspiral within General Relativity, however, one knows that the dominant mode is the $\ell=m=2$ one. In such a case, the observation of this complex mode by itself allows one to extract the mass and spin angular momentum of the black hole. Then, the detection of the real frequency in an additional mode can be used to test the no-hair theorem~\cite{Berti:2005ys,Berti:2007zu}. 

Although the logic behind these tests is clear, one must study them carefully to determine whether all systematic and statistical errors are sufficiently under control so that they are feasible. Berti et al.~\cite{Berti:2005ys,Berti:2007zu} investigated such tests carefully through a frequentist approach. First, they found that a matched-filtering type analysis with two-mode ringdown templates would increase the volume of the template manifold by roughly three orders of magnitude. A better strategy then is perhaps to carry out a Bayesian analysis, like that of Gossan et al.~\cite{Kamaretsos:2011um,Gossan:2011ha}; through such a study one can determine whether a given detection is consistent with a two-mode or a one-mode hypothesis. Berti et al.~\cite{Berti:2005ys,Berti:2007zu} also calculated that a signal-to-noise ratio of ${\cal{O}}(10^{2})$ would be sufficient to detect the presence of two modes in the ringdown signal and to resolve their frequencies, so that no-hair tests would be possible. Strong signals are necessary because one must be able to distinguish at least two modes in the signal. Unfortunately, however, whether the ringdown leads to such strong signal-to-noise ratios and whether the sub-dominant ringdown modes are of a sufficiently large amplitude depends on a plethora of conditions: the location of the source in the sky, the mass of the final black hole, which depends on the rest mass fraction that is converted into ringdown gravitational waves (the ringdown efficiency), the mass ratio of the progenitor, the magnitude and direction of the spin angular momentum of the final remanent and probably also of the progenitor and the initial conditions that lead to ringdown. Thus, although such tests are possible, one would have to be quite fortunate to detect a signal with the right properties so that a two-mode extraction and a test of the no-hair theorems is feasible. 

%------------------------------------------------------------------------------------------------------------------------------
\subsubsection{The Hairy Search for Exotica}

Another way to test General Relativity is to modify the matter-sector of the theory through the introduction of matter corrections to the Einstein--Hilbert action that violate the assumptions made in the no-hair theorems. More precisely, one can study whether gravitational waves emitted by binaries composed of strange stars, like quark stars, or horizonless objects, such as boson stars or gravastars, are different from waves emitted by more traditional neutron star or black hole binaries. In what follows, we will describe such hairy tests of the existence of compact exotica.  

Boson stars are a classic example of a compact object that is essentially indistinguishable from a black hole in the weak field, but which differs drastically from one in the strong field due to its lack of an event horizon. A boson star is a coherent scalar-field configuration supported against gravitational collapse by its self-interaction. One can construct several Lagrangian densities that would allow for the existence of such an object, including mini-boson stars~\cite{Friedberg:1986tp,Friedberg:1986tq}, axially-symmetric solitons~\cite{Ryan:1996nk}, and nonsolitonic stars supported by a non-canonical scalar potential energy~\cite{Colpi:1986ye}. Boson stars are well-motivated from fundamental theory, since they are the gravitationally-coupled limit of q-balls~\cite{Coleman:1985ki,Kusenko:1998em}, a coherent scalar condensate that can be described classically as a non-topological soliton and that arises unavoidably in viable supersymmetric extensions of the standard model~\cite{Kusenko:1997zq}. In all studies carried out to date, boson stars have been studied within General Relativity, but they are also allowed in scalar-tensor theories~\cite{Balakrishna:1997ek}.

At this junction, one should point out that the choice of a boson star is by no means special; the key point here is to select a {\emph{straw-man}} to determine whether gravitational waves emitted during the coalescence of compact binaries are sensitive to the presence of an event horizon or the evasion of the no-hair theorems induced by a non-vacuum spacetime. Of course, depending on the specific model chosen, it is possible that the exotic object will be unstable to binary evolution or even to its own rotation. For example, in the case of an extreme mass-ratio inspiral, one could imagine that as the small compact object enters the boson star's surface, it will accrete scalar field, forcing the boson star to collapse into a black hole. Alternatively, one can imagine that as two supermassive boson stars merge, the remnant might collapse into a black hole, emitting the usual General Relativity quasinormal modes. What is worse, even when such objects are in isolation, they are unstable under small perturbations if their angular momentum is large, possibly leading to gravitational collapse into a black hole or possibly a scalar explosion~\cite{Cardoso:2007az,Cardoso:2008kj}. Since most astrophysical black hole candidates are believed to have high spins, such instabilities somewhat limit the interest of horizonless objects. Even then, however, the existence of slowly spinning or non spinning horizonless compact objects cannot be currently ruled out by observation. 

Boson stars evade the no-hair theorems within General Relativity because they are not vacuum spacetimes, and thus, their metric and quasinormal mode spectrum cannot be described by just their mass and spin angular momentum; one also requires other quantities intrinsic to the scalar-field energy momentum tensor, {\emph{scalar hair}}. Therefore, as before, two types of gravitational wave tests for scalar hair have been proposed: extreme mass-ratio inspiral tests and ringdown tests. As for the former, several studies have been carried out that considered a supermassive boson star background. Kesden, et al.~\cite{Kesden:2004qx} showed that stable circular orbits exist both outside and inside of the surface of the boson star, provided the small compact object interacts with the background only gravitationally. This is because the effective potential for geodesic motion in such a boson star background lacks the Schwarzschild-like singular behavior at small radius, instead turning over and allowing for a new minimum. Gravitational waves emitted in such a system would then stably continue beyond what one would expect if the background had been a supermassive black hole; in the latter case the small compact object would simply disappear into the horizon. Kesden et al.~\cite{Kesden:2004qx} found that orbits inside the boson star exhibit strong precession, exciting high frequency harmonics in the waveform, and thus allowing one to easily distinguish between such boson stars from black hole backgrounds.  

Just like the inspiral phase is modified by the presence of a boson star, the merger phase is also greatly altered, but this must be treated fully numerically. A few studies have found that the merger of boson stars leads to a spinning bar configuration that either fragments or collapses into a Kerr black hole~\cite{Palenzuela:2006wp,Palenzuela:2007dm}. Of course, the gravitational waves emitted during such a merger will be drastically different from those produced when black holes merge. Unfortunately, the complexity of such simulations makes predictions difficult for any one given example, and the generalization to other more complicated scenarios, such as theories with modified field equations, is currently not feasible.  

Recently, Pani et al.~\cite{Pani:2009ss,Pani:2010em} revisited this problem, but instead of considering a supermassive boson star, they considered a gravastar. This object consists of a Schwarzschild exterior and a de Sitter interior, separated by an infinitely thin shell with finite tension~\cite{Mazur:2001fv,Chapline:2000en}. Pani et al.~\cite{Pani:2010em} calculated the gravitational waves emitted by a stellar-mass compact object in a quasi-circular orbit around such a gravastar background. In addition to considering a different background, Pani et al used a radiative-adiabatic waveform generation model to describe the gravitational waves~\cite{Poisson:1993vp,Hughes:1999bq,Hughes:2001jr,Yunes:2009ef,2009GWN.....2....3Y,Yunes:2010zj}, instead of the kludge scheme used by Kesden et al.~\cite{Barack:2003fp,Babak:2006uv,2009GWN.....2....3Y}. Pani el al~\cite{Pani:2010em} concluded that the waves emitted during such inspirals are sufficiently different that they could be used to discern between a Kerr black hole and a gravastar. 

On the ringdown side of no-hair tests, several studies have been carried out. Berti and Cardoso~\cite{Berti:2006qt} calculated the quasi-normal mode spectrum of boson stars. Chirenti and Rezzolla~\cite{Chirenti:2007mk} studied the non-radial, axial perturbations of gravastars, and Pani, et al.~\cite{Pani:2009ss} the non-radial, axial and polar oscillations of gravastars. Medved, et al.~\cite{Medved:2003rga,Medved:2003pr} considered the quasinormal ringdown spectrum of skyrmion black holes~\cite{Shiiki:2005pb}. In all cases, it was found that the quasi-normal mode spectrum of such objects could be used to discern between them and Kerr black holes. Of course, such tests still require the detection of ringdown gravitational waves with the right properties, such that more than one mode can be discerned and extracted from the signal (see Section~\ref{sec:rd}).

%% file: musings.tex
% Nico and Xavi write this
%------------------------------------------------------------------------------------------------------------------------------
% Where do the research trends discussed point to?
%------------------------------------------------------------------------------------------------------------------------------

Gravitational waves hold the key to testing Einstein's theory of
General Relativity to new exciting levels in the previously unexplored
strong-field regime. Depending on the type of wave that is detected,
e.g., compact binary inspirals, mergers, ringdowns, continuous sources, supernovae, etc, different tests will be possible. Irrespective of the type of wave detected, two research trends seem currently to be arising: direct tests and generic tests. These trends aim at answering different questions. With direct tests, one wishes to determine whether a certain modified theory is consistent with the data. Generic tests, on the other hand, ask whether the data is statistically consistent with our canonical beliefs. Or put another way: are there any statistically significant deviations present in the data from what we expected to observe? This approach is currently used in cosmological observations, for example by the WMAP team, and it is particularly well-suited when one tries to remain agnostic as to which is the correct theory of Nature. Given that we currently have no data in the strong-field, it might be too restrictive to assume General Relativity is correct prior to verifying that this is the case. 

%------------------------------------------------------------------------------------------------------------------------------
%	What are we confident of, and what are the accomplishments?
%------------------------------------------------------------------------------------------------------------------------------

What one would like to believe is that gravitational waves will be detected by the end of this decade, either through ground-based detectors or through pulsar timing arrays. Given this, there is a concrete effort to develop the proper formalism and implementation pipelines to test Einstein's theory once data becomes available. Currently, the research groups separate into two distinct classes: theory and implementation. The theory part of the research load is being carried out at a variety of institutions without a given focal point. The implementation part is being done mostly within the LIGO Scientific Collaboration and the pulsar timing consortia. Cross-communication between the theory and implementation groups has recently flourished and one expects more interdisciplinary work in the future. 

Many accomplishments have been made in the past 50 years that it is almost impossible to list them here. From the implementation side, perhaps one of the most important ones is the actual construction and operation of the initial LIGO instruments at design sensitivity in all of its frequency domain. This is a tremendously important engineering and physics challenge. Similarly, the construction of impressive pulsar timing arrays, and the timing of these pulses to nano-second precision is an instrumental and data analysis feat to be admired. Without these observatories no waves would be detectable in the future, and of course, no tests of Einstein's theory would be feasible. On the theory side, perhaps the most important accomplishment has been the understanding of the inspiral phase to really high post-Newtonian order and the merger phase with numerical simulations. The latter, in particular, had been an unsolved problem for over 50 years, until very recently. It is these accomplishments that then allow us to postulate modified inspiral template families, since we understand what the General Relativity expectation is. This is particularly true if one is considering small deformations away from Einstein's theory, as it would be impossible to perturb about an unknown solution.  

%------------------------------------------------------------------------------------------------------------------------------
%	What are the main questions that are now at the forefront?
%------------------------------------------------------------------------------------------------------------------------------

The main questions that are currently at the forefront are the following. On the theory side of things, one would wish to understand the inspiral to high post-Newtonian order in certain strong-field modifications to General Relativity, like dynamical Chern--Simons gravity or Einstein-Dilaton-Gauss--Bonnet theory. One would also like to investigate theories with preferred frames, such as Einstein-Aether theory or Horava--Lifshitz gravity, which will lead to Lorentz violating observables. Understanding these theories to high post-Newtonian order is particularly important for those that predict dipolar gravitational emission, such as Einstein-Dilaton-Gauss--Bonnet theory. Such corrections dominate over Einstein's quadrupole emission at sufficiently low velocities. 

%The ringdown phase is also fertile ground to continue research. Many ringdown studies exist that suggest a modification of the quasi-normal mode spectrum, but such studies should be extended to strong-field modified theories. It is currently unclear whether these theories will modify all quasi-normal modes, or whether they will leave the low $\ell$ modes unmodified and correct higher order ones due to their strong-field nature. This is certainly the case for stationary solutions of strong-field modified gravity theories, but it need not be the case for the ringdown phase. Moreover, most ringdown studies have concentrated on modifying the background metric (the potential in the Teukolsky equation), with very few modifying the field equations themselves. It would be interesting to see how the ringdown modes are modified in theories that excite propagating vector and scalar modes, in addition to the usual tensor ones. Of course, this might not be feasible, as the separability of the Einstein equations is not guaranteed once the field equations are modified. 

A full inspiral-merger-ringdown template is of course not complete unless we also understand the merger. This would require full numerical simulations, which are of course very taxing even within General Relativity. Once one modifies the Einstein field equations, the characteristic structure of the evolution equations will also likely change, and it is unclear whether the standard evolution methods will continue to work. Moreover, when dealing with the merger phase, one is usually forced to treat the modified theory as exact, instead of as an effective theory. Without the latter, it is likely that certain modified theories will not have a well-posed initial value problem, which would force any numerical evolution to fail. Of course, one could order-reduce these equations and then use these to evolve black hole spacetimes. Much work still remains to be done to understand whether this is feasible. 

On the implementation side of things, there is also much work that remains to be done. Currently, efforts are only beginning on the implementation of Bayesian frameworks for hypothesis testing. This seems today like one of the most promising approaches to testing Einstein's theory with gravitational waves. Current studies concentrate mostly on single-detectors, but by the beginning of the next decade we expect 4 or 5 detectors to be online, and thus, one would like to see these implementations extended. The use of multiple detectors also opens the door to the extraction of new information, such as multiple polarization modes, a precise location of the source in the sky, etc. Moreover, the evidence for a given model increases dramatically if the event is observed in several detectors. One therefore expects that the strongest tests of General Relativity will come from leveraging the data from all detectors in a multiply coincident event. 

%------------------------------------------------------------------------------------------------------------------------------
%	In which direction(s) is/should the research field be moving?
%------------------------------------------------------------------------------------------------------------------------------

Ultimately, research is moving toward the construction of robust techniques to test Einstein's theory. A general push is currently observed toward the testing of general principles that serve as foundations of General Relativity. This allows one to answer general questions like for example: Does the graviton have a mass? Are compact objects represented by the Kerr metric and the no-hair theorems satisfied? Does the propagating metric perturbation possess only two transverse-traceless polarization modes? What is the rate of change of a binary's binding energy? Do naked singularities exist in Nature and are orbits chaotic? Is Lorentz-violation present in the propagation of gravitons? These are examples of questions that can be answered once gravitational waves are detected. The more questions of this type that are generated and the more robust the methods to answer them are, the more stringent the test of Einstein's theories and the more information we will obtain about the gravitational interaction in a previously unexplored regime.